\documentclass[12pt]{article}
\usepackage{pdflscape}
\usepackage[T1]{fontenc}
\usepackage{geometry}
\usepackage{graphicx}
\usepackage{mathtools}
\usepackage{amsmath}
\usepackage{amssymb} 
\usepackage{dsfont} 
\usepackage[small]{caption} 
\usepackage{mathrsfs}	
\usepackage{upgreek}
\usepackage{eqparbox}
\usepackage{braket}
\usepackage{xfrac}
\usepackage{xspace}
\usepackage{booktabs}
\usepackage{subcaption}
\usepackage{letltxmacro}
\usepackage{multirow}
\usepackage{xcolor}
\usepackage{pgfornament}
\usepackage{tikz}
\usepackage{cancel}
\usepackage{float}
\restylefloat{table}
\usepackage{dashrule}
\usepackage{longtable}
\usepackage{array}
\usepackage[numbers, compress, elide]{natbib}
\usepackage[no-natbib-sort]{jheppub}
\usepackage{comment}
\usepackage{soul}
\usepackage{cleveref} 
\makeatletter
\g@addto@macro\bfseries{\boldmath}
\makeatother
\allowdisplaybreaks
\newcommand{\ctw}{c_W}
\newcommand{\stw}{s_W}

\newcommand{\four}{\mathbf{4}}
\newcommand{\three}{\mathbf{3}}

\newcommand{\ellA}{(\lambda^A)_{\alpha}^{\;\beta}}
\newcommand{\fourptamp}[4]{\mathcal{A}\left({#1}{#2}{#3}{#4}\right)}

\newcommand{\SU}[1]{\ensuremath{\mathrm{SU}(#1)}}
\newcommand{\bs}[1]{\boldsymbol{#1}}
\newcommand{\sqb}[1]{[#1]}
\newcommand{\anb}[1]{\langle #1\rangle}
\newcommand{\sab}[1]{[#1\rangle}
\newcommand{\asb}[1]{\langle #1]}
\newcommand{\Sqb}[1]{[\bs{#1}]}
\newcommand{\Anb}[1]{\langle\bs{#1}\rangle}

\newcommand{\sQbl}[2]{[\bs{#1} #2]}
\newcommand{\sQbr}[2]{[#1 \bs{#2}]}
\newcommand{\aNbl}[2]{\langle\bs{#1} #2\rangle}
\newcommand{\aNbr}[2]{\langle #1 \bs{#2}\rangle}
\newcommand{\Eqref}[1]{Equation~\eqref{#1}}
\newcommand{\beq}{\begin{equation}}
\newcommand{\eeq}{\end{equation}}
\def\beqa{\begin{eqnarray}}
\def\eeqa{\end{eqnarray}} 

\newcommand{\Cone}{c_1^{2W_+2W_-}}
\newcommand{\Conetwo}{c_2^{2W_+2W_-}}
\newcommand{\Ctwo}{c^{2B_+2B_-}}
\newcommand{\Cthree}{c^{2W_+2B_-}}
\newcommand{\Cfour}{c^{W_+W_-B_+B_-}}
\newcommand{\Cfive}{c_1^{4W_+}}
\newcommand{\Csix}{c_2^{4W_+}}
\newcommand{\Csev}{c^{4B_+}}
\newcommand{\Ceight}{c_1^{2W_+2B_+}}
\newcommand{\Ceightwo}{c_2^{2W_+2B_+}}
\newcommand{\Cnine}{c_+^{2W_+2H}}
\newcommand{\Cten}{c_-^{2W_+2H}}
\newcommand{\Celeven}{c^{2B_+2H}}
\newcommand{\Ctwelve}{c_1^{W_+B_+2H}}
\newcommand{\Cthirteen}{c_2^{W_+B_+2H}}
\newcommand{\Cfourteen}{c_1^{W_+W_-2H}}
\newcommand{\Cfifteen}{c_2^{W_+W_-2H}}
\newcommand{\Csixteen}{c^{B_+B_-2H}}
\newcommand{\Cseventeen}{c^{W_+B_-2H}}
\newcommand{\Ceighteen}{c_{+,1}^{4H}}
\newcommand{\Cnineteen}{c_{+,2}^{4H}}
\newcommand{\Ctwenty}{c_-^{4H}}
\newcommand{\Ctwentyone}{c_1^{2g_+2W_+}}
\newcommand{\Ctwentytwo}{c_2^{2g_+2W_+}}
\newcommand{\Ctwentythree}{c^{2g_+2W_-}}
\newcommand{\Ctwentyfour}{c^{g_+g_-W_+W_-}}
\newcommand{\Ctwentyfive}{c^{2g_+2H}}
\newcommand{\Ctwentysix}{c^{g_+g_-2H}}
\newcommand{\Ctwentyseven}{c_1^{2g_+2B_+}}
\newcommand{\Ctwentyseventag}{c_2^{2g_+2B_+}}
\newcommand{\Ctwentyeight}{c^{2g_+2B_-}}
\newcommand{\Ctwentynine}{c^{g_+g_-B_+B_-}}
\newcommand{\Cthirtytwo}{c^{g_+g_-2H}}
\newcommand{\Cthirtyone}{c^{2g_+g_-B_-}}
\newcommand{\Cthirty}{c^{3g_+B_+}}
\newcommand{\Cfone}{c_1^{Q^cUW_+H}}
\newcommand{\Cftwo}{c_2^{Q^cUW_+H}}
\newcommand{\Cfthree}{c^{Q^cUW_-H}}
\newcommand{\Cffour}{c_1^{Q^cUB_+H}}
\newcommand{\Cffourtag}{c_2^{Q^cUB_+H}}
\newcommand{\Cffive}{c^{Q^cUB_-H}}
\newcommand{\Cfsix}{c_1^{Q^cUg_+H}}
\newcommand{\Cfsixtwo}{c_2^{Q^cUg_+H}}
\newcommand{\Cfsixtag}{c^{Q^cUg_-H}}
\newcommand{\Cfseven}{c_1^{Q^cDW_+H}}
\newcommand{\Cfseventwo}{c_2^{Q^cDW_+H}}
\newcommand{\Cfseventag}{c^{Q^cDW_-H}}
\newcommand{\Cfeight}{c_1^{Q^cDB_+H}}
\newcommand{\Cfeighttag}{c_2^{Q^cDB_+H}}
\newcommand{\Cfnine}{c^{Q^cDB_-H}}
\newcommand{\Cften}{c_1^{Q^cDg_+H}}
\newcommand{\Cftentag}{c_2^{Q^cDg_+H}}
\newcommand{\Cfeleven}{c^{Q^cDg_-H}}

\newcommand{\Cffifteen}{c^{Q^cQ2W_+}}
\newcommand{\Cfsixteen}{c^{Q^cQW_+W_-}_+}
\newcommand{\Cfseventeen}{c^{Q^cQW_+W_-}_-}
\newcommand{\Cfeighteen}{c^{Q^cQB_+B_-}}

\newcommand{\Cftwenty}{c_1^{Q^cQg_+g_-}}

\newcommand{\Cftwentythree}{c^{Q^cQg_+b_+}}
\newcommand{\Cftwentyfour}{c^{Q^cQg_+B_-}}
\newcommand{\Cftwentyfive}{c^{Q^cQg_+W_+}}
\newcommand{\Cftwentysix}{c^{Q^cQg_+W_-}}
\newcommand{\Cftwentyseven}{c^{Q^cQW_+B_+}}
\newcommand{\Cftwentyseventag}{c^{Q^cQW_+B_-}}
\newcommand{\Cftwentyeight}{c^{U^cUW_+W_-}}
\newcommand{\Cftwentynine}{c^{U^cUB_+B_-}}

\newcommand{\Cfthirtyfour}{c^{U^cUg_+B_+}}
\newcommand{\Cfthirtyfive}{c^{U^cUg_+B_-}}
\newcommand{\Cfthirtysix}{c^{Q^cU3H}_1}
\newcommand{\Cfthirtyseven}{c^{Q^cU3H}_2}
\newcommand{\Cfthirtyeight}{c^{Q^cD3H}_1}
\newcommand{\Cfthirtynine}{c^{Q^cD3H}_2}
\newcommand{\Cfforty}{c^{Q^cQW_+2H}_1}
\newcommand{\Cffortyone}{c^{Q^cQW_+2H}_2}
\newcommand{\Cffortytwo}{c^{Q^cQW_+2H}_3}
\newcommand{\Cffortysix}{c^{Q^cQW_+2H}_4}
\newcommand{\Cffortyseven}{c^{Q^cQW_+2H}_5}
\newcommand{\Cffortyeight}{c^{Q^cQW_+2H}_6}
\newcommand{\Cffiftytwo}{c^{Q^cQB_+2H}_1}
\newcommand{\Cffiftythree}{c^{Q^cQB_+2H}_2}
\newcommand{\Cffiftyfour}{c^{Q^cQB_+2H}_3}
\newcommand{\Cffiftyfive}{c^{Q^cQB_+2H}_4}
\newcommand{\Cffiftysix}{c^{Q^cQg_+2H}_1}
\newcommand{\Cffiftyseven}{c^{Q^cQg_+2H}_2}
\newcommand{\Cffiftyeight}{c^{Q^cQg_+2H}_3}
\newcommand{\Cffiftynine}{c^{Q^cQg_+2H}_4}
\newcommand{\Cfsixty}{c^{U^cUW_+2H}_1}
\newcommand{\Cfsixtyone}{c^{U^cUW_+2H}_2}
\newcommand{\Cfsixtytwo}{c^{U^cUB_+2H}_1}
\newcommand{\Cfsixtythree}{c^{U^cUB_+2H}_2}
\newcommand{\Cfsixtytwotag}{c^{U^cUg_+2H}_1}
\newcommand{\Cfsixtythreetag}{c^{U^cUg_+2H}_2}

\newcommand{\Cfsixtyfive}{c^{U^cDW_+2H}_+}
\newcommand{\Cfsixtysix}{c^{U^cDB_+2H}}
\newcommand{\Cfsixtyseven}{c^{U^cDg_+2H}}
\newcommand{\Cfseventy}{c^{QU2W_+H}_1}
\newcommand{\Cfseventyone}{c^{QU2W_+H}_2}
\newcommand{\Cfseventytwo}{c^{QU2W_-H}}
\newcommand{\Cfseventythree}{c^{QUW_+B_+H}_1}
\newcommand{\Cfseventyfive}{c^{QUW_+B_+H}_2}
\newcommand{\Cfseventyseven}{c^{QUW_-B_-H}_1}
\newcommand{\Cfseventyeight}{c^{QUW_-B_-H}_2}
\newcommand{\Cfseventynine}{c^{QU2B_+H}}
\newcommand{\Cfeighty}{c^{QU2B_-H}}

\newcommand{\Cfeightysix}{c^{QU2W_+g_+H}_1}
\newcommand{\Cfeightyeight}{c^{QU2W_+g_+H}_2}
\newcommand{\Cfninety}{c^{QU2W_-g_-H}_1}

\newcommand{\Cfninetytwo}{c_1^{QU2B_+g_+H}}
\newcommand{\Cfninetytwotag}{c_2^{QU2B_+g_+H}}
\newcommand{\Cfninetythree}{c^{QU2B_-g_-H}}
\newcommand{\Cfninetyfour}{c_1^{Q^cQ2H}}
\newcommand{\Cfninetyfive}{c_2^{Q^cQ2H}}
\newcommand{\Cfninetyfourtag}{c_3^{Q^cQ2H}}
\newcommand{\Cfninetyfivetag}{c_4^{Q^cQ2H}}
\newcommand{\Cfninetysix}{c^{U^cD2H}}
\newcommand{\Cfninetyseven}{c^{D^cDW_+W_-}}
\newcommand{\Cfninetyeight}{c^{D^cDB_+B_-}}

\newcommand{\Cfhundredthree}{c^{D^cDg_+B_+}}
\newcommand{\Cfhundredfour}{c^{D^cDg_+B_-}}
\newcommand{\Cfhundredfive}{c^{D^cDW_+2H}_1}
\newcommand{\Cfhundredsix}{c^{D^cDW_+2H}_2}
\newcommand{\Cfhundredseven}{c^{D^cDB_+2H}_1}
\newcommand{\Cfhundredeight}{c^{D^cDB_+2H}_2}
\newcommand{\Cfhundrednine}{c^{D^cDg_+2H}_1}
\newcommand{\Cfhundredten}{c^{D^cDg_+2H}_2}
\newcommand{\Cfhundredeleven}{c^{QD2W_+H}_1}
\newcommand{\Cfhundredtwelve}{c^{QD2W_+H}_2}
\newcommand{\Cfhundredthirteen}{c^{QD2W_-H}}
\newcommand{\Cfhundredfourteen}{c^{QDW_+B_+H}_1}
\newcommand{\Cfhundredfifteen}{c^{QDW_+B_+H}_2}
\newcommand{\Cfhundredsixteen}{c^{QDW_-B_-H}}
\newcommand{\Cfhundredseventeen}{c^{QD2B_+H}}
\newcommand{\Cfhundredeighteen}{c^{QD2B_-H}}

\newcommand{\Cfhundredtwentyfour}{c^{QDW_+g_+H}_1}
\newcommand{\Cfhundredtwentyfive}{c^{QDW_+g_+H}_2}
\newcommand{\Cfhundredtwentysix}{c^{QDW_-g_-H}}
\newcommand{\Cfhundredtwentyseven}{c^{QDB_+g_+H}_1}
\newcommand{\Cfhundredtwentyseventag}{c_2^{QDB_+g_+H}}
\newcommand{\Cfhundredtwentyeight}{c^{QDB_-g_-H}}

\newcommand{\Cfhundredthirty}{c^{D^cUW_+2H}_+}
\newcommand{\Cfhundredthirtyone}{c^{D^cUB_+2H}}
\newcommand{\Cfhundredthirtytwo}{c^{D^cUg_+2H}}
\newcommand{\Cfthirtyeighttag}{c_3^{Q^cD3H}}
\newcommand{\Cfthirtyninetag}{c_4^{Q^cD3H}}
\newcommand{\Cfthirtyeighttwo}{c_5^{Q^cD3H}}
\newcommand{\Cfthirtyninetwo}{c_6^{Q^cD3H}}
\newcommand{\Cfthirtysixtag}{c_3^{Q^cU3H}}
\newcommand{\Cfthirtyseventag}{c_4^{Q^cU3H}}
\newcommand{\Cfthirtysixtagtwo}{c_5^{Q^cU3H}}
\newcommand{\Cfthirtyseventagtwo}{c_6^{Q^cU3H}}
\newcommand{\Cfhundredthirtythree}{c_1^{U^cU2H}}
\newcommand{\Cfhundredthirtyfour}{c_2^{U^cU2H}}
\newcommand{\Cfhundredthirtyfive}{c_1^{D^cD2H}}
\newcommand{\Cfhundredthirtysix}{c_2^{D^cD2H}}
\newcommand{\Ctwohundred}{(\Cthree)^*}
\newcommand{\Ctwohundredone}{(\Cseventeen)^*}
\title{Dimension-8 SMEFT contact-terms for vector-pair production via on-shell Higgsing}
\author[a]{Jared~M.~Goldberg,}
\author[a]{Hongkai~Liu,}
\author[a]{Yael~Shadmi}
\affiliation[a]{Physics Department, Technion - Israel Institute of Technology, Haifa 3200003, Israel}
\emailAdd{jmg116@campus.technion.ac.il}
\emailAdd{liu.hongkai@campus.technion.ac.il}
\emailAdd{yshadmi@physics.technion.ac.il}
\abstract{
We derive the dimension-8 standard-model effective theory (SMEFT) contact terms relevant for vector-pair production at the LHC and lepton colliders.
 We first list the relevant dimension-8 massless SMEFT amplitudes, and then obtain the low-energy amplitudes using on-shell Higgsing. 
 In all cases, the contributions we calculate are the leading-order contributions to 4-point contact-terms; the dimension-6 SMEFT merely corrects the three-point couplings entering the amplitudes. 
 Since they are given in terms of physical quantities, namely momenta and polarizations, the results
 allow for a direct mapping of EFT effects to low-energy observables.
 The vector amplitudes are sensitive to both anomalous vector couplings and Higgs self-couplings.
 The left-handed fermion amplitudes feature SU(2) violating effects first generated at dimension-8.
 We also compare our results to HEFT predictions. 
 Interestingly, the dimension-8 SMEFT populates almost all the novel structures generated by the dimension-8
 HEFT. 
}
\begin{document}
%%%%%%%%%%%%%%%%%%%%%%%%%%%%%%%%%%%%%%%%%%%%%%%%%%%%%%%%%%%%%%%%%%%%%%%
\titlepage
\maketitle
%%%%%%%%%%%%%%%%%%%%%%%%%%%
\flushbottom
%%%%%%%%%%%%%%%%%%%%%%%%%%%
%%%%%%%%%%%%%%%%%%%%%%%%%%%%%%%%%%%%%%%%%%%%%%%%%%%%%%%
\section{Introduction}
%%%%%%%%%%%%%%%%%%%%%%%%%%%%%%%%%%%%%%%%%%%%%%%%%%%%%%%
The LHC experiments are gradually confirming the details of the standard-model Higgs mechanism~\cite{ATLAS:2022vkf,CMS:2022dwd}. 
This mechanism simply and elegantly accounts for electroweak symmetry breaking (EWSB),  
but its  essential features---the symmetry-breaking minimum, the origin of the 246~GeV scale, 
and the stability of this scale against radiative corrections---are left unexplained, 
motivating new particles near the TeV, and possibly somewhat above the direct reach of the LHC.
The low-energy (LE) signatures of such new particles can be parameterized in the
Standard Model Effective Theory (SMEFT) framework~\cite{Buchmuller:1985jz,  Jenkins:2009dy, Grzadkowski:2010es}, 
which is usually formulated in terms of higher-dimension operators
that the new particles induce. There are many independent SMEFT operators at 
dimension-six~\cite{Grzadkowski:2010es},
and the number of dimension-8 operators 
is larger by roughly an order of magnitude~\cite{Jenkins:2009dy,Lehman:2015coa, Henning:2015alf}.
Over the next decades, the LHC experiments will supply a wealth of data on electroweak
processes. 
Interpreting the data in the EFT framework amounts to a mapping between EFT operators and observables.
This is a difficult task owing to the large dimensionalities of the spaces of operators and observables,
with multiple operators affecting a single observable, and conversely, 
a single operator affecting multiple observables.
Operators are furthermore altered by field redefinitions
which leave the physics unchanged, and different operator bases can be better suited
to describe different sets of  observables or UV models~\cite{Giudice:2007fh,Gupta:2014rxa,Falkowski:2015wza}. 

To streamline the mapping, the EFT can be formulated directly in terms of physical 
observables --- namely the scattering amplitudes of the physical SM particles.
In the on-shell formulation, new heavy physics manifests itself as new 
contact-interactions among SM particles, with a one-to-one correspondence between 
contact terms (CTs) and EFT operators~\cite{Shadmi:2018xan,Ma:2019gtx,Aoude:2019tzn,Durieux:2019eor,Durieux:2019siw,
Durieux:2020gip,Falkowski:2020fsu,AccettulliHuber:2021uoa,Balkin:2021dko,DeAngelis:2022qco,
Chang:2022crb,Liu:2023jbq,Bradshaw:2023wco,Arzate:2023swz,
Bertuzzo:2023slg}. 
Writing the most general amplitude consistent with the assumed symmetries, one obtains a 
parameterization of all possible new physics effects.
Indeed, the simplicity of scattering amplitudes 
has been extensively utilized recently to derive various results
on EFTs and EFT Lagrangians, including the construction of operator bases~\cite{Li:2020gnx,Li:2020xlh,Low:2022iim,Sun:2022ssa,Sun:2022snw,Harlander:2023psl}, deriving their anomalous dimensions~\cite{AccettulliHuber:2021uoa,Cheung:2015aba,EliasMiro:2020tdv,Baratella:2020lzz,Bern:2020ikv,Jiang:2020mhe,Jin:2020pwh,Baratella:2021guc,EliasMiro:2021jgu,Baratella:2022nog,Machado:2022ozb,Chala:2023xjy,Bresciani:2023jsu}, 
and matching to specific UV models~\cite{Carrasco:2019qwr,DeAngelis:2023bmd,Li:2023pfw}.

In this work, we focus on processes of central importance for EWSB, namely vector-pair production 
at the LHC or future lepton colliders,
and present the relevant LE SMEFT amplitude basis.
We assume baryon and lepton number conservation.  
The required amplitudes are of two types: four-vector amplitudes, which we refer to as $VVVV$, and two-fermion two-vector amplitudes,
which we label as $ffVV$ and in which we take all fermions to be massless. 
In each case, the 4-point CTs we obtain are first generated at dimension-8 
in the SMEFT.
This is easy to see using dimensional analysis. 
The amplitudes can be written in terms of
massless spinors of mass dimension 1/2. An external vector polarization is written as a product of two spinors, 
while an external fermion corresponds to a single spinor. 
Thus, $VVVV$ amplitudes involve spinor structures  of dimension $\geq4$, and
since  4-point amplitudes are dimensionless,
these come with a $\Lambda^{-4}$ or higher  suppression.
The $ffVV$ amplitudes involve spinor structures of dimension $\geq3$. Since the (B- and L-conserving) SMEFT
only features even powers of $\Lambda$, the $ffVV$ CTs also appear at order $\Lambda^{-4}$~\cite{Degrande:2012wf}. In contrast, at dimension-6, the only effect of the SMEFT  on the LE $VVVV$ and $ffVV$ 
amplitudes is to  shift the 3-point SM couplings\footnote{The
$d=6$ SMEFT also generates new 3-point couplings, namely the dipole couplings, but these do not interfere 
with the SM amplitudes. Recall that we neglect fermion masses.}. Thus, our results are the leading SMEFT effects that can generate novel kinematic structures in vector-pair production. The effects of dimension-8 SMEFT operators were discussed by many authors, see for example~\cite{Davidson:2003ha,Degrande:2013kka,Berthier:2015oma,Berthier:2015gja,Ellis:2019zex,Alioli:2020kez,Hays:2020scx,Corbett:2021eux,Boughezal:2021tih,Dawson:2021xei,Martin:2021cvs}.

The dimension-six SMEFT predictions for all SM 3-points were presented in~\cite{Durieux:2019eor}, and for all SM 4-point CTs
in~\cite{Liu:2023jbq}, whose results we extend here to dimension-8. The only 4-point CTs generated by the dimension-six SMEFT are four fermion CTs and CTs involving Higgs legs.
To derive the LE CTs spanning a given LE amplitude, we use on-shell Higgsing~\cite{Balkin:2021dko}.
We first identify the dimension-8 SMEFT amplitudes of the unbroken theory that feed into the LE amplitude.
These are massless amplitudes with the full SU(3)$\times$SU(2)$\times$U(1) symmetry.
For each of these, we write down a basis of massless CTs.
Bases of dimension-8 SMEFT operators were derived in~\cite{Li:2020gnx, Murphy:2020rsh},
and our massless SMEFT CTs match the counting in these references.
We then ``Higgs" the massless CTs into massive CTs following~\cite{Balkin:2021dko}. The basic elements of this Higgsing are reviewed in Section~\ref{boson}.
The result is a set of independent CTs contributing to the LE amplitude, written in terms of
the little group-covariant massive spinors of Ref.~\cite{Arkani-Hamed:2017jhn}.
Concretely, each CT is just a complex number, and depends on the external particle momenta
and heavy boson polarizations.
We also discuss the CP properties of the CTs.  These can be easily read off from the structure of the amplitude.

One can alternatively construct  generic $VVVV$
and $ffVV$ amplitudes imposing just the  SU(3)$\times$U(1) symmetry of the LE theory~\cite{Shadmi:2018xan,Aoude:2019tzn,Durieux:2019eor,
Chang:2022crb,Liu:2023jbq,Bradshaw:2023wco,Arzate:2023swz}. 
Such on-shell constructions correspond to the EFT framework known as the Higgs EFT (HEFT)~\cite{Feruglio:1992wf, Bagger:1993zf, Koulovassilopoulos:1993pw, Burgess:2000kt, Grinstein:2007iv, Alonso:2012px,Espriu:2013fia,Buchalla:2013rka,Brivio:2013pma,Alonso:2015fsp, Alonso:2016oah,Buchalla:2017jlu,Alonso:2017tdy,deBlas:2018tjm,Cohen:2020xca}. 
In fact, the on-shell construction gives a concrete and clear
formulation of the HEFT, 
with an unambiguous prescription for identifying the dimensions
of HEFT operators~\cite{Liu:2023jbq}.
The full set of dimension-8 HEFT 4-point CTs was derived in~\cite{Liu:2023jbq}.
We compare the SMEFT and HEFT predictions for the LE CTs.
Obviously, the SMEFT implies various relations among the Wilson coefficients
of the CTs. Somewhat surprisingly, most of the structures generated
by the HEFT are also generated in the dimension-8 SMEFT.

This paper is organized as follows. 
In Section~\ref{boson} we briefly review the on-shell construction of massless contact terms
and their Higgsing, and use this to derive the LE $VVVV$ CTs.
We also comment on the CP properties of the CTs.
Our results for the CTS are relegated to 
Appendix~\ref{tablesApp}. In Section~\ref{fermions} we repeat this for two-fermion two-vector amplitudes. 
In Section~\ref{Discussion}, we discuss the results and compare them to the dimension-8 CTs generated in 
the HEFT. We conclude in Section~\ref{conclusions}.
Appendix~\ref{tablesApp} contains our results for the massless and massive CTs. 
The massless bosonic SMEFT CTs are given in Table~\ref{table:vvvv4}, and the LE bosonic CTs are listed in Table~\ref{T1}. The massless fermionic CTs are given in Table~\ref{Tferm},
and the corresponding LE  CTs are listed in Table~\ref{ChiYes}.
In Appendix~\ref{AppCount}, we give further details and examples of the relations between CTs following from complex conjugation and CP. 
For completeness, we list the chirality violating fermionic amplitudes in Appendix~\ref{AppCVA}. 
%%%%%%%%%%%%%%%%%%%%%%%%%%%%%%%%%%%%%%%%%%%%%%%%%%%%%%%
\section{Four-vector amplitudes}
\label{boson}
%%%%%%%%%%%%%%%%%%%%%%%%%%%%%%%%%%%%%%%%%%%%%%%%%%%%%%%
\subsection{Basics of the construction}
%%%%%%%%%%%%%%%%%%%%%%%%%%%%%%%%%%%%%%%%%%%%%%%%%%%%%%%
We aim to derive bases for the LE massive CTs with vector boson final states at dimension-8 in the SMEFT. 
In this section, we will construct the four-vector amplitudes, and in the next section we will construct two-fermion two-vector amplitudes. We start with the massless SMEFT CTs and Higgs them to obtain the LE amplitudes as described in Refs.~\cite{Balkin:2021dko,Liu:2023jbq}. We review the essentials of the construction here and refer the reader to Refs.~\cite{Durieux:2020gip,Balkin:2021dko,Liu:2023jbq} for more detail. 
Our conventions for massive spinors follow~\cite{Durieux:2019eor}\footnote{
For details of the massive spinor formalism of~\cite{Arkani-Hamed:2017jhn},
the high-energy limit and
and explicit expressions for the spinors, we refer the reader to the Appendices of refs.~\cite{Shadmi:2018xan,Durieux:2019eor}.}.

The LE four-vector amplitudes are obtained from the massless $4+n_H$ amplitudes of the SMEFT, with four legs of external vectors, namely gluons and electroweak gauge bosons, and/or Higgses,
and $n_H$ additional Higgs legs. 
The massless high-energy (HE) CTs can be assembled from Mandelstam invariants and Stripped CTs (SCTs)~\cite{Durieux:2020gip}. Each SCT is a product of positive powers of massless spinor contractions, with no additional powers of Mandelstams.
 An SCT basis for a given massless  amplitude with particles of nonzero spin is a set of independent SCTs, such that any other SCT is given as a linear combination of the SCT basis elements, with coefficients that are polynomials in the Mandelstam
invariants. These SCT basis elements,
multiplied by positive powers of the Mandelstams, span the contact term part of the amplitude which is 
therefore manifestly local\footnote{In some cases,
SCTs multiplied by powers of Mandelstams may become redundant at higher dimensions~\cite{Durieux:2020gip}. 
This will not occur in the amplitudes constructed here.}.
Obviously, massless scalar CTs are spanned by positive powers of Mandelstam invariants only.

With the massless contact-term basis in hand, the construction of the massive basis proceeds as follows~\cite{Balkin:2021dko}:
\begin{itemize}
    \item Massless spinor structures are bolded into massive spinor structures. A LE vector external leg can come
    either from a massless vector leg or from a massless scalar leg. In the latter case,
    the massless CT must involve the scalar momentum $p$. The LE amplitude is
    obtained by replacing
    $p\to \bf{p\rangle[p}$.
    \item The $n_H$ Higgs momenta are taken to zero, with each giving a factor of the Higgs vacuum expectation value (VeV)~$v$.
\end{itemize}
Note that the polynomials in the Mandelstam invariants multiplying the SCTs correspond to the derivative expansion.
Amplitudes with $n_H>0$ Higgs legs generate  $v/\Lambda$ corrections to the LE Wilson coefficients. 
As we will see, in the massless SMEFT four-vector amplitudes of this section, $n_H=0$ at dimension-8. Thus the Wilson coefficients in these LE amplitudes
are order $v^0$ at dimension-8. In the $ffVV$ amplitudes of the next section, $n_H>0$ is allowed at dimension-8, and we
will see $v/\Lambda$ corrections to the Wilson coefficients.
%%%%%%%%%%%%%%%%%%%%%%%%%%%%%%%%%%%%%%%%%%%%%%%%%%%%%%%
\subsection{Massless contact terms}\label{sec:masslesscts}
%%%%%%%%%%%%%%%%%%%%%%%%%%%%%%%%%%%%%%%%%%%%%%%%%%%%%%%
The relevant 4-point SCTs are~\cite{Durieux:2019siw},
\beqa
\label{eq:massless4}
A_4(V_+,V_+,V_+,V_+) &=& \sqb{12}^2\sqb{34}^2,\,\sqb{13}^2\sqb{24}^2,
\quad(d=8),\nonumber\\
A_4(V_-,V_-,V_-,V_-) &=& \anb{12}^2\anb{34}^2,\,\anb{13}^2\anb{24}^2,
\quad(d=8),\nonumber \\
A_4(V_+,V_+,V_-,V_-) &=& \sqb{12}^2\anb{34}^2,\quad(d=8),\nonumber\\
A_4(V_+,V_-,s,s) &=& \sab{132}^2,\quad(d=8),\nonumber\\
A_4(V_+,V_+,s,s) &=& \sqb{12}^2,\quad(d=6),\nonumber\\
A_4(V_-,V_-,s,s) &=& \anb{12}^2,\quad(d=6),\nonumber\\
A_4(s,s,s,s) &=& \text{constant},\quad(d=4),
\eeqa
where $V$ stands for a generic vector,  $s$ stands for a generic scalar, the helicity of the particle is indicated by the $\pm$ subscript, and $d$ in the parentheses is the dimension at which the structure is generated.

Higher-point amplitudes do not generate additional  LE 4-vector SCTs. 
The only ones with even numbers of scalars at $d\leq8$ are~\cite{Durieux:2019siw},
\beqa
A_5(V_+,V_+,V_+,s,s) &=& \sqb{12}\sqb{13}\sqb{23},\quad(d=8), \nonumber\\
A_5(V_+,s,s,s,s) &=& \sqb{1341},\,\sqb{1241},\,\sqb{1231},\quad(d=8),\nonumber\\
A_6(V_+,V_+,s,s,s,s) &=& \sqb{12}^2,\quad(d=8)\nonumber\\
A_6(s,s,s,s,s,s) &=& \text{constant},\quad(d=6). 
\eeqa
To bold these  into four-vector amplitudes requires additional powers of the Mandelstam invariants, 
and/or additional momentum insertions, which only arise at $d>8$. 
Thus, only the four-point amplitudes in Eq.~(\ref{eq:massless4}) feed into the massive four vector amplitudes.
As a result, at dimension-8, there are no corrections to the Wilson coefficients proportional to the Higgs VeV.
These require additional soft Higgs legs, and are first generated at $d=10$.

The relevant massless gauge boson and Higgs contact-term amplitudes are assembled from the SCTs of Eq.~(\ref{eq:massless4}),
multiplied by the appropriate group theory structures, and symmetrized over identical bosons. 
We collect these in Table~\ref{table:vvvv4}, with the spinor and group-theory structures
shown in the second column, and the corresponding Wilson coefficients in the third column. 
These Wilson coefficients will enter the LE amplitudes via the on-shell Higgsing. Note that massless CTs with Higgs legs only contribute to the LE amplitudes
of interest if they feature the Higgs momenta; otherwise, they cannot be bolded into vector CTs.
Furthermore, since they vanish as the Higgs momentum goes to zero, these contact
terms do not generate any LE 3-point CTs
which could feed into the factorizable part of the amplitude.
Indeed, the distinction between Goldstone CTs and Higgs CTs is simply encoded
in the amplitude. 
The 4-point (or higher point) CTs with an $H$ leg of momentum $p_H$
can be divided into two groups.
The first feature $p_H$ and go to zero as $p_H\to0$.
These correspond to Goldstone CTs, and do not generate any 3-point CTs.
The second type of CTs do not vanish at $p_H\to0$.
These map to LE 4-point CTs with a physical
Higgs leg $h$, and furthermore give rise to 3-point CTs obtained by taking $p_H\to0$.\footnote{See discussion in Appendix~A of~\cite{Balkin:2021dko}.}
Only the Goldstone CTs are relevant for our analysis.
We have checked that the number of independent massless CTs matches the counting of  dimension-8 SMEFT operators in
Refs.~\cite{Li:2020gnx, Remmen:2019cyz, Murphy:2020rsh}. 
%%%%%%%%%%%%%%%%%%%%%%%%%%%%%%%%%%%%%%%%%%%%%%%%%%%%%%%%%%%%%%%%%%%%%%%%%%%%%%%%%%%%%%%%%%%%%%%%%%%%%%%%%%%%%%%%%%%%%%%%%%%%%%%%%
\subsection{Massive contact terms}
%%%%%%%%%%%%%%%%%%%%%%%%%%%%%%%%%%%%%%%%%%%%%%%%%%%%%%%
To obtain the LE CTs, the results of Table~\ref{table:vvvv4} are first written 
in terms of the 
broken phase fields---$W^\pm$, $Z$, $\gamma$, $g$, and the Goldstone bosons.
This involves the Weinberg mixing angle, which at dimension-8 can be taken to be just the SM value. 
Finally, the spinor structures are bolded into massive spinor structures.
The physical dimension-8 EFT amplitude is then given as a sum of these contributions.
It is useful to label the different terms by their helicity category --  
the four external helicities of the source HE amplitude.
The results are listed in Table~\ref{T1}. 
This Table, along with Table~\ref{ChiYes}, contains the main results of this work, 
namely the LE four-point CTs derived from the dimension-8 SMEFT.
%%%%%%%%%%%%%%%%%%%%%%%%%%%%%%%%%%%%%%%%%%%%%%%%%%%%%%%
\subsection{Independent parameters and CP}\label{sec:cp}
%%%%%%%%%%%%%%%%%%%%%%%%%%%%%%%%%%%%%%%%%%%%%%%%%%%%%%%
%%%%%%%%%%%%%%%%%%%%%%%%%%%%%%%%%%%%%%%%%
Let us now discuss the properties of the CTs under complex conjugation and CP transformations.
Consider a CT contributing to the $n$-point amplitude $M(1,..,n)$.
We can write the CT as $c\, S$, where $c$ is a Wilson coefficient and $S$ is a 
combination of spinor products and group theory structures.
Complex conjugation yields the structure $c^* S^*$,
which is a CT of the amplitude with particles 
replaced by antiparticles, $M(1^c,..,n^c)$. 
Since complex conjugation flips angle 
and square spinor products, with $[ij]^*=\anb{ji}$, the two amplitudes are related by the product of parity and charge conjugation.
In some cases, the complex-conjugated structure $S^*$ is the same
as $S$, possibly after a relabeling of the external momenta.
The coefficient $c$ is then either real or pure imaginary,
and corresponds to the coefficient of a Hermitian
(or anti-Hermitian) operator.
Otherwise, complex conjugation yields a new CT,
with the 
Wilson coefficient $c^*$. 
For brevity, we do not show these CTs in Tables~\ref{table:vvvv4} and \ref{Tferm}.
In particular, their Wilson coefficients can be obtained
from the Wilson coefficients listed in the Tables by complex conjugation.
For example, for the $W^-W^-W^+W^+$ amplitude of Table~\ref{T1}, there are two CTs in the  $++++$ helicity category,
\beqa\label{cpexample1}
  2c_2^{4W_+}\,  \Sqb{12}^2\Sqb{34}^2 ~~\text{and} ~~\left( c_1^{4W_+}+c_2^{4W_+}\right)\,\left(\Sqb{13}^2\Sqb{24}^2 + \Sqb{14}^2\Sqb{23}^2 \right).
\eeqa
The complex conjugates of these structures,
\beqa\label{cpexample2}
   2(c_2^{4W_+})^*\, \Anb{12}^2\Anb{34}^2 ~~\text{and} ~~~ \left(c_1^{4W_+}+c_2^{4W_+}\right)^*\,  \left(\Anb{13}^2\Anb{24}^2 + \Anb{14}^2\Anb{23}^2
  \right),
\eeqa
are not shown in~Table~\ref{T1}.
Similarly, $\Sqb{12}^2 \Anb{34}^2$ appears in Table~\ref{T1} with the coefficient $c_2^{2W_+2W_-}$,
but its complex conjugate $\Anb{12}^2 \Sqb{34}^2$ is not shown.
In this case however, the two HE CTs which generate these structures are equivalent after a relabeling of the external particles, so
$c_2^{2W_+2W_-}$ is real.

Under a CP transformation, the amplitude $M(1,..,n) = c\, S\to c\, S^*$. 
Therefore, if CP is conserved, the coupling $c$ must be real. Of course, a complex  $c$ does not immediately imply CP violation, since its phase may be eliminated by redefinitions of the 
external spinors, just as in the Lagrangian case. 
The CTs of~Eqs.~(\ref{cpexample1}) and (\ref{cpexample2}) for instance,  are related
by a CP transformation.
In a CP-invariant theory, the coefficients of these CTs would be real, and therefore identical. 
We will not study CP violation further in this work.

Note that the CTs of Tables~\ref{table:vvvv4} and \ref{Tferm} are helicity amplitudes,
and therefore do not form bases of CP eigenstates.
 In terms of operators, each external vector
leg of definite helicity corresponds to a self-dual or anti-self-dual field strength, $V_{\mu\nu}\pm i\tilde V_{\mu\nu}$, which are exchanged by parity (and CP). Schematically then, the CTs correspond to operators of the type $O\pm i\tilde O$, where $\tilde O$  contains a single power of $\tilde V_{\mu\nu}$. If $M(1,..,n)$ is generated by $c\, ( O\pm i\tilde O)$,  
$M(1^c,..,n^c)$ is generated by $c^*\, ( O\mp i\tilde O)$. A basis of CP-eigenstates can be obtained by forming symmetric and anti-symmetric combinations of the structures of Table~\ref{table:vvvv4} and their complex conjugates. Thus, for example, the six independent four-$W$ CTs of Table~\ref{table:vvvv4}, supplemented by the two Hermitian conjugates of the 4$W_+$ CTs, can be 
combined to give four CP-even and two CP-odd combinations, reproducing the operator counting of~\cite{Remmen:2019cyz}.

We discuss a few additional examples of Hermitian and non-Hermitian CTs, 
their CP properties and the corresponding operators in Appendix~\ref{AppCount}.  
%%%%%%%%%%%%%%%%%%%%%%%%%%%%%%%%%%%%%%%%%%%%%%%%%%%%%%%%%%%%%%%%%%%%%%%%%%%%%%%%%%%%%%%%%%%%%%%%%%%%%%%%%
%%%%%%%%%%%%%%%%%%%%%%%%%%%%%%%%%%%%%%%%%%%%%%%%%%%%%%%%%%%%%%%%%%%%%%%%%%%%%%%%%%%%%%%%%%%%%%%%%%%%%%%%%
\section{Two-fermion two-vector amplitudes}
\label{fermions}
%%%%%%%%%%%%%%%%%%%%%%%%%%%%%%%%%%%%%%%%%%%%%%%%%%%%%%%%%%%%%%%%%%%%%%%%%%%%%%%%%%%%%%%%%%%%%%%%%%%%%%%%%
%%%%%%%%%%%%%%%%%%%%%%%%%%%%%%%%%%%%%%%%%%%%%%%%%%%%%%%%%%%%%%%%%%%%%%%%%%%%%%%%%%%%%%%%%%%%%%%%%%%%%%%%%
We now turn to two-fermion two-vector amplitudes, relevant for diboson production at the LHC or lepton colliders.
Since we are mainly interested in $\bar f f\to VV$ processes, we will treat the fermions as massless.
The relevant SCTs in this case at $d\leq8$ are:
\beqa
\label{12}
A_4(f_+,f_-, s,s) &=& \sab{132}, \quad (d=6),\; \nonumber\\
A_4(f_+,f_-,V_+,V_+) &=& \sqb{34}^2 \sab{132} , \quad (d=8), \nonumber \label{fpfm2Vp} \\
A_5(f_+,f_-,V_+,s,s) &=& \sqb{13}\sab{342}, \sqb{13}\sab{352}\quad (d=8), \; \nonumber\\
A_4(f_+,f_-,V_+,V_-) &=& \sqb{13}\anb{24}\sab{3(1-2)4} , \quad (d=8), \nonumber\\
A_4(f_+,f_+,V_+,s) &=& \sqb{13}\sqb{23}, \quad (d=6), \; \nonumber\\
A_4(f_+,f_+,V_-,s) &=& \sqb{12}\anb{3143},\quad (d=8), \nonumber\\ 
A_4(f_+,f_+, V_+,V_+) &=& \sqb{12}\sqb{34}^2,\,\sqb{34}(\sqb{13}\sqb{24}+\sqb{14}\sqb{23}), \quad (d=7),\; \nonumber\\
A_4(f_+,f_+, V_-,V_-) &=& \sqb{12}\anb{34}^2, \quad (d=7),\; \nonumber\\
A_5(f_+,f_+,s,s,s) &=& \sqb{12}, \quad (d=6),\; \nonumber\\
A_5(f_+,f_+,V_+,V_+,s) &=& \sqb{34}\sqb{13}\sqb{24},\;\sqb{34}\sqb{14}\sqb{23}, \quad (d=8),\nonumber\\ 
\label{22}
A_5(f_+,f_+, V_-,V_-,s) &=& \sqb{12}\anb{34}^2, \quad (d=8).
\label{dim6no}
\eeqa
The dimension in parentheses following the spinor structure is the minimal dimension of the operator which generates the structure. 
The $d=7$ SCTs cannot contribute to baryon- and lepton-number conserving SMEFT amplitudes, since 
the latter are associated with operators of  even dimensions~\cite{Degrande:2012wf}\footnote{Since Mandelstam invariants are even-dimensional, these SCTs never appear, at any dimension.}. As opposed to the purely bosonic case, here there are five-point amplitudes which contribute to the Higgsed LE amplitude. These contributions are generated by taking the soft limit of one of the Higgs legs, and lead to $v E^3/\Lambda^4$ terms in the EFT expansion. Note that for $VV$ production, only chirality conserving amplitudes contribute at dimension-8, since the SM amplitudes are chirality conserving. Thus, $f_\pm f_\pm$ amplitudes can be neglected at this order. We show both the chirality-conserving and chirality-violating SCTs in Eq.~(\ref{12}) for completeness. The LE chirality-conserving CTs are discussed in this Section. The chirality-violating HE SMEFT bases and LE CTs are listed in Appendix~\ref{AppCVA}. 

For simplicity, we consider a single SM generation. We denote SU(2) doublet quarks and leptons by $Q$ and $L$ respectively, and SU(2) singlet fermions by $U$, $D$, and $E$ in obvious notation. Taking all the particles as incoming, an SU(2)-doublet fermion of momentum $p$ is associated with a $p\rangle$ spinor, and its antifermion is associated with a $[p$ spinor. SU(2)-singlet fermions correspond to $p]$ spinors, and their antiparticles to $\langle p$ spinors. The HE quark amplitudes are collected in Table~\ref{Tferm}. For brevity, we do not show here $D^cD$ amplitudes. Their structure is identical to the $U^cU$ amplitudes. We also suppress color indices in amplitudes that do not include gluons. These indices will just be trivially contracted.

It is easy to read off the leptonic amplitudes from the quark amplitudes of Table~\ref{Tferm} by omitting \SU{3} indices and color-structures. $L^cL$ CTs are the same as in $Q^cQ$ amplitudes. From hypercharge considerations, $Q^cU$ CTs map into $E^cL$ CTs, while $Q^cD$ CTs map into $L^cE$ ones. Since leptons are color-neutral, lepton pair amplitudes with a single gluon vanish. Leptonic amplitudes with two same-helicity gluons vanish as well. For example, the only available $\SU{3}$ structure for $E^CEg_+^Ag_+^B$ is $\delta^{AB}$. 
Therefore, the kinematic structure must be symmetric under gluon exchange. 
However, since the SCT for this helicity category is given by the second line of~\Eqref{fpfm2Vp}, the only symmetric structure we can write is:
\beq
\sqb{34}^2\asb{1(3+4)2} = \sqb{34}^2\asb{1(-1-2)2} = 0
\eeq
For opposite helicity gluons, only structures proportional to $\delta^{AB}$ are generated. 
We have checked that the number of independent massless CTs matches the counting of  dimension-8 SMEFT operators of~\cite{Remmen:2019cyz, Murphy:2020rsh,Li:2020gnx}.

We denote the elements of the $\SU{2}$ doublets as:
\beq
Q_L\equiv \begin{pmatrix}
    u \\
    d
\end{pmatrix}, \quad L_L\equiv\begin{pmatrix}
    e \\
    \nu
\end{pmatrix}\;,
\eeq
which correspond to the left-handed (or negative helicity) LE quark/lepton fields. 
Since the $U/D/E$ fields map directly to the LE right handed (positive helicity) quark/lepton fields, we will continue to use the same notation for the right-handed LE fermions.  
As above, we do not show structures related by complex conjugation/CP.  
Note that some Hermitian fermion CTs may appear at first glance to have complex coefficients. 
For example,  the LE amplitude $\bar u u W^+W^-$ receives a contribution from the HE amplitudes $\bar Q Q W_{\pm}^aW_{\mp}^b$,
\begin{equation}
   \left(c_+^{Q^cQW_+W_-} + \frac{i}{2} c_-^{Q^cQW_+W_-}\right)\, \sqb{1\three}\anb{2\four}\sab{\three (1-2)\four}.
\label{uuWWeq}
\end{equation} 
Although this coefficient might seem complex, it is actually real. We can see this in two ways. From the amplitudes in Table~\ref{Tferm}, the contributing structures are,
\begin{equation}
\sqb{13}\anb{24}\sab{3(1-2)4}\delta^{ab}\delta^{j}_i,
\end{equation}
\begin{equation}
\sqb{14}\anb{23}\sab{4(1-2)3}\epsilon^{abc}(\tau^c)^j_i.
\end{equation}
Complex conjugation returns the first structure to itself up to a relabeling (for a more thorough explanation, see Appendix~\ref{AppCount}), meaning its coefficient $c_+^{Q^cQW_+W_-}$ is real. The second term returns to itself \textit{with opposite sign} (again, up to relabeling), which follows from switching the indices $a,b$ in the Levi-Civita tensor. This implies that the coefficient $c_-^{Q^cQW_+W_-}$ is \textit{purely imaginary}. It follows that the coefficient in Eq.~\eqref{uuWWeq} is real. The complex conjugate term,
\beqa
\left(c_+^{Q^cQW_+W_-}+\frac{i}{2}c_-^{Q^cQW_+W_-}\right)\anb{1\three}\sqb{2\four}\asb{\three (1-2)\four},
\eeqa
in which we need to switch $1\leftrightarrow 2$ due to the chirality of the fermions, is not an independent structure.

One can also show this by considering the corresponding HE operators,
\beqa
c_+Q^{c,j}Q_iW_+^aW_-^b\delta_{ab}\delta_j^i,\\
c_-Q^{c,j}Q_iW_+^aW_-^b\epsilon_{abc}(\tau^c)_j^i,
\eeqa
which are Hermitian and anti-Hermitian, respectively. It follows that $c_+$ is purely real and $c_-$ is purely imaginary, and therefore $c_++ic_-$ is real, 
as expected.

The chirality conserving LE fermionic CTs are listed in Table~\ref{ChiYes}. For completeness, the chirality violating LE amplitudes are also listed in Table~\ref{ChVio} of Appendix~\ref{AppCVA}. 
Amplitudes with two-gluon final states are not listed as they are in one-to-one correspondence with their HE counterparts 
in Table~\ref{Tferm}, with additional Higgs legs taken to be soft incurring powers of the Higgs VeV.
%%%%%%%%%%%%%%%%%%%%%%%%%%%%%%%%%%%%%%%%%%%%%%%%%%%%%%%%%%%%%%%%%%%%%%%%%%%%%%%%%%%%%%%%%%%%%%%%%%%%%%%%%%%%%%%%%%%%%%%%%%%%%%%%%%%%%%%%%%%%%%%%%%%%%%%%%%%%%%%%%%%%%%%%%%%%%%%%%%%%%%%%%%%%%%%%%%%%%%%%%%%%%%%%%%%%%%%%%%%%%%%%%%%%%%%%%%%%%%%%%%%%%%%%%%%%%%%%%%%%%%%%%%%%%%%%%%%%%%%%%%%%%%%%%%%%%%%%%%%%%%%%%%%%%
\section{Discussion}
\label{Discussion}
%%%%%%%%%%%%%%%%%%%%%%%%%%%%%%%%%%%%%%%%%%%%%%%%%%%%%%%%%%%%%%%%%%%%%%%%%%%%%%%%%%%%%%%%%%%%%%%%%%%%%%%%%%%%%%%%%%%%%%%%%%%%%%%%%%%%%%%%%%%%%%%%%%%%%%%%%%%%%%%%%%%%%%%%%%%%%%%%%%%%%%%%%%%%%%%%%%%%%%%%%%%%%%%%%%%%%%%%%%%%%%%%%%%%%%%%%%%%%%%%%%%%%%%%%%%%%%%%%%%%%%%%%%%%%%%%%%%%%%%%%%%%%%%%%%%%%%%%%%%%%%%%%%%%%
The leading SMEFT contributions to the 4-point CTs entering the LE $VVVV$ and $ffVV$ amplitudes are of order $1/\Lambda^4$~(dimension-8). 
The naive energy growth of the CTs is easy to read off.
Each spinor product scales as a single power of the energy, since $\vert (ij)\vert=\sqrt{s_{ij}}$. Most of the CTs we derived contain four spinor products and scale as $E^4$. The remaining ones, featured only in the fermionic amplitudes, contain three spinor products and scale as $vE^3$.
Some of these CTs---e.g. CTs in the $+-+-$ helicity categories of $ffVV$---interfere with the tree-level SM amplitudes even in the massless limit. The remaining chirality conserving amplitudes interfere with the  SM at tree-level when $M_V$ effects are taken into account\footnote{Recall that we neglect fermion masses throughout.}.
As is well known, EFT effects are most pronounced at the high energy tails of distributions, but this is also where the EFT expansion starts to break down.
Since our results include mass effects, 
they are applicable also for intermediate energies, 
$M_Z  \lesssim E \ll\Lambda$, where mass effects are sizable and
the EFT expansion can be trusted.

Many of the four-vector  CTs involve Higgs-vector couplings, and some involve Higgs self-couplings. It would be interesting to explore the sensitivity of the amplitudes to these couplings, compared to amplitudes with external Higgses.

CTs featuring left-handed up and down quarks are related by SU(2) 
in the HE theory.
The LE $\bar u u$ and $\bar d d$ CTs have the same kinematic structure,
but their Wilson coefficients 
can be different, which is a manifestation of the SU(2) symmetry breaking. In Table~\ref{uuvddWW}, we show several examples of $\overline{u}u W^+W^-$ vs $\overline{d}dW^+W^-$.
%%%%%%%%%%%%%%%%%%%%%%%%%%%%%%%%%%%%%%%%%%%%%%%%%%%%%%%%%%%%%%%%%%%%%%%%%%%%%%%%%%%%%%%%%%%%%%%%%%%%%%%
 \begin{longtable}{|>{\centering\arraybackslash}p{2in}|c|c|}
     \hline
     Structure & $\overline{u}uW^+W^-$&$\overline{d}dW^+W^-$\\
     \hline \hline
     \endfirsthead
     \hline
     \multicolumn{3}{|c|}{Continued from~\ref{uuvddWW}}\\
     \hline 
     Structure & $\overline{u}uW^+W^-$&$\overline{d}dW^+W^-$\\
     \hline \hline
     \endhead
     \hline
     \endfoot
    \endlastfoot
         \multirow{1}{*}{$\sQbr{1}{3}\Sqb{34}\anb{\four 2} $}&$\frac{v}{\sqrt{2}}(\Cffortyone+\Cffortytwo)$&$-\frac{v}{\sqrt{2}}\Cffortyone$\\
         \hline
         %%%%%%%%%%%%%%%
         \multirow{1}{*}{$\sQbr{1}{4}\Sqb{43}\anb{\three 2} $}&$\frac{v}{\sqrt{2}} \Cffortyseven $&$-\frac{v}{\sqrt{2}}(\Cffortyseven + \Cffortyeight)$\\
         \hline
         %%%%%%%%%%%%%%%
         $\sab{\four \three \four}\sqb{1\mathbf{3}}\anb{\mathbf{3}2}$ & $2(\Cfninetyfour+\Cfninetyfive)$&$2\Cfninetyfour$\\
         \hline
         %%%%%%%%%%%%%%%
         $\sab{\four 1\four}\sqb{1\mathbf{3}}\anb{\mathbf{3}2}$ & $2(\Cfninetyfourtag+\Cfninetyfivetag)$&$2\Cfninetyfourtag$\\
         \hline
         %%%%%%%%%%%%%%%
         \caption{Comparison of the coefficients of LE $\overline{u}uW^+W^-$ and $\overline{d}dW^+W^-$ amplitudes.}
     \label{uuvddWW}
 \end{longtable}
%%%%%%%%%%%%%%%%%%%%%%%%%%%%%%%%%%%%%%%%%%%%%%%%%%%%%%%%%%%%%%%%%%%%%%%%%%%%%%%%%%%%%%%%%%%%%%%%%%%%%%%
These amplitudes involve either the Goldstones or the Higgs VeV and are therefore sensitive to the
Higgs ``direction" in SU(2) space.
Note that this is the first time that these SU(2) breaking effects
appear---at dimension-6, the $\overline{u}u$ and $\overline{d}d$
Wilson coefficients are equal~\cite{Liu:2023jbq}.
%%%%%%%%%%%%%%%%%%%%%%%%%%%%%%%%%%%%%%%%%%%%%%%%%%%%%%%%%%%%%%%%%%%%%%%%%%%%%%%%%%%%%%%%%%%%%%%%%%%%%%%%%%%55555
\subsection{HEFT SMEFT comparison}
%%%%%%%%%%%%%%%%%%%%%%%%%%%%%%%%%%%%%%%%%%%%%%%%%%%%%%%%%%%%%%%%%%%%%%%%%%%%%%%%%%%%%%%%%%%%%%%%%%%%%%%
It is interesting 
to compare the results to HEFT predictions.
The full list of HEFT LE CTs at $d\leq8$ was derived in~\cite{Liu:2023jbq}.
Since the HEFT assumes only the unbroken $\SU{3}\times \text{U}(1)$ symmetry, it is more generic than the SMEFT and we would therefore expect to find there more independent structures. 
As it turns out, there is only one bosonic contact-term
that is generated in the HEFT but not in the SMEFT, namely the $W^-W^-W^+W^+$ CT,
\beqa
\Sqb{12}^2\Sqb{34}\Anb{34}\;,
\eeqa
(and its conjugate structure).
The reason is hypercharge: this CT 
 would have to come from a HE $WWHH$ (or $WWH^\dagger H^\dagger$) amplitude, which violates hypercharge.

In~\cref{Discreps}, we collect the structures that are not generated 
in dimension-8 SMEFT $f^cfW^+W^-$ and $f^cf'W^{\pm}Z$ amplitudes.
These structures may still be generated in the SMEFT at higher dimensions,
if they are allowed by symmetry.
All of these structures are generated in the dimension-8 HEFT.
%%%%%%%%%%%%%%%%%%%%%%%%%%%%%%%%%%%%%%%%%%%%%%%%%%%%%%%%%%%%%%%%%%%%%%%%%%%%%%%%
\begin{longtable}{|c|c|c|c|c|}
    \hline
         %%%%%%%%%%%%%%%
         \multicolumn{5}{|c|}{$f^cfW^+W^-$ and $f^cf'W^{\pm}Z$}\\
      \hline \hline
\multirow{1}{*}{Structure/Amplitude} & $\overline{U}UW^+W^-$ & $\overline{D}DW^+W^-$&$\overline{U}DW^-Z$&$\overline{D}UW^+Z$\\
&or $\bar UU ZZ$&or $\bar DDZZ$&&\\
\hline
         $\sQbr{2}{3}\aNbr{1}{4}\sab{\three (1-2) \four}$
         &$\checkmark$&$\checkmark$&$-$&$-$\\
         \hline
       $\sQbl{3}{4}^2 \sab{2(\three -\four) 1} $&$-$&$-$&$-$&$-$\\
        \hline
         \multicolumn{5}{|c|}{$f^cfZ(\gamma/g)$ and $f^cf'W^{\pm}(\gamma/g)$}\\
         \hline \hline
         Structure & $\overline{U}UZ(\gamma/g)$&$\overline{D}DZ(\gamma /g)$&$\overline{U}DW^-(\gamma /g)$&$\overline{D}UW^+(\gamma /g)$\\
         \hline
         %%%%%%%%%%%%%%%     
        $\aNbr{1}{3}\sqb{24}\sab{4(1-2)\three}$&$\checkmark$&$\checkmark$&$-$&$-$\\     %
        \hline
        %%%%%%%%%%%%%%%%%%%%%%%%%%%%%%%
         $\sQbl{3}{4}^2\sab{2(\textbf{3}-4)1}$&$-/\checkmark$&$-/\checkmark$&$-$&$-$\\
         \hline
    \caption{
    Missing kinematic structures in the dimension-8 SMEFT  $f^cfW^+W^-$ and $f^cf'W^{\pm}Z$ amplitudes. A
    $\checkmark$ means a structure is generated at dimension-8 for the given amplitude, while $-$ means it is not.} 
    \label{Discreps}
\end{longtable}
%%%%%%%%%%%%%%%%%%%%%%%%%%
%%%%%%%%%%%%%%%%%%%%%%%%%%%%%%%%%%%% 
The contact term $\sQbr{2}{3}\aNbr{1}{4}\sab{\three 2 \four}$, generated in the HEFT for all listed particle configurations, is absent in the SMEFT for amplitudes whose fermionic state has non-zero hypercharge. 
The reason for this is clear from the HE symmetry: this kinematic structure is in the $f_-^c f_+^\prime  V_+V_-$ helicity category. 
Since the HE vector bosons 
have zero hypercharge,  
and the fermion pair has non-zero total hypercharge, these amplitudes are forbidden. 
The same argument explains the discrepancy in the last line of Table~\ref{Discreps}.

Next, we turn to the structures
\beq
\Sqb{34}^2\sab{2(\three-\four) 1}~~~\text{and}~~~\sqb{3\four}^2 \sab{2(\three-4)1},
\eeq
which are generated in the HEFT for all particle configurations except $f^cfZZ$ due to Bose symmetry. 
In the SMEFT however, these structures are generated only for amplitudes with 
zero total fermionic hypercharge whose final state includes a gluon.  
These structures are in the $f^c_-f_+V_+V_+$ helicity category. 
Examining the HE CTs in Table~\ref{Tferm}, we see that the contributions to 
$\bar UU W_+W_+$, $\bar UUB_+B_+$ and $\bar UU W_+B_{\pm}$ vanish. 
The first two vanish due to  Bose symmetry.
The contribution to $\bar UU W^a_+B_{\pm}$ is forbidden by SU(2). 
On the other hand, the gluon amplitude $f^cf Zg$  
originates from $\bar UUgB$ which respects the HE symmetry.

One also expects fewer independent parameters in the Wilson coefficients of the SMEFT
compared to the HEFT. Indeed, in the LE $4W$ amplitudes, we find 13 independent parameters,
whereas the HEFT $4W$ contact terms of~\cite{Liu:2023jbq} feature 21 parameters. 
Similarly, in the
$u^cuW^+W^-$ amplitude for example,  we find 14 (16) free parameters in the chirality-conserving (violating) CTs. The HEFT contains 20~(22) free parameters in this case. 
In the HEFT, each independent CT corresponds to an independent Wilson coefficient, so one
could naively expect a sharper difference in the numbers of parameters featured in the two
theories. However, LE SMEFT CTs in transverse and longitudinal helicity categories originate from 
different HE SMEFT amplitudes, with vectors and Higgses respectively, and these feature independent Wilson
coefficients.

Finally, we note that, although not presented as part of the main results of this paper, 
there are also numerous chirality violating structures which appear in the HEFT but not in the 
SMEFT. These are collected in Appendix~\ref{AppCVA}, Table~\ref{T12}. 
\section{Conclusions}
\label{conclusions}
%%%%%%%%%%%%%%%%%%%%%%%%%%%%%%%%%%%%%%%%%%%%%%%%%%%%%%%%%
%%%%%%%%%%%%%%%%%%%%%%%%%%%%%%%%%%%%%%%%%%%%%%%%%%%%%%%%%
In this paper, we used the on-shell EFT formulation for a direct construction of
observables predicted by the dimension-8 SMEFT.
We listed bases of contact-terms for amplitudes with four SM vector particles, as well as
amplitudes with two fermions and two vectors.
In all cases, the contact-terms are first generated at dimension-8.
Each contact term is a combination of spinor products, and depends on the scattering angle and polarization
direction of the spin-1 particles. 
The results give a convenient starting point for mapping new physics effects to experimental
data. 
%%%%%%%%%%%%%%%%%%%%%%%%%%%%%%%%%%%%%%%%%%%%%%%%%%%%%%%%%%%%%%%%%%%%%%%%%%%%%%%%%%%%%%%%%%%%%%%%%%%%%%%%%%%%%%%%%%%%%%%%%%%%%%%%%%%%%%%%%%%%%%%%%%%%%%%%%%%%%%%%%%%%%%%%%%%%%%%%
\acknowledgments
 We thank Yotam Soreq for useful discussions. HL would like to thank the Institute for Nuclear Theory at the University of Washington for its hospitality
and the Department of Energy for partial support during the completion of this work.
Research supported by the Israel Science Foundation (Grant No.~1002/23), and by the NSF-BSF (Grant No.~2020-785). 
HL was also supported  in part by the Azrieli Foundation. 
%%%%%%%%%%%%%%%%%%%%%%%%%%%%%%%%%%%%%%%%%%%%%%%%%%%%%%%%%%%%%%%%%%%%%%%%%%%%%%%%%%%%%%%%%%%%%%%%%%%%%%%
\appendix 
%%%%%%%%%%%%%%%%%%%%%%%%%%%%%%%%%%%%%%%%%%%%%%%%%%%%%%%%%%%%%%%%
%%%%%%%%%%%%%%%%%%%%%%%%%%%%%%%%%%%%%%%%%%%%%%%%%%%%%%%%%%%%%%%%
\newpage
\section{Results}
\label{tablesApp}
\subsection{High-energy bosonic contact terms}
%%%%%%%%%%%%%%%%%%%%%%%%%%%%%%%%%%%%%%%%%%%%%%%%%%%%%%%%%%%%%%%%%%%%%%%%%%% 
\begin{longtable}{ |c|c|c| } 
 \hline
 Amplitude & Massless Contact-Term & Wilson Coefficient \\ 
 \hline \hline 
 \endfirsthead
 \hline
 \multicolumn{3}{|c|}{Continued from Table~\ref{table:vvvv4}}\\
 \hline
 Amplitude & Massless Contact-Term & Wilson Coefficient \\ 
 \hline \hline 
 \endhead
 \hline
 \endfoot
 \endlastfoot
\multirow{1}{*}{$W_+^aW_+^bW_+^cW_+^d$}  & $\sqb{12}^2\sqb{34}^2\delta^{ab}\delta^{cd} + \text{Perm}(1234)$ & $c_1^{4W_+}$ \\ 
\cline{2-3}
& $(\sqb{13}^2\sqb{24}^2+\sqb{14}^2\sqb{23}^2)\delta^{ab}\delta^{cd} + \text{Perm}(1234)$ & $c_2^{4W_+}$ \\ 
 \hline
 %%%%%%%%%%%%%%%
\multirow{1}{*}{$W_+^aW_+^bW_-^cW_-^d$}  & $\sqb{12}^2\anb{34}^2\delta^{ab}\delta^{cd}$ & $c_1^{2W_+2W_-}$ \\ 
\cline{2-3} & $\sqb{12}^2\anb{34}^2(\delta^{ad}\delta^{bc}+\delta^{ac}\delta^{bd})$ & $c_2^{2W_+2W_-}$ \\ 
\hline
%%%%%%%%%%%%%%%
\multirow{1}{*}{$W_+^aW_+^bB_+B_+$} & $\sqb{12}^2\sqb{34}^2\delta^{ab} $ & $c_1^{2W_+2B_+}$\\
\cline{2-3} & $(\sqb{13}^2\sqb{24}^2+\sqb{14}^2\sqb{23}^2)\delta^{ab} $ & $c_2^{2W_+2B_+}$ \\ 
\hline
%%%%%%%%%%%%%%%
$W_+^aW_+^bB_-B_-$ & $\sqb{12}^2\anb{34}^2\delta^{ab}$ & $c^{2W_+2B_-}$\\
\hline
%%%%%%%%%%%%%%%
$W_+^aW_-^bB_+B_-$ & $\sqb{13}^2\anb{24}^2\delta^{ab}$ & $c^{W_+W_-B_+B_-}$\\
\hline
%%%%%%%%%%%%%%%
$B_+B_+B_+B_+$ & $\sqb{12}^2\sqb{34}^2 + \text{Perm}(1234)$ & $c^{4B}$\\
\hline
%%%%%%%%%%%%%%%
$B_+B_+B_-B_-$ & $\sqb{12}^2\anb{34}^2 $ & $c^{2B_+2B_-}$\\
\hline
%%%%%%%%%%%%%%%
\multirow{1}{*}{$W_+^aW_+^bH^{\dagger,j}H_i$}  & $\sqb{12}(\sqb{1342} - \sqb{2341})\delta^{ab}\delta^j_i$ & $c_+^{2W_+2H}$ \\ 
\cline{2-3}
& $\sqb{12}(\sqb{1342} + \sqb{2341})\epsilon^{abc}(\tau_c)_i^{\;j}$ & $c_-^{2W_+2H}$ \\ 
 \hline
 %%%%%%%%%%%%%%%
\multirow{1}{*}{$W_+^aW_-^bH^{\dagger,j}H_i$}  & $\sab{132}\sab{142}\delta^{ab}\delta^j_i$ & $c_1^{W_+W_-2H}$ \\ 
\cline{2-3}
& $\sab{132}\sab{142}\epsilon^{abc}(\tau_c)_i^{\;j}$ & $c_2^{W_+W_-2H}$ \\ 
\hline
%%%%%%%%%%%%%%%
\multirow{1}{*}{$W_+^aB_+H^{\dagger,j}H_i$}  & $\sqb{12}\sqb{1432}(\tau^a)_i^{\;j}$ & $c_1^{W_+B_+2H}$ \\ 
\cline{2-3}
& $\sqb{12}\sqb{1342}(\tau^a)_i^{\;j}$ & $c_2^{W_+B_+2H}$ \\ 
\hline
%%%%%%%%%%%%%%%
$W_+^aB_-H^{\dagger,j}H_i$ & $\sab{132}\sab{142}(\tau^a)_i^{\;j}$ & $c^{W_+B_-2H}$ \\ 
\hline
%%%%%%%%%%%%%%%
$B_+B_+H^{\dagger,j}H_i$ & $\sqb{12}(\sqb{1342} - \sqb{2341})\delta^j_i$ & $c^{2B_+2H}$ \\ 
\hline
%%%%%%%%%%%%%%%
$B_+B_-H^{\dagger,j}H_i$ & $\sab{132}\sab{142}\delta^j_i$ & $c^{B_+B_-2H}$ \\ 
\hline
%%%%%%%%%%%%%%%
\multirow{1}{*}{$H^{\dagger,i}H^{\dagger,j}H_kH_l$} & $s_{12}s_{34}(\delta_i^k\delta_j^l+\delta_i^l\delta_j^k)$ & $c_{+,1}^{4H}$ \\ 
\cline{2-3}
& $(s_{13}s_{24}+s_{14}s_{23})(\delta_i^k\delta_j^l+\delta_i^l\delta_j^k)$ & $c_{+,2}^{4H}$ \\
\cline{2-3}
& $(s_{13}s_{24}-s_{14}s_{23})\epsilon^{ij}\epsilon_{kl}$ & $c_-^{4H}$ \\
\hline
%%%%%%%%%%%%%%%
\multirow{1}{*}{$g^A_+g^B_+g^C_+g^D_+$} & $\sqb{12}^2\sqb{34}^2 \delta^{ab}\delta^{cd}+\text{Perm}(1234)$ & $c_1^{4g_+}$\\
		\cline{2-3} & $\sqb{12}^2\sqb{34}^2 d_{ABE}d_{CDE}+\text{Perm}(1234)$ & $c_2^{4g_+}$\\
		\cline{2-3} & $\sqb{12}^2\sqb{34}^2 (f_{ACE}f_{BDE} +f_{BCE}f_{ADE} )+\text{Perm}(1234)$ & $c_3^{4g_+}$\\ 
		\hline
  %%%%%%%%%%%%%%%
		\multirow{1}{*}{$g^A_+g^B_+g^C_-g^D_-$} & $\sqb{12}^2\anb{34}^2 \delta^{AB}\delta^{CD}$ & $c_1^{2g_+2g_-}$\\
\cline{2-3} & $\sqb{12}^2\anb{34}^2 d_{ABE}d_{CDE}$ & $c_2^{2g_+2g_-}$\\
\cline{2-3} & $\sqb{12}^2\anb{34}^2 (f_{ACE}f_{BDE} +f_{BCE}f_{ADE} )$ & $c_3^{2g_+2g_-}$\\ 
\hline
%%%%%%%%%%%%%%%
$g^A_+g^B_+g^C_+B_+$ & $(\sqb{12}^2\sqb{34}^2+\sqb{13}^2\sqb{24}^2+\sqb{14}^2\sqb{23}^2)d^{ABC}$ & $c^{3g_+B_+}$ \\ 
\hline
%%%%%%%%%%%%%%%
$g^A_+g^B_+g^C_-B_-$ & $\sqb{12}^2\anb{34}^2d^{ABC}$ & $c^{2g_+g_-B_-}$ \\ 
\hline
%%%%%%%%%%%%%%%
\multirow{1}{*}{$g^A_+g^B_+B_+B_+$} & $\sqb{12}^2\sqb{34}^2\delta^{ab}$ & $c_1^{2g_+2B_+}$ \\ 
\cline{2-3}
& $(\sqb{13}^2\sqb{24}^2+\sqb{14}^2\sqb{23}^2)\delta^{ab}$ & $c_2^{2g_+2B_+}$\\
\hline
%%%%%%%%%%%%%%%
$g^A_+g^B_+B_-B_-$ & $\sqb{12}^2\anb{34}^2\delta^{ab}$ & $c^{2g_+2B_-}$ \\ 
\hline
%%%%%%%%%%%%%%%
$g^A_+g^B_-B_+B_-$ & $\sqb{13}^2\anb{24}^2\delta^{ab}$ & $c^{g_+g_-B_+B_-}$ \\ 
\hline
%%%%%%%%%%%%%%%
\multirow{1}{*}{$g^A_+g^B_+W^a_+W^b_+$} & $\sqb{12}^2\sqb{34}^2\delta^{ab}\delta^{ab}$ & $c_1^{2g_+2W_+}$ \\ 
\cline{2-3}
&$(\sqb{13}^2\sqb{24}^2+\sqb{14}^2\sqb{23}^2)\delta^{ab}\delta^{ab}$&$c_2^{2g^+2W^+}$\\
\hline
%%%%%%%%%%%%%%%
$g^A_+g^B_+W^a_-W^b_-$ & $\sqb{12}^2\anb{34}^2\delta^{ab}\delta^{ab}$ & $c^{2g_+2W_-}$ \\ 
\hline
%%%%%%%%%%%%%%%
$g^A_+g^B_-W^a_+W^b_-$ & $\sqb{13}^2\anb{24}^2\delta^{ab}\delta^{ab}$ & $c^{g_+g_-W_+W_-}$ \\ 
\hline
%%%%%%%%%%%%%%%
$g_+^Ag_+^BH_i H^{\dagger, j}$& $\sqb{12}\left(\sqb{1342}+\sqb{1432}\right)\delta^{ab}\delta_i^{\; j}$& $c_+^{2g_+2H}$\\
\hline
%%%%%%%%%%%%%%%
$g^A_+g^B_-H_iH^{\dagger,j}$&$\sab{132}\sab{142}\delta^{ab}\delta_i^{\; j}$ & $c_+^{g_+g_-2H}$\\
\hline
%%%%%%%%%%%%%%%
%%%%%%%%%%%%%%%%%%%%%%%%%%%%%%%%%%%%%%
\caption{High energy 4-pt CTs for $W$, $B$, gluon ($g$) and Higgs SMEFT helicity amplitudes. The vector helicity is indicated by $\pm$ subscripts.
SU(2) adjoint indices are denoted by $a,b,..$ and (anti)fundamental indices by $i,j,..$. \SU{3} adjoint indices are denoted by $A,B,..$. 
 $\tau_a$ and $\lambda^A$ 
 are the SU(2) and \SU{3} generators respectively. $f^{ABC}$ and $d^{ABC}$ are the \SU{3} structure constants and symmetric tensor respectively. The third column defines the Wilson coefficient of each CT, which contain an implicit factor of $1/\Lambda^4$. }
\label{table:vvvv4}
\end{longtable}
%%%%%%%%%%%%%%%%%%%%%%%%%%%%%%%%%%%%%%%%%%%%%%%%%%%%%%%%%%%%%%%%%%%%%%%%%%%%%%%
To illustrate the correspondence between amplitude contact terms and operator bases, let us comment on these bosonic examples.
$VVVV$ amplitudes are associated with the four field-strength operators of Table~2 of~\cite{Murphy:2020rsh}. A positive-helicity vector $i]i]$ corresponds to $F_{\mu\nu}+i\tilde F_{\mu\nu}$, while a negative-helicity vector $i\rangle i\rangle$ corresponds to $F_{\mu\nu}-i\tilde F_{\mu\nu}$ (see Sec~\ref{sec:cp}). VVss amplitudes correspond to amplitudes from the $X^2H^2D^2$ operators of Table 7 of~\cite{Murphy:2020rsh}, where only the partial derivative parts of the covariant derivatives contribute to the 4-point contact terms. The 
gauge boson parts of the covariant derivative generate 5-point and higher contact terms, which, as discussed in Sec~\ref{sec:masslesscts}, do not contribute to the LE contact terms at dimension-8.
%%%%%%%%%%%%%%%%%%%%%%%%%%%%%%%%%%%%%%%%%%%%%%%%%%%%%%%%%%%%%%%%%%%%%%%%%%% 
%%%%%%%%%%%%%%%%%%%%%%%%%%%%%%%%%%%%%%%%%%%%%%%%%%%%%%%%%%%%%
%%%%%%%%%%%%%%%%%%%%%%%%%%%%%%%%%%%%%%%%%%%%%%%%%%%%%%%%%%%%%%%%%%%%%%%%%%%%%%%%%%%%%%%%%%%%%
%%%%%%%%%%%%%%%%%%%%%%%%%%%%%%%%%%%%%%%%%%%%%%%%%%%%%%%%%%%%%
%%%%%%%%%%%%%%%%%%%%%%%%%%%%%%%%%%%%%%%%%%%%%%%%%%%%%%%%%%%%%%%%%%%%%%%%%%%%%%%%%%%%%%%%%%%%%%%%%%%%%%%%%
\newpage
%%%%%%%%%%%%%%%%%%%%%%%%%%%%%%%%%%%%%%%%%%%%%%%%%%%%%%%%%%%%%
%%%%%%%%%%%%%%%%%%%%%%%%%%%%%%%%%%%%%%%%%%%%%%%%%%%%%%%%%%%%%%%%%%%%%%%%%%%%%%%%%%%%%%%%%%%%%%%%%%%%%%%%%
\begin{landscape}
\subsection{Low-energy bosonic contact terms}
\begin{longtable}[c]{ |c|c|>{\centering\arraybackslash}p{4.2in}| } 
 \hline
 Amplitude & Massive Contact-Term & Wilson Coefficient \\ 
 \hline \hline 
 \endfirsthead
 \hline
 \multicolumn{3}{|c|}{Continued from Table~\ref{T1}}\\
 \hline
 Amplitude & Massive Contact-Term & Wilson Coefficient \\ 
 \hline \hline 
 \endhead
 \hline
 \endfoot
 \endlastfoot
\multirow{1}{*}{$A_4\left(W^-W^-W^+W^+\right)$} & $\Sqb{12}^2\Sqb{34}^2 $ & $2\Csix$ \\
\cline{2-3}
&$\Sqb{13}^2\Sqb{24}^2+\Sqb{14}^2\Sqb{23}^2  $ & $ \Cfive +\Csix$\\ 
\cline{2-3}
 & $\Sqb{12}^2\Anb{34}^2 $&  $ \frac{3}{2}\Conetwo$\\
 \cline{2-3}
 &$\Sqb{13}^2\Anb{24}^2+\Sqb{14}^2\Anb{23}^2 $&$\Cone + \frac{1}{2}\Conetwo$\\
\cline{2-3}
 & $\Sqb{12}\Sqb{34}(\Sqb{13}\Anb{24}+\Sqb{24}\Anb{13}-(3\leftrightarrow 4)) $ & $-(2\Cnine+i\Cten)$ \\
 \cline{2-3}
& $\Sqb{14}\Sqb{23}(\Sqb{13}\Anb{24}+ \Sqb{24}\Anb{13})+(3\leftrightarrow 4)   $&$-(2\Cnine-i\Cten)$  \\
\cline{2-3}
 & $\Sqb{12}\Anb{34}(\Sqb{14}\Anb{23}+\Anb{14}\Sqb{23}-(3\leftrightarrow 4))  $ & $ -(2\Cfourteen-i \Cfifteen)$\\
\cline{2-3}
 & $\Sqb{12}\Sqb{34}\Anb{12}\Anb{34}$ & $8c_{+,1}^{4H}$\\
 \cline{2-3}
&$\Sqb{13}\Sqb{24}\Anb{13}\Anb{24} + (3\leftrightarrow 4)$ &$8c_{+,2}^{4H}$\\
\hline 
%%%%%%%%%%%%%%%
\multirow{1}{*}{$\mathcal{A}\left(W^+W^-ZZ\right)$}   & $\Sqb{12}^2\Sqb{34}^2 $& $\ctw^2 c_1^{4W_+}+ \stw^2 c_1^{2W_+2B_+}$\\
      \cline{2-3}
      &$\Sqb{14}^2\Sqb{32}^2+\Sqb{13}^2\Sqb{24}^2 $&$\ctw^2 c_2^{4W_+}+ \stw^2 c_2^{2W_+2B_+}$ \\
      \cline{2-3} 
         & $ \Sqb{12}^2\Anb{34}^2 $& $\Cthree \stw^2+\ctw^2 \Cone$\\
       \cline{2-3}
        & $\Sqb{13}^2\Anb{24}^2+\Sqb{14}^2\Anb{23}^2  $ & $\Cfour \stw^2 + \ctw^2\Conetwo $\\
       \cline{2-3}
       & $\Sqb{12}\Sqb{13}\Sqb{34}\Anb{24}+(3\leftrightarrow 4) $ &$-\ctw \Cten - i\stw \Ctwelve$ \\
       \cline{2-3}
       & $\Sqb{12}\Sqb{23}\Sqb{34}\Anb{14}+(3\leftrightarrow 4) $ &$-\ctw \Cten + i\stw \Ctwelve$ \\
       \cline{2-3}
       &$\Sqb{13}\Sqb{14}\Sqb{23}\Anb{24}+(3\leftrightarrow 4) $ & $\ctw \Cten - i\stw \Cthirteen$\\
       \cline{2-3}
       &$\Sqb{23}\Sqb{24}\Sqb{13}\Anb{14}+(3\leftrightarrow 4) $ & $\ctw \Cten + i\stw \Cthirteen$\\
       \cline{2-3}
       & \multirow{1}{*}{$\Sqb{34}^2\Sqb{12}\Anb{12} $} &$2\stw^2 c^{2B_+2H}+ 2\ctw^2 c_+^{2W_+2H}+\ctw \stw
(c_1^{W_+B_+2H}+c_2^{W_+B_+2H}) $\\
       \cline{2-3} 
        & $ \Sqb{12}^2\Sqb{34}\Anb{34} $ &$-2 c_+^{2W_+2H}$\\
       \cline{2-3}
       & $\Sqb{12}\Sqb{14}\Anb{23}\Anb{34}+(3\leftrightarrow 4) $&$-\ctw c_2^{W_+W_-2H}+\stw i \Cseventeen$\\
       \cline{2-3}
       &$\Sqb{13}\Sqb{14}\Anb{23}\Anb{24}  $&$2\Cfourteen-i\Cfifteen$\\
       \cline{2-3}
       &\multirow{1}{*}{$\Sqb{13}\Sqb{23}\Anb{14}\Anb{24}+(3\leftrightarrow 4)$}&$\ctw^2 \Cfourteen +\stw^2 \Csixteen +\ctw\stw \Cseventeen$\\
       \cline{2-3}
       &$ \Sqb{12}\Sqb{34}\Anb{12}\Anb{34}$ & $4(\Cnineteen+\Ctwenty)$\\
       \cline{2-3}
       &$\Sqb{13}\Sqb{24}\Anb{13}\Anb{24}+ (3\leftrightarrow 4)$&$2(\Ceighteen+\Cnineteen-\Ctwenty)$\\
       \hline
       %%%%%%%%%%%%%%%
\multirow{1}{*}{$\mathcal{A}\left(W^+W^-\gamma\gamma\right)$} & $\Sqb{12}^2\sqb{34}^2  $&$\stw^2 \Cfive +\ctw^2\Ceight $\\
        \cline{2-3}
        &$\sqb{\mathbf{1}3}^2\sqb{\mathbf{2}4}^2+\sqb{\mathbf{2}3}^2\sqb{\mathbf{1}4}^2 $ &$\stw^2 \Csix +\ctw^2 \Ceightwo$\\
        \cline{2-3} 
        & \multirow{1}{*}{$\sqb{34}^2\Sqb{12}\Anb{12} $} &$- 2\stw^2 \Cnine -2\ctw^2 \Celeven+\ctw \stw (\Ctwelve + \Cthirteen) $\\
       \cline{2-3}
        & $\Sqb{12}^2\anb{34}^2 $&$c_1^{2W_+2W_-} \stw^2 + c^{2W_+2B_-}\ctw^2$\\
        \cline{2-3}
        &$\sqb{\mathbf{1}3}^2\anb{\mathbf{2}4}^2+(3\leftrightarrow 4)  $ &$\stw^2 \Conetwo +\ctw^2 \Cfour$\\
        \cline{2-3}
        & \multirow{1}{*}{$\sqb{\mathbf{1}3}\sqb{\mathbf{2}3}\anb{\mathbf{1}4}\anb{\mathbf{2}4}+(3\leftrightarrow 4)$} &$2c^{B_+B_-2H} c_W^2 + 2s_W^2 c_1^{W_+W_-2H}-( c^{W_+B_-2H}+ \Ctwohundredone) s_W c_W$\\
        \hline
        %%%%%%%%%%%%%%%
\multirow{1}{*}{$\mathcal{A}\left(ZZZZ\right)$} & \multirow{1}{*}{$ \Sqb{12}^2\Sqb{34}^2 + (1\leftrightarrow 3) + (1 \leftrightarrow 4) $} & $\ctw^4(\Cfive+2\Csix)+\stw^4 \Csev+2\ctw^2\stw^2 (\Ceight +2\Ceightwo)$ \\
        \cline{2-3}
         & \multirow{1}{*}{$\Sqb{12}^2\Anb{34}^2+\text{Perm}(1234)$} &$\ctw^4 (\Cone + 2 \Conetwo) + \stw^4 \Ctwo + 2 \ctw^2 \stw^2 (\Cthree + 2 \Cfour)$\\
        \cline{2-3} 
        &  \multirow{1}{*}{$\Sqb{12}^2\Sqb{34}\Anb{34}+\text{Perm}(1234) $}&$-2\stw^2 c^{2B_+2H}$$- 2\ctw^2 c_+^{2W_+2H}+\stw \ctw(c_1^{W_+B_+2H}+c_2^{W_+B_+2H})$\\
        \cline{2-3} 
        & \multirow{1}{*}{$\Sqb{13}\Anb{32}\Sqb{14}\Anb{42}+\text{Perm}(1234)$}&$2(\ctw^2 \Cfourteen+\stw^2 \Csixteen-\ctw \stw \Cseventeen)$ \\
        \cline{2-3} 
        & $\Sqb{12}\Anb{12}\Sqb{34}\Anb{34}+ \text{Perm}(1234)$  &$16 \left(c_{+,1}^{H^2H^{\dagger2}}+2c_{+,2}^{H^2H^{\dagger2}}\right)$\\
        \hline 
        %%%%%%%%%%%%%%%
\multirow{1}{*}{$\mathcal{A}\left(ZZ\gamma\gamma\right)$} &\multirow{1}{*}{$\Sqb{12}^2\sqb{34}^2 $} &$(\ctw^4 + \stw^4) \Ceight +\ctw^2\stw^2(\Cfive+2\Csix + \Csev-4\Ceightwo)$\\
        \cline{2-3}
        & \multirow{1}{*}{$\sqb{\mathbf{1}3}^2\sqb{\mathbf{2}4}^2 + (1\leftrightarrow 2) $} &$(\ctw^2 - \stw^2)^2 \Ceightwo+\ctw^2\stw^2(\Cfive+2\Csix + \Csev-2\Ceight)$ \\
        \cline{2-3}
        & \multirow{1}{*}{$\sqb{34}^2\Sqb{12}\Anb{12} $} &$-2 \ctw^2 \Celeven- 2 \stw^2 \Cnine-\ctw \stw(\Ctwelve + \Cthirteen) $\\
        \cline{2-3}
        %%%%%%%%%%%%%%%
        & \multirow{1}{*}{$\Sqb{12}^2\anb{34}^2 $} & $(\Ctwohundred\stw^4+\Cthree\ctw^4)
        +\ctw^2\stw^2(c_1^{2W_+2W_-}+2c_2^{2W_+2W_-}+ c^{2B_+2B_-})-4\ctw^2\stw^2c^{W_+W_-B_+B_-}$\\
        \cline{2-3}
        & \multirow{1}{*}{$\sqb{\mathbf{1}3}^2\anb{\mathbf{2}4}^2+\sqb{\mathbf{2}3}^2\anb{\mathbf{1}4}^2  $} &$\left(\ctw^2-\stw^2\right)^2 \Cfour+ \ctw^2 \stw^2 (\Cone + 2\Conetwo + \Ctwo -\Cthree-\Ctwohundred) $\\
        \cline{2-3}
        & \multirow{1}{*}{$\sqb{\mathbf{1}3}\sqb{\mathbf{2}3}\anb{\mathbf{1}4}\anb{\mathbf{2}4}+(3\leftrightarrow 4)$} &$2\stw^2 \Cfourteen+2 \ctw^2 \Csixteen+\stw\ctw(c^{W_+B_-2H}+\Ctwohundredone)$\\
        \hline 
        %%%%%%%%%%%%%%%
        \multirow{1}{*}{$\mathcal{A}\left(ZZZ\gamma\right) $}  & \multirow{1}{*}{$\Sqb{12}^2\sqb{\mathbf{3}4}^2 + \text{Perm}(123) $}&$ -\stw \ctw^3(\Cfive+2\Csix-\Ceight-2\Ceightwo)-\stw^3 \ctw(\Ceight+2\Ceightwo -\Csev)$\\
        \cline{2-3}
        & $\sQbl{3}{4}^2\Sqb{12}\Anb{12}+\text{Perm}(123) $&$\frac{1}{2}\left(\ctw^2-\stw^2 \right)(c_1^{W_+B_+2H} + 2c_2^{W_+B_+2H}) +2 \ctw \stw (-c^{2B_+2H} + c_+^{2W_+2H})$\\
        \cline{2-3}
       & \multirow{1}{*}{$ \Sqb{12}^2\anb{\mathbf{3}4}^2 +\text{Perm}(123) $} &$\ctw^3\stw (c^{2W_+2B_-}+2\Cfour-\Cone-2\Conetwo)+\stw^3 \ctw(\Ctwo-\Ctwohundred -2\Cfour)$\\
         \cline{2-3}
         & \multirow{1}{*}{$\Sqb{13}\Sqb{23}\anb{\mathbf{1}4}\anb{\mathbf{2}4}+ \text{Perm}(123) $}& $2 \ctw \stw (-\Cfourteen + \Csixteen) +\stw^2\Ctwohundredone -\ctw^2 \Cseventeen$\\
         \hline
         %%%%%%%%%%%%%%%
         \multirow{1}{*}{$\fourptamp{W^-}{W^+}{Z}{\gamma}$}& $\Sqb{12}^2\sqb{\mathbf{3}4}^2  $&$\ctw \stw (\Cfive - \Ceight)$\\
         \cline{2-3}
         &$ \Sqb{13}^2\sqb{\mathbf{2}4}^2+\sqb{\mathbf{1}4}^2\Sqb{23}^2 $ &$ \ctw \stw (\Csix-\Ceightwo)$\\
         \cline{2-3} 
          & $\Sqb{12}^2\anb{\mathbf{3}4}^2 $& $\ctw \stw (c^{2W_+2B_-}-c_1^{2W_+2W_-})$ \\
         \cline{2-3}
          & $\Sqb{13}^2\anb{\mathbf{2}4}^2+\Sqb{23}^2\aNbl{1}{4}^2  $& $\ctw \stw (c^{W_+W_-B_+B_-}-c_2^{2W_+2W_-})$ \\
         \cline{2-3}
         & $\sqb{\mathbf{3}4}\Sqb{23}\Anb{12}\sqb{\mathbf{1}4} $ & $\ctw^2\Cthirteen - \stw^2 \Ctwelve+2\ctw \stw (\Celeven - \Cnine)$\\
         \cline{2-3}
         & $\sqb{\mathbf{3}4}\Sqb{13}\Anb{12}\sqb{\mathbf{2}4} $  &$\stw^2 \Cthirteen - \ctw^2 \Ctwelve- 2\ctw\stw (\Celeven-\Cnine)$ \\
         \cline{2-3}
         &$\sqb{\mathbf{2}4}^2\Sqb{13}\Anb{13} $ &$-\stw \Cten +i \ctw \Ctwelve$\\
         \cline{2-3}
         &$\sqb{\mathbf{2}4}\Sqb{23}\Anb{13}\sqb{\mathbf{1}4} $ &$\stw \Cten +i \ctw \Cthirteen$\\
         \cline{2-3}
         &$\sqb{\mathbf{1}4}^2\Sqb{23}\Anb{23} $ &$-\stw \Cten +i\ctw \Cthirteen$\\
         \cline{2-3}
         &$\sqb{\mathbf{1}4}\Sqb{13}\Anb{23}\sqb{\mathbf{2}4} $ &$-\stw \Cten -i\ctw \Ctwelve$\\
         \cline{2-3}
         &$\Sqb{13}\anb{\mathbf{1}4}\Sqb{23}\anb{\mathbf{2}4} $&$\ctw^2c^{W_+B_-2H}-\stw^2\Ctwohundredone + 2\ctw \stw (\Csixteen-\Cfourteen)$\\
         \cline{2-3}
         &$\Sqb{12}\anb{\mathbf{1}4}\Sqb{23}\anb{\mathbf{3}4} $&$-\stw \Cfifteen + i\ctw \Cseventeen$\\
         \cline{2-3}
         &$\Sqb{12}\anb{\mathbf{2}4}\Sqb{13}\anb{\mathbf{3}4} $&$\stw \Cfifteen +i \ctw \Cseventeen$\\
         \hline
         %%%%%%%%%%%%%%%
         \multirow{1}{*}{$\fourptamp{\gamma}{\gamma}{\gamma}{Z}$}& \multirow{1}{*}{$\sqb{12}^2\sqb{3\mathbf{4}}^2+ \text{Perm}(123) $}&$\stw^3 \ctw (\Ceight +2\Ceightwo -\Cfive -2\Csix+\ctw^3 \stw (\Csev-\Ceight-2\Ceightwo)$\\
         \cline{2-3}
         & \multirow{1}{*}{$ \sqb{12}^2\anb{3\mathbf{4}}^2+\text{Perm}(123) $}&$ \ctw^3 \stw (\Ctwo - \Ctwohundred - 2 \Cfour)+\stw^3 \ctw(-\Cone - 2 \Conetwo + \Cthree + 2 \Cfour)$\\
         \hline 
         %%%%%%%%%%%%%%%
         \multirow{1}{*}{$\fourptamp{\gamma}{\gamma}{\gamma}{\gamma}$}& \multirow{1}{*}{$\sqb{12}^2\sqb{34}^2 +\text{Perm}(1234)  $}& $\stw^4(c_1^{4W_+}+2c_2^{4W_+}) + \ctw^4 c^{4B_+} + 2\ctw^2\stw^2(\Ceight+2\Ceightwo)$\\
         \cline{2-3} 
         & \multirow{1}{*}{$ \sqb{12}^2\anb{34}^2+\text{Perm}(1234)$}& $\stw^4 c^{2W_+2W_-} + \ctw^4 c^{2B_+2B_-}+2\ctw^2\stw^2(c^{2W_+2B_-}+2c^{W_+W_-B_+B_-})$\\
         \hline
         %%%%%%%%%%%%%%%
         \multirow{1}{*}{$\mathcal{A}(g^Ag^BW^-W^+)$} & $\sqb{12}^2\Sqb{34}^2 \delta^{AB} $ &$\Ctwentyone$\\
        \cline{2-3}
        &$(\sqb{1\mathbf{4}}^2\sqb{2\mathbf{3}}^2+\sqb{1\mathbf{3}}^2\sqb{2\mathbf{4}}^2)\delta^{AB}\; $&$\Ctwentytwo$\\
        \cline{2-3} 
         & $\sqb{12}^2\Anb{34}^2\delta^{AB} $&$\Ctwentythree$\\
         \cline{2-3}
         & $(\sqb{1\mathbf{3}}^2\anb{2\mathbf{4}}^2+(1\leftrightarrow 2)  )\delta^{AB}$&$\Ctwentyfour$\\
         \cline{2-3}
         & $\sqb{12}\Anb{34}\left(\sqb{1\mathbf{4}}\sqb{2\mathbf{3}}-\sqb{1\mathbf{3}}\sqb{2\mathbf{4}}\right)\delta^{AB} $&$2\Ctwentyfive$\\
         \cline{2-3}
         & $(\sqb{1\mathbf{3}}\anb{2\mathbf{3}}\sqb{1\mathbf{4}}\anb{2\mathbf{4}}+(1\leftrightarrow 2))\delta^{AB}$&$2\Cthirtytwo$\\
         \hline
         %%%%%%%%%%%%%%%
         \multirow{1}{*}{$\mathcal{A}(g^Ag^BZZ)$}&$\sqb{12}^2\Sqb{34}^2\delta^{AB} $&$\ctw^2 \Ctwentyone + \stw^2 \Ctwentyseven$\\
         \cline{2-3}
         &$(\sqb{1\mathbf{3}}^2\sqb{2\mathbf{4}}^2+\sqb{1\mathbf{4}}^2\sqb{2\mathbf{3}}^2)\delta^{AB} $&$\stw^2 \Ctwentytwo +\stw^2 \Ctwentyseventag$\\
         \cline{2-3}
         &$\sqb{12}^2\Anb{34}^2\delta^{AB} $&$\ctw^2 \Ctwentythree+\stw^2 \Ctwentyeight$\\
         \cline{2-3}
         &$(\sqb{1\mathbf{3}}^2\anb{2\mathbf{4}}^2+(1\leftrightarrow 2)   )\delta^{AB}$&$\ctw^2 \Ctwentyfour+\stw^2 \Ctwentynine$\\
         \cline{2-3}
         &$\sqb{12}\Anb{34}(\sqb{1\mathbf{4}}\sqb{2\mathbf{3}}-\sqb{1\mathbf{3}}\sqb{2\mathbf{4}})\delta^{AB}$&$2\Ctwentyfive$\\
         \cline{2-3}
         &$(\sqb{1\mathbf{3}}\sqb{1\mathbf{4}}\anb{2\mathbf{3}}\anb{2\mathbf{4}}+(1\leftrightarrow 2))\delta^{AB}$&$2\Ctwentysix$\\
         \hline
         %%%%%%%%%%%%%%%
         \multirow{1}{*}{$\mathcal{A}(g^Ag^B\gamma Z)$}&$\sqb{12}^2\sQbr{3}{4}^2\delta^{AB} $&$\ctw \stw (-\Ctwentyone+\Ctwentyseven)$\\
         \cline{2-3}
         &$(\sqb{13}^2\sqb{2\mathbf{4}}^2+\sqb{1\mathbf{4}}^2\sqb{23}^2)\delta^{AB} $&$\stw\ctw(-\Ctwentytwo + \Ctwentyseventag)$\\
         \cline{2-3}
         &$\sqb{12}^2\aNbr{3}{4}^2\delta^{AB} $&$\ctw\stw(-\Ctwentythree+ \Ctwentyeight)$\\
         \cline{2-3}
         &$(\sqb{13}^2\anb{2\mathbf{4}}^2+(1\leftrightarrow 2)  )\delta^{AB}$&$\ctw\stw(-\Ctwentyfour+ \Ctwentynine)$\\
         \hline
         %%%%%%%%%%%%%%%
         $\fourptamp{g^A}{g^B}{g^C}{Z}$&$(\sqb{12}^2\sqb{3\mathbf{4}}^2+\sqb{13}^2\sqb{2\mathbf{4}}^2+\sqb{1\mathbf{4}}^2\sqb{23}^2)d^{ABC} $ &$\stw \Cthirty$\\
         \cline{2-3}
        &$(\sqb{12}^2\anb{3\mathbf{4}}^2+\text{Perm}(123))d^{ABC} $ &$\stw \Cthirtyone$\\
        \hline
        %%%%%%%%%%%%%%%
         $\fourptamp{g^A}{g^B}{g^C}{\gamma}$&$(\sqb{12}^2\sqb{3\mathbf{4}}^2+\sqb{13}^2\sqb{2\mathbf{4}}^2+\sqb{1\mathbf{4}}^2\sqb{23}^2)d^{ABC} $ &$\ctw \Cthirty$\\
         \cline{2-3}
        &$(\sqb{12}^2\anb{3\mathbf{4}}^2+\text{Perm}(123))d^{ABC} $ &$\ctw \Cthirtyone$\\
         \hline
         %%%%%%%%%%%%%%%
         $\mathcal{A}(g^Ag^B\gamma\gamma)$&$\sqb{12}^2\sqb{34}^2 \delta^{AB} $ &$\stw^2 \Ctwentyone + \ctw^2 \Ctwentyseven$\\
         \cline{2-3}
         &$(\sqb{13}^2\sqb{24}^2+\sqb{14}^2\sqb{23}^2)\delta^{AB} $ &$\stw^2 \Ctwentytwo + \ctw^2 \Ctwentyseventag$\\
         \cline{2-3}
         &$\sqb{12}^2\anb{34}^2 \delta^{AB} $&$\stw^2 \Ctwentythree + \ctw^2 \Ctwentyeight$\\
         \cline{2-3}
         &$(\sqb{13}^2\anb{24}^2+(1\leftrightarrow 2)  )\delta^{AB}$&$\stw^2 \Ctwentyfour + \stw^2 \Ctwentynine$\\
         \hline
         \caption{Dimension-8 LE four-vector amplitudes. 
         $c_W$ ($s_W$) denotes the cosine (sine) of the Weinberg mixing angle. 
         Four-gluon amplitudes are not shown here, since they remain as in Table~\ref{table:vvvv4}}
\label{T1}
\end{longtable}
\end{landscape}
%%%%%%%%%%%%%%%%%%%%%%%%%%%%%%%%%%%%%%%%%%%%%%%%%%%%%%%%%%%%%%%%%%%%%%%%%%%%%%%%%%%%%%%%%%%%%%%%%%%%%%%%%
\subsection{High-energy fermion contact-terms}
%%%%%%%%%%%%%%%%%%%%%%%%%%%%%%%%%%%%%%%%%%%%%%%%%%%%%%%%%%%%%%%%%%%%%%%%%%%%%%%%%%%%%%%%%%%%%%%
\begin{longtable}[|c|]{|c|c|c|}
       \hline
       Amplitude & Massless Contact-Term & Wilson Coefficient \\
       \hline \hline
       \endfirsthead
       \hline
       \multicolumn{3}{|c|}{Continued from Table~\ref{Tferm}}\\
       \hline
       Amplitude & Massless Contact-Term & Wilson Coefficient \\
       \hline \hline
       \endhead
       \hline
       \endfoot
       \endlastfoot
        \multicolumn{3}{|c|}{\textbf{$f_+f_-ss$}}\\
        \hline
        %%%%%%%%%%%%%%%
        \multirow{1}{*}{$Q^{c, j}Q_iH_kH^{\dagger,l}$}&$\sab{132}s_{34}\delta_i^{j}\delta_k^l$ &$c_1^{Q^cQ2H}$\\
        \cline{2-3}
        &$\sab{132}s_{34}\delta_i^{k}\delta^j_{l}$ &$c_2^{Q^cQ2H}$\\
        \cline{2-3}
        &$\sab{132}s_{14}\delta_i^{j}\delta_k^l$ &$c_3^{Q^cQ2H}$\\
        \cline{2-3}
        &$\sab{132}s_{14}\delta_i^{k}\delta^j_{l}$ &$c_4^{Q^cQ2H}$\\
        \hline
        %%%%%%%%%%%%%%%
        $U^cUH_iH^{\dagger,j}$&$\asb{132}s_{34}\delta^j_i$&$c_1^{U^cU2H}$\\
        \cline{2-3}
        &$\asb{132}s_{14}\delta^j_i$&$c_2^{U^cU2H}$\\
        \hline
        %%%%%%%%%%%%%%%
        $U^cDH_iH_j$&$\asb{1(3-4)2}s_{34} \epsilon_{ij}$&$c^{U^cD2H}$\\
        \hline
        %%%%%%%%%%%%%%%
        \multicolumn{3}{|c|}{\textbf{$f_+f_-V_+V_{\pm}$ and $f_-f_+V_+V_{\pm}$}}\\
        \hline
        %%%%%%%%%%%%%%%
        $Q^{c,j}Q_iW^a_+W^b_+$ & $\sqb{34}^2\sab{1(3-4)2}\epsilon^{abc}(\tau^c)_{i}^{\;j}$ & $c^{Q^cQ2W_+}$\\
        \hline
        %%%%%%%%%%%%%%%
        \multirow{1}{*}{$Q^{c,j}Q_ iW^a_+W^b_-$}&$\sqb{13}\anb{24}\sab{3(1-2)4}\delta^{ab}\delta_i^j$&$c_+^{Q^cQW_+W_-}$\\
        \cline{2-3}
        &$\sqb{13}\anb{24}\sab{3(1-2)4} \epsilon^{abc}(\tau^c)_{i}^{\; j}$&$c_-^{Q^cQW_+W_-}$\\
        \hline
        %%%%%%%%%%%%%%%
        $Q^{c,j}Q_iB_+B_+$ &$0$ & $-$\\
        \hline
        %%%%%%%%%%%%%%%
        $Q^{c,j}Q_iB_+B_-$ &$\sqb{13}\anb{24}\sab{3(1-2)4}\delta_i^j$ & $c^{Q^cQB_+B_-}$\\
        \hline
        %%%%%%%%%%%%%%%
        $Q^{c,\beta j}Q_{\alpha i}g^A_+g^B_+$
        &$\sqb{34}^2\sab{1(3-4)2}f^{ABC}(\lambda^C)_{\alpha}^{ \;\beta}\delta_i^j$ &$c^{Q^cQ2g_+}$\\
         \hline
         %%%%%%%%%%%%%%%
         \multirow{1}{*}{$Q^{c,\beta j}Q_{\alpha i}g^A_+g^B_-$}
        &$\sqb{13}\anb{24}\sab{3(1-2)4}\delta^{ab}\delta_{\beta}^{\alpha}\delta_i^j$ &$c_1^{Q^cQg_+g_-}$\\
         \cline{2-3}
        &$\sqb{13}\anb{24}\sab{3(1-2)4}f^{ABC}(\lambda^C)_{\alpha}^{ \;\beta} \delta_i^j$ &$c_2^{Q^cQg_+g_-}$\\
         \cline{2-3}
        &$\sqb{13}\anb{24}\sab{3(1-2)4}d^{ABC}(\lambda^C)_{\alpha}^{ \;\beta} \delta_i^j$ &$c_3^{Q^cQg_+g_-}$\\
         \hline
         %%%%%%%%%%%%%%%
         $Q^{c,\beta j}Q_{\alpha i}g^A_+B_+$
        &$\sqb{34}^2\sab{1(3-4)2}(\lambda^A)^{\;\beta}_{\alpha} \delta_i^j$ &$c^{Q^cQg_+B_+}$\\
        \hline
        %%%%%%%%%%%%%%%
         $Q^{c,\beta j}Q_{\alpha i}g^A_+B_-$
        &$\sqb{13}\anb{24}\sab{3(1-2)4}(\lambda^A)^{\;\beta}_{\alpha}\delta_i^j$ &$c^{Q^cQg_+B_-}$\\
        \hline
        %%%%%%%%%%%%%%%
        $Q^{c,\beta j}Q_{\alpha i}g^A_+W^a_+$ & $\sqb{34}^2\sab{1(3-4)2}(\lambda^A)^{\;\beta}_{\alpha} (\tau^a)_{i}^{\;j}$ &$c^{Q^cQg_+W_+}$\\
        \hline
        %%%%%%%%%%%%%%%
        $Q^{c,\beta j}Q_{\alpha i}g^A_+W^a_-$&$\sqb{13}\anb{24}\sab{3(1-2)4}(\lambda^A)^{\;\beta}_{\alpha} (\tau^a)^{\; j}_i$ &$c^{Q^cQg_+W_-}$\\
        \hline
        %%%%%%%%%%%%%%%
        $Q^{c, j}Q_iW^a_+B_+$&$\sqb{34}^2\sab{1(3-4)2} (\tau^a)_{i}^{\;j}$ &$c^{Q^cQW_+B_+}$\\
        \hline
        %%%%%%%%%%%%%%%
        $Q^{c, j}Q_iW^a_+B_-$&$\sqb{13}\anb{24}\sab{3(1-2)4} (\tau^a)_{i}^{\;j}$ &$c^{Q^cQW_+B_-}$\\
        \hline
        %%%%%%%%%%%%%%%
        $U^cUW^a_+W^b_+$ & $0$ & $-$\\
        \hline
        %%%%%%%%%%%%%%%
        $U^cUW^a_+W^b_-$&$\sqb{23}\anb{14}\sab{3(1-2)4}\delta^{ab}$&$c^{U^cUW_+W_-}$\\
        \hline
        %%%%%%%%%%%%%%%
        $U^cUW^a_+B_{\pm}$&$0$&$-$\\
        \hline
        %%%%%%%%%%%%%%%
        $U^cUB_+B_+$ &$0$ & $-$\\
        \hline
        %%%%%%%%%%%%%%%
        $U^cUB_+B_-$ &$\sqb{23}\anb{14}\sab{3(1-2)4}$ & $c^{U^cUB_+B_-}$\\
        \hline
        %%%%%%%%%%%%%%%
        $U^{c,\beta}U_{\alpha}g^A_+g^B_+$
        &$\sqb{34}^2\asb{1(3-4)2}f^{ABC}(\lambda^C)_{\; \beta}^{\alpha}$ &$c^{U^cU2g_+}$\\
         \hline
         %%%%%%%%%%%%%%%
         \multirow{1}{*}{$U^{c,\beta}U_{\alpha}g^A_+g^B_-$}
        &$\sqb{23}\anb{14}\sab{3(1-2)4}\delta^{ab}\delta_{\beta}^{\alpha}$ &$c_1^{U^cUg_+g_-}$\\
         \cline{2-3}
        &$\sqb{23}\anb{14}\sab{3(1-2)4}f^{ABC}(\lambda^C)_{\alpha}^{ \;\beta}$ &$c_2^{U^cUg_+g_-}$\\
         \cline{2-3}
        &$\sqb{23}\anb{14}\sab{3(1-2)4}d^{ABC}(\lambda^C)_{\alpha}^{ \;\beta} $ &$c_3^{U^cUg_+g_-}$\\
         \hline
         %%%%%%%%%%%%%%%
         $U^{c,\beta}U_{\alpha}g^A_+B_+$
        &$\sqb{34}^2\asb{1(3-4)2}(\lambda^A)_{\alpha}^{\;\beta} $ &$c^{U^cUg_+B_+}$\\
        \hline
        %%%%%%%%%%%%%%%
         $U^{c,\beta}U_{\alpha}g^A_+B_-$
        &$\sqb{23}\anb{14}\sab{3(1-2)4}(\lambda^A)_{\alpha}^{\;\beta}$ &$c^{U^cUg_+B_-}$\\
        \hline
        %%%%%%%%%%%%%%%
        $U^cUg^A_+W^a_{\pm}$&$0$ &$-$\\
        \hline
        %%%%%%%%%%%%%%%
        \multicolumn{3}{|c|}{\textbf{$f_{\mp}f_{\pm}V_+ss$}}\\
        \hline
        %%%%%%%%%%%%%%%
        \multirow{1}{*}{$Q^{c,j}Q_iW^a_+H^{\dagger k}H_l$}&$\sqb{13}\sab{342}(\tau^a)_{i}^{\;j}\delta^k_l$ &$c_1^{Q^cQW_+2H}$\\
        \cline{2-3}
        &$\sqb{13}\sab{342}(\tau^a)^{\; k}_{l}\delta_i^j$&$c_2^{Q^cQW_+2H}$\\
        \cline{2-3}
        &$\sqb{13}\sab{342}\epsilon^{abc}(\tau^b)^{ j}_{\;i}(\tau^c)^{ k}_{\;l}$&$c_3^{Q^cQW_+2H}$\\
        \cline{2-3}
       &$\sqb{13}\sab{352}(\tau^a)_{i}^{\;j}\delta^k_l$ &$c_4^{Q^cQW_+2H}$\\
        \cline{2-3}
        &$\sqb{13}\sab{352}(\tau^a)^{\; k}_{l}\delta_i^j$&$c_5^{Q^cQW_+2H}$\\
        \cline{2-3}
        &$\sqb{13}\sab{352}\epsilon^{abc}(\tau^b)^{ j}_{\;i}(\tau^c)^{ k}_{\;l}$&$c_6^{Q^cQW_+2H}$\\
        \hline
        %%%%%%%%%%%%%%%
        \multirow{1}{*}{$Q^{c,j}Q_iB_+H^{\dagger k}H_l$}&$\sqb{13}\sab{342}\delta_i^j\delta^k_l$&$c_1^{Q^cQB_+2H}$\\
        \cline{2-3}
        &$\sqb{13}\sab{342}\delta^j_l\delta^k_i$&$c_2^{Q^cQB_+2H}$\\
        \cline{2-3}
        &$\sqb{13}\sab{352}\delta_i^j\delta^k_l$&$c_3^{Q^cQB_+2H}$\\
        \cline{2-3}
        &$\sqb{13}\sab{352}\delta^k_i\delta^j_l$&$c_4^{Q^cQB_+2H}$\\
        \hline
        %%%%%%%%%%%%%%%
        \multirow{1}{*}{$Q^{c,\beta j}Q_{\alpha i}g^A_+H^{\dagger k}H_l$}&$\sqb{13}\sab{342}\delta_i^j\delta^k_l(\lambda^A)_{\alpha}^{\;\beta}$&$c_1^{Q^cQg_+2H}$\\
        \cline{2-3}
        &$\sqb{13}\sab{342}\delta^k_i\delta^j_l(\lambda^A)_{\alpha}^{\;\beta}$&$c_2^{Q^cQg_+2H}$\\
        \cline{2-3}
        &$\sqb{13}\sab{352}\delta_i^j\delta^k_l(\lambda^A)_{\alpha}^{\;\beta}$&$c_3^{Q^cQg_+2H}$\\
        \cline{2-3}
        &$\sqb{13}\sab{352}\delta^k_i\delta^j_l(\lambda^A)_{\alpha}^{\;\beta}$&$c_4^{Q^cQg_+2H}$\\
        \hline
        %%%%%%%%%%%%%%%
        \multirow{1}{*}{$U^cUW^a_+H^{\dagger,j}H_i$}&$\sqb{23}\sab{341}(\tau^a)_{\; i}^j$&$c_1^{U^cUW_+2H}$\\
        \cline{2-3}
        &$\sqb{23}\sab{351}(\tau^a)_{ i}^{\;j}$&$c_2^{U^cUW_+2H}$\\
        \hline
        %%%%%%%%%%%%%%%
        \multirow{1}{*}{$U^cUB_+H^{\dagger,j}H_i$}&$\sqb{23}\sab{341}\delta_i^j$&$c_1^{U^cUB_+2H}$\\
        \cline{2-3}
        &$\sqb{23}\sab{351}\delta_i^j$&$c_2^{U^cUB_+2H}$\\
        \hline
        %%%%%%%%%%%%%%%
        \multirow{1}{*}{$U^{c,\beta}U_{\alpha}g^A_+H^{\dagger,j}H_i$}&$\sqb{23}\sab{341}\delta_i^j(\lambda^A)_{\alpha}^{\;\beta}$&$c_1^{U^cUg_+2H}$\\
        \cline{2-3}
        &$\sqb{23}\sab{351}\delta_i^j(\lambda^A)_{\alpha}^{\;\beta}$&$c_2^{U^cUg_+2H}$\\
        \hline
        %%%%%%%%%%%%%%%
        \multirow{1}{*}{$U^cDW^a_+H_iH_i$}&$\sqb{23}(\sab{341}+\sab{351})(\epsilon_{ik}(\tau^a)_{j}^{\;k}+ \epsilon_{jk}(\tau^a)_{i}^{\;k})$  & $c_+^{U^cDW_+2H}$\\
        \hline
        %%%%%%%%%%%%%%%
        $U^cDB_+H_iH_i$& $\sqb{23}(\sab{341}-\sab{351})\epsilon_{ij}$ &$c^{U^cDB_+2H
        }$\\
        \hline
        %%%%%%%%%%%%%%%
        $(U^{c})^{\beta}D_{\alpha}g^A_+H_iH_i$&$\sqb{23}(\sab{341}-\sab{351})\epsilon_{ij}(\lambda^A)_{\alpha}^{\;\beta}$ &$c^{U^cDg_+2H
        }$\\
        \hline
        %%%%%%%%%%%%%%%
        \multirow{1}{*}{$D^cUW^a_+H^{\dagger,i}H^{\dagger,j}$}&$\sqb{23}(\sab{341}+\sab{351})(\epsilon^{ik}(\tau^a)_{k}^{\;j}+ \epsilon^{jk}(\tau^a)_{k}^{\;j})$  & $c_+^{D^cUW_+2H}$\\
        \hline
        %%%%%%%%%%%%%%%
        $D^cUB_+H^{\dagger,i}H^{\dagger,j}$& $\sqb{23}(\sab{341}-\sab{351})\epsilon_{ij}$ &$c^{D^cUB_+2H
        }$\\
        \hline
        %%%%%%%%%%%%%%%
        $(D^{c})^{\beta}U_{\alpha}g^A_+H^{\dagger,i}H^{\dagger,j}$&$\sqb{23}(\sab{341}-\sab{351})\epsilon_{ij}(\lambda^A)_{\alpha}^{\;\beta}$ &$c^{D^cUg_+2H
        }$\\
        \caption{Dimension-8 HE structures for amplitudes with chirality-conserving fermionic initial states. Amplitudes with $D^cD$ have the same structures as their $U^cU$ counterparts and are omitted for brevity. Their coefficients are defined by replacing $U^cU$ in the respective coefficients with $D^cD$. When the color index is omitted, the color factor is just $\delta^{\beta}_{\alpha}$. The table is organized by helicity categories, which are listed as full lines and in bold. 0 means the structure cannot be generated at $d\leq8$. }
    \label{Tferm}
\end{longtable}
%%%%%%%%%%%%%%%%%%%%%%%%%%%%%%%%%%%%%%%%%%%%%%%%%%%%%%%%%%%%%%%%%%%%%%%%%%%%%%%%%%%%%%%
%%%%%%%%%%%%%%%%%%%%%%%%%%%%%%%%%%%%%%%%%%%%%%%%%%%%%%%%%%%%%%%%%%%%%%%%%%%%%%%%%%%%%%%
%%%%%%%%%%%%%%%%%%%%%%%%%%%%%%%%%%%%%%%%%%%%%%%%%%%
\begin{landscape}
\subsection{Low-energy fermion contact terms}
%%%%%%%%%%%%%%%%%%%%%%%%%%%%%%%%%%%%%%%%%%%%%%%%%%%%%%%%%%%%%%%%%%%%%%%%%%%%%%
%%%%%%%%%%%%%%%%%%%%%%%%%%%%%%%%%%%%%%%%%%%%%%%%%%%%%%%%%%%%%%%%%%%%%%%%%%%%%%%%%%%%%%%%%%%%%%
\begin{longtable}{|c|c|>{\centering\arraybackslash}p{5in}|}  
\hline
    Amplitude & Massive Contact-Term & Wilson Coefficient \\
    \hline
    \endfirsthead
    \hline
    \multicolumn{3}{|c|}{Continued from Table~\ref{ChiYes}}\\
    \hline
    Amplitude & Massive Contact-Term & Wilson Coefficient \\
    \hline
    \endhead
    \hline
    \endfoot
    \endlastfoot
\multirow{1}{*}{$\mathcal{A}(\overline{u}uW^+W^-)$} &$\Sqb{34}^2(\sab{1(\three-\four) 2}) $&$\frac{i}{2}\Cffifteen$\\
         \cline{2-3}
         &$\sQbr{1}{3}\aNbr{2}{4}(\sab{\three (1-2) \four})$&$\Cfsixteen+\frac{i}{2}\Cfseventeen$\\
         \cline{2-3}
         &\multirow{1}{*}{$\sQbr{1}{3}\Sqb{34}\anb{\four 2} $}&$\frac{v}{\sqrt{2}}(\Cffortyone+\Cffortytwo)$\\
         \cline{2-3}
         &\multirow{1}{*}{$\sQbr{1}{4}\Sqb{43}\anb{\three 2} $}&$\frac{v}{\sqrt{2}} \Cffortyseven $\\
         \cline{2-3}
         &$\Sqb{34}\Anb{34}\sqb{1\three }\anb{\three 2}$ & $2(\Cfninetyfour+\Cfninetyfive)$\\
         \cline{2-3}
         &$\sab{\four 1 \four}\sqb{1\three }\anb{\three 2}$ & $2(\Cfninetyfourtag+\Cfninetyfivetag)$\\
         \hline
         %%%%%%%%%%%%%%%
         \multirow{1}{*}{$\mathcal{A}(\overline{U}UW^+W^-)$ }
         &$\sqb{2\mathbf{3}}\anb{1\mathbf{4}}(\sab{\three (1-2) \four})$ &$\Cftwentyeight$ \\
         \cline{2-3}
         &$\sQbr{2}{3}\Sqb{34}\anb{\four 1} $&$-\frac{v}{\sqrt{2}}\Cfsixty$\\
         \cline{2-3}
         &$\sQbr{2}{4}\Sqb{43}\anb{\three 1} $&$\frac{v}{\sqrt{2}}\Cfsixtyone$\\
         \cline{2-3}
         &$\sab{\four \three \four}\anb{1\three}\sqb{\three 2}$&$2\Cfhundredthirtythree$\\
         \cline{2-3}
         &$\sab{\four 1 \four}\anb{1\three}\sqb{\three 2}$&$2\Cfhundredthirtyfour$\\
         \hline
         %%%%%%%%%%%%%%%
         \multirow{1}{*}{$\mathcal{A}(\overline{d}dW^+W^-)$} &$\Sqb{34}^2(\sab{1(\three-\four) 2}) $ &$-\frac{i}{2}\Cffifteen$ \\
         \cline{2-3}
         &$\sQbr{1}{3}\aNbr{2}{4}(\sab{\three (1-2) \four})$&$\frac{1}{2}(2\Cfsixteen-i \Cfseventeen)$\\
         \cline{2-3}
         &\multirow{1}{*}{$\sQbr{1}{4}\Sqb{43}\anb{\three 2} $}&$-\frac{v}{\sqrt{2}}(\Cffortyseven + \Cffortyeight)$\\
         \cline{2-3}
         &\multirow{1}{*}{$\sQbr{1}{3}\Sqb{34}\anb{\four 2} $}&$\frac{v}{\sqrt{2}}\Cffortyone$\\
         \cline{2-3}
         &$\sQbr{1}{3}\aNbr{2}{3}\sab{\four \three \four}$&$2\Cfninetyfour$\\
         \cline{2-3}
         &$\sQbr{1}{3}\aNbr{2}{3}\sab{\four 1 \four}$&$2\Cfninetyfourtag$\\
         \hline
         %%%%%%%%%%%%%%%
         \multirow{1}{*}{$\mathcal{A}(\overline{D}DW^+W^-)$} &$\aNbr{1}{3}\sQbr{2}{4}(\sab{\four (1-2) \three})$ &$\Cfninetyseven$ \\
         \cline{2-3}
         &$\sQbr{2}{4}\Sqb{43}\anb{\three 1} $&$\frac{v}{\sqrt{2}}\Cfhundredsix$\\
         \cline{2-3}
         &$\aNbr{1}{3}\sQbl{3}{2}\sab{\four \three \four}$&$2\Cfhundredthirtyfive$\\
         \cline{2-3}
         &$\aNbr{1}{3}\sQbl{3}{2}\sab{\four 1 \four}$&$2\Cfhundredthirtysix$\\
         \hline
         %%%%%%%%%%%%%%%
         \multirow{1}{*}{$\mathcal{A}(\overline{u}uZZ)$}&$\sqb{1\mathbf{3}}\anb{2\mathbf{4}}(\sab{\three (1-2) \four})+(3\leftrightarrow 4)$ &$\ctw^2 \Cfsixteen+\stw^2 \Cfeighteen+\ctw \stw \Cftwentyseventag$\\
         \cline{2-3}
         &\multirow{1}{*}{$\sqb{1\mathbf{3}}\Sqb{34}\anb{\four 2}+(3\leftrightarrow 4) $}&$-\frac{iv}{2\sqrt{2}}\big(\ctw (\Cfforty-\Cffortyone-\Cffortysix+\Cffortyseven)+2\stw (\Cffiftytwo-\Cffiftyfour)\big)$\\
         \cline{2-3}
         &$\sQbr{1}{3}\aNbl{3}{2}\sab{\four 1 \four}+(3\leftrightarrow 4)$&$2\Cfninetyfourtag$\\
         \hline
         %%%%%%%%%%%%%%%
         \multirow{1}{*}{$\mathcal{A}(\overline{U}UZZ)$}&$\anb{1\mathbf{4}}\sqb{2\mathbf{3}}(\sab{\three (1-2) \four})+(3\leftrightarrow 4)$ &$\ctw^2 \Cftwentyeight+\stw^2 \Cftwentynine$\\
         \cline{2-3}
         &\multirow{1}{*}{$\sqb{2\mathbf{3}}\Sqb{34}\anb{\four 1}+(3\leftrightarrow 4) $}&$-\frac{iv}{2\sqrt{2}}\big(\ctw(\Cfsixty-\Cfsixtyone)+2\stw (\Cfsixtythree-\Cfsixtytwo)\big)$\\
         \cline{2-3}
         &$\anb{1\three}\sqb{\three 2}\sab{\four 1 \four}+(3\leftrightarrow 4)$&$-2\Cfhundredthirtyfour$\\
         \hline
         %%%%%%%%%%%%%%%
          \multirow{1}{*}{$\mathcal{A}(\overline{d}dZZ)$}&$\sQbr{1}{3}\aNbr{2}{4}(\sab{\three (1-2) \four})+(3\leftrightarrow 4)$&$\ctw^2 \Cfsixteen + \stw^2 \Cfeighteen - \ctw\stw \Cftwentyseventag$\\
         \cline{2-3}
         &\multirow{1}{*}{$\sQbr{1}{3}\Sqb{34}\anb{\four 2}+(3\leftrightarrow 4) $}&$-\frac{iv}{2\sqrt{2}}\big(\ctw (\Cfforty + \Cffortyone + \Cffortytwo - \Cffortysix - \Cffortyseven - \Cffortyeight) + 2 \stw (\Cffiftyfour + \Cffiftyfive-\Cffiftytwo - \Cffiftythree)\big)$\\
         \cline{2-3}
         &$\sQbr{1}{3}\aNbl{3}{2}\sab{\four 1 \four}+(3\leftrightarrow 4)$&$\Cfninetyfourtag+\Cfninetyfivetag$\\
         \hline
         %%%%%%%%%%%%%%%
         \multirow{1}{*}{$\mathcal{A}(\overline{D}DZZ)$}&$\aNbr{1}{4}\sQbr{2}{3}(\sab{\three (1-2) \four})+(3\leftrightarrow 4)$ &$\ctw^2 \Cfninetyseven + \stw^2 \Cfninetyeight$\\
         \cline{2-3}
         &\multirow{1}{*}{$\sQbr{2}{4}\Sqb{43}\anb{\three 1}+(3\leftrightarrow 4) $}&$\frac{iv}{2\sqrt{2}}(\ctw (\Cfhundredfive - \Cfhundredsix)+ 2 \stw (\Cfhundredeight-\Cfhundredseven))$\\
         \cline{2-3}
         &$\anb{1\three}\sqb{\three 2}\sab{\four 1 \four}+(3\leftrightarrow 4) $&$\Cfhundredthirtysix$\\
         \hline
         %%%%%%%%%%%%%%%
         \multirow{1}{*}{$\mathcal{A}(\overline{u}uZ\gamma_+)$}&$\sQbl{3}{4}^2(\sab{1(\three-4)2}) $&$(\ctw^2-\stw^2)\Cftwentyseven$\\
        \cline{2-3}
        &\multirow{1}{*}{$\sqb{14}\sqb{4 \three}\anb{\three 2} $}&$\frac{iv}{4}\big(\stw (\Cfforty - \Cffortyone + \Cffortysix + \Cffortyseven ) + 2 \ctw (-\Cffiftytwo + \Cffiftyfour)\big)$\\
        \cline{2-3}
        &\multirow{1}{*}{$\sqb{14}\aNbr{2}{3}(\sab{4(1-2)\three})$}&$\frac{1}{2}\big(2 \ctw \stw (-\Cfsixteen + \Cfeighteen)+ (\ctw^2- \stw^2) \Cftwentyseventag\big)$\\
        \hline
        %%%%%%%%%%%%%%%
        \multirow{1}{*}{$\mathcal{A}(\overline{u}uZ\gamma_-)$}&\multirow{1}{*}{$\sQbr{1}{3}\anb{24}(\sab{\three (1-2)4})$}&$\frac{1}{2}\big((\ctw^2-\stw^2)\Cftwentyseventag +2\ctw \stw (\Cfeighteen-\Cfsixteen)\big)$\\
        \hline
        %%%%%%%%%%%%%%%
        \multirow{1}{*}{$\mathcal{A}(\overline{U}UZ\gamma_+)$}&$\aNbr{1}{3}\sqb{24}(\sab{4(1-2)\three})$&$\ctw \stw (\Cftwentynine-\Cftwentyeight)$\\
        \cline{2-3}
        &\multirow{1}{*}{$\sqb{24}\sqb{4\three}\anb{\three 1} $}&$\frac{iv}{2\sqrt{2}}\big(\stw (\Cfsixty - \Cfsixtyone) + 2 \ctw (\Cfsixtytwo - \Cfsixtythree)\big)$\\
        \hline
        %%%%%%%%%%%%%%%
        $\mathcal{A}(\overline{U}UZ\gamma_-)$&$\anb{14}\sQbr{2}{3}(\sab{\three 1 4 }-\sab{\three 2 4})$&$\ctw \stw (\Cftwentynine-\Cftwentyeight)$\\
        \hline
        %%%%%%%%%%%%%%%
        \multirow{1}{*}{$\mathcal{A}(\overline{d}dZ\gamma_+)$}&$\sQbl{3}{4}^2(\sqb{1\three}\anb{\three 2} - \sab{142})$&$-\frac{1}{2}\Cftwentyseven$\\
        \cline{2-3}
        &\multirow{1}{*}{$\sqb{14}\aNbr{2}{3}(\sab{4(1-2)\three})$}&$\frac{1}{2}(2\ctw \stw(\Cfeighteen-\Cfsixteen)+(\stw^2-\ctw^2)\Cftwentyseventag$\\
        \cline{2-3}
        &\multirow{1}{*}{$\sqb{14}\sqb{4 \three}\anb{\three 2} $}&$\frac{iv}{2\sqrt{2}}\big(\stw (\Cfforty + \Cffortyone + \Cffortytwo - \Cffortysix - \Cffortyseven - \Cffortyeight)+ 2 \ctw (\Cffiftytwo + \Cffiftythree - \Cffiftyfour - \Cffiftyfive))$\\
        \hline
        %%%%%%%%%%%%%%%
        \multirow{1}{*}{$\mathcal{A}(\overline{d}dZ \gamma_-)$}
        &\multirow{1}{*}{$\sqb{1\three}\anb{24}(\sab{\three (1-2) 4})$}&$\frac{1}{2}(2\ctw \stw(\Cfeighteen-\Cfsixteen)+(\stw^2-\ctw^2)\Cftwentyseventag$\\
        \hline
        %%%%%%%%%%%%%%%
        \multirow{1}{*}{$\mathcal{A}(\overline{D}DZ \gamma_+)$}&$\aNbr{1}{3}\sqb{24}(\sab{4(1-2)\three})$&$-\ctw \stw \Cfninetyseven$\\
        \cline{2-3}
        &\multirow{1}{*}{$\sqb{24}\sqb{4\three}\anb{\three 1} $}&$\frac{iv}{2\sqrt{2}}\big(\stw (\Cfhundredfive - \Cfhundredsix) + 2 \ctw (\Cfhundredseven - \Cfhundredeight))$\\
        \hline
        %%%%%%%%%%%%%%%
        $\mathcal{A}(\overline{D}DZ \gamma_-)$&$\anb{14}\sQbr{2}{3}(\sab{\three (1-2) 4})$&$-\ctw \stw \Cfninetyseven$\\
        \hline
        %%%%%%%%%%%%%%%
        \multirow{1}{*}{$\mathcal{A}(\overline{u}^{\beta}u_{\alpha}Zg^A_+)$}&$\sQbl{3}{4}^2(\sab{1(\three-4) 2} )(\lambda^A)_{\alpha}^{\;\beta}$&$\frac{1}{2}(2 \stw \Cftwentythree + \ctw \Cftwentyfive)$\\
        \cline{2-3}
        &$\sqb{14}\aNbr{2}{3}(\sab{4(1-2)\three})(\lambda^A)_{\alpha}^{\;\beta}$&$\frac{1}{2}(2 \stw \Cftwentyfour + \ctw \Cftwentysix)$\\
        \cline{2-3}
        &$\sqb{14}\sqb{4\three}\anb{\three 2}(\lambda^A)_{\alpha}^{\;\beta} $&$\frac{iv}{2\sqrt{2}}(\Cffiftysix - \Cffiftyeight)$\\
        \hline
        %%%%%%%%%%%%%%%
        $\mathcal{A}(\overline{u}^{\beta}u_{\alpha}Zg^A_-)$&$\sQbr{1}{3}\anb{24}(\sab{\three (1-2) 4})(\lambda^A)_{\alpha}^{\;\beta}$&$\frac{1}{2}(2 \stw \Cftwentyfour + \ctw \Cftwentysix)$\\
        \hline
        %%%%%%%%%%%%%%%
        \multirow{1}{*}{$\mathcal{A}(\overline{U}^{\beta}U_{\alpha}Zg^A_+)$}&$\sQbl{3}{4}^2(\asb{1(4-\three)2})(\lambda^A)_{\alpha}^{\;\beta}$&$\stw \Cfthirtyfour$\\
        \cline{2-3}
        &$\sqb{24}\aNbr{1}{3}(\sab{4(1-2)\three})(\lambda^A)_{\alpha}^{\;\beta}$&$\stw \Cfthirtyfive$\\
        \cline{2-3}
        &$\sqb{24}\sqb{4\three}\anb{\three 1}(\lambda^A)_{\alpha}^{\;\beta} $&$\frac{iv}{2\sqrt{2}}(\Cfsixtytwotag-\Cfsixtythreetag)$\\
        \hline
        %%%%%%%%%%%%%%%
        $\mathcal{A}(\overline{U}^{\beta}U_{\alpha}Zg^A_-)$&$\anb{14}\sQbr{2}{3}(\sab{\three (1-2) 4})\ellA$&$\stw \Cfthirtyfive$\\
        \hline
        %%%%%%%%%%%%%%%
        \multirow{1}{*}{$\mathcal{A}(\overline{d}^{\beta}d_{\alpha}Zg_+)$}&$\sQbl{3}{4}^2(\sab{1(\three-4) 2} )\ellA$&$\frac{1}{2}(2\stw \Cftwentythree-\ctw \Cftwentyfive)$\\
         \cline{2-3}
         &$\sqb{14}\aNbr{2}{3}(\sab{4(1-2)\three})\ellA$&$\frac{1}{2}(2\stw \Cftwentyfour-\ctw \Cftwentysix)$\\
         \cline{2-3}
         &$\sqb{14}\sqb{4 \three}\anb{\three 2}\ellA$&$\frac{iv}{\sqrt{2}}(\Cffiftysix+\Cffiftyseven-\Cffiftyeight-\Cffiftynine)$\\
         \hline
         %%%%%%%%%%%%%%%
         \multirow{1}{*}{$\mathcal{A}(\overline{d}^{\beta}d_{\alpha}Zg_-)$}&$\anb{24}\sQbr{1}{3}(\sab{\three (1-2) 4})\ellA$&$\frac{1}{2}(2\stw \Cftwentyfour-\ctw \Cftwentysix)$\\
         \hline
         %%%%%%%%%%%%%%%
         \multirow{1}{*}{$\mathcal{A}(\overline{D}^{\beta}D_{\alpha}Zg_+)$}&$(\asb{1(4-\three)2})\sQbl{3}{4}^2\ellA$&$\stw \Cfhundredthree$\\
         \cline{2-3}
         &$\sqb{24}\aNbr{1}{3}(\sab{4(1-2)\three})\ellA$&$\stw \Cfhundredfour$\\
         \cline{2-3}
         &$\sqb{24}\sqb{4\three}\anb{\three 1}\ellA  $&$\frac{iv}{\sqrt{2}}(\Cfhundrednine-\Cfhundredten)$\\
         \hline
         %%%%%%%%%%%%%%%
         $\mathcal{A}(\overline{D}^{\beta}D_{\alpha}Zg_-)$&$\sQbr{2}{3}\anb{14}(\sab{\three (1-2) 4})\ellA$&$\stw \Cfhundredfour$\\
        \hline
        %%%%%%%%%%%%%%%
         $\mathcal{A}(\overline{u}u\gamma_+\gamma_-)$&$\sqb{13}\anb{24}(\sab{3(1-2)4})$&$\stw^2 \Cfsixteen + \ctw^2 \Cfeighteen - \ctw \stw \Cftwentyseventag$\\
        \hline
        %%%%%%%%%%%%%%%
        $\mathcal{A}(\overline{U}U\gamma_+\gamma_-)$&$\sqb{23}\anb{14}(\sab{3(1-2)4})$&$\stw^2 \Cftwentyeight + \ctw^2 \Cftwentynine$\\
        \hline
        %%%%%%%%%%%%%%%
        $\mathcal{A}(\overline{U}U\gamma_-\gamma_+)$&$\sqb{24}\anb{13}(\sab{4(1-2)3})$&$\stw^2 \Cftwentyeight + \ctw^2 \Cftwentynine$\\
        \hline
        %%%%%%%%%%%%%%%
        $\mathcal{A}(\overline{u}_{\pm}u_{\mp}\gamma_{\pm}\gamma_{\pm})$&$0$&$-$\\
        \hline
        %%%%%%%%%%%%%%%
        $\mathcal{A}(\overline{d}d\gamma_+ \gamma_-)$&$\sqb{13}\anb{24}(\sab{3(1-2)4})$&$\stw^2\Cfsixteen+\ctw^2 \Cfeighteen+\ctw\stw \Cftwentyseventag$\\
         \hline
         %%%%%%%%%%%%%%%
         $\mathcal{A}(\overline{d}d\gamma_- \gamma_+)$&$\sqb{14}\anb{23}(\sab{4(1-2)3})$&$\stw^2\Cfsixteen+\ctw^2 \Cfeighteen+\ctw\stw \Cftwentyseventag$\\
         \hline
         %%%%%%%%%%%%%%%
         $\mathcal{A}(\overline{D}D\gamma_+ \gamma_-)$&$\anb{14}\sqb{23}(\sab{3(1-2)4})$&$\stw^2 \Cfninetyseven+\ctw^2 \Cfninetyeight$\\
         \hline
         %%%%%%%%%%%%%%%
         $\mathcal{A}(\overline{d}_{\pm}d_{\mp}\gamma_{\pm}\gamma_{\pm})$&$0$&$-$\\
        \hline
        %%%%%%%%%%%%%%%
        $\mathcal{A}(\overline{u}^{\beta}u_{\alpha}\gamma_+g_+^A)$&$\sqb{34}^2(\sab{1(3-4)2})(\lambda^A)_{\alpha}^{\;\beta} $&$\frac{1}{2}(2 \ctw \Cftwentythree - \stw \Cftwenty)$\\
        \hline 
        %%%%%%%%%%%%%%%
        $\mathcal{A}(\overline{u}^{\beta}u_{\alpha}\gamma_+g_-^A)$&$\sqb{13}\anb{24}(\sab{3(1-2)4})(\lambda^A)_{\alpha}^{\;\beta}$&$\frac{1}{2}(2 \ctw \Cftwentyfour - \stw \Cftwentysix)$\\
        \hline 
        %%%%%%%%%%%%%%%
        $\mathcal{A}(\overline{u}^{\beta}u_{\alpha}\gamma_-g_+^A)$&$\sqb{14}\anb{23}(\sab{4(1-2)3})(\lambda^A)_{\alpha}^{\;\beta}$&$\frac{1}{2}(2 \ctw \Cftwentyfour - \stw \Cftwentysix)$\\
        \hline
        %%%%%%%%%%%%%%%
        $\mathcal{A}(\overline{U}U\gamma_+g^A_+)$&$\sqb{34}^2(\asb{1\three 2} - \asb{142})\ellA  $&$\ctw \Cfthirtyfour$\\
        \hline
        %%%%%%%%%%%%%%%
        $\mathcal{A}(\overline{U}^{\beta}U_{\alpha}\gamma_-g_+^A)$&$\sqb{14}\anb{23}(\sab{4(1-2)3}(\lambda^A)_{\alpha}^{\;\beta}$&$\ctw \Cfthirtyfive$\\
        \hline 
        %%%%%%%%%%%%%%%
        $\mathcal{A}(\overline{U}^{\beta}U_{\alpha}\gamma_+g_-^A)$&$\sqb{13}\anb{24}(\sab{3(1-2)4}(\lambda^A)_{\alpha}^{\;\beta}$&$\ctw \Cfthirtyfive$\\
        \hline
        %%%%%%%%%%%%%%%
         $\mathcal{A}(\overline{d}d\gamma_+ g_+)$&$\sqb{34}^2(\sab{1(3-4)2})\ellA  $&$\frac{1}{2}(2\ctw \Cftwentythree+\stw \Cftwentyfive)$\\
         \hline
         %%%%%%%%%%%%%%%
         $\mathcal{A}(\overline{d}d\gamma_+ g_-)$&$\sqb{13}\anb{24}(\sab{3(1-2)4}_\ellA$&$\frac{1}{2}(2\ctw \Cftwentyfour+\stw \Cftwentysix)$\\
         \hline
         %%%%%%%%%%%%%%%
         $\mathcal{A}(\overline{d}d\gamma_- g_+)$&$\sqb{14}\anb{23}(\sab{4(1-2)3}_\ellA$&$\frac{1}{2}(2\ctw \Cftwentyfour+\stw \Cftwentysix)$\\
         \hline
         %%%%%%%%%%%%%%%
         $\mathcal{A}(\overline{D}D\gamma_+ g_+)$&$\sqb{34}^2(\asb{142}-\asb{132})\ellA  $&$\ctw \Cfhundredthree$\\
         \hline
         %%%%%%%%%%%%%%%
         $\mathcal{A}(\overline{D}D\gamma_+ g_-)$&$\sqb{24}\anb{13}(\sab{4(1-2)3})$&$\ctw \Cfhundredfour$\\
         \hline
         %%%%%%%%%%%%%%%
         $\mathcal{A}(\overline{D}D\gamma_- g_+)$&$\sqb{23}\anb{14}(\sab{3(1-2)4})$&$\ctw \Cfhundredfour$\\
         \hline
         %%%%%%%%%%%%%%%
         \multirow{1}{*}{$\mathcal{A}(\overline{u}dW^+Z)$}&$\Sqb{34}^2
         \sab{1(\three-\four )2}
          $&$\frac{1}{\sqrt{2}}(-i \ctw \Cffifteen + \stw \Cftwentyseven)$\\
        \cline{2-3}
        &$\sQbr{1}{3}\aNbr{2}{4}(\sab{\three (1-2) \four})+(3\leftrightarrow 4)$&$-\frac{i}{\sqrt{2}}(\ctw \Cfseventeen)$\\
        \cline{2-3}
        &\multirow{1}{*}{$\sQbr{1}{3}\Sqb{34}\aNbl{4}{2} $}&$\frac{iv}{2}(2i\Cfforty+\Cffortytwo-2i \Cffortysix-\Cffortyeight)$\\
        \cline{2-3}
        &$\sQbr{1}{4}\Sqb{43}\anb{\three 2} $&$v\stw \Cffiftyfive$\\
        \cline{2-3}
        &$\sab{\four \three \four}\sQbr{1}{3}\aNbl{3}{2}$&$i\sqrt{2}\Cfninetyfive$\\
        \cline{2-3}
        &$\sab{\four 1 \four}\sQbr{1}{3}\aNbl{3}{2}$&$i\sqrt{2}\Cfninetyfivetag$\\
        \hline
        %%%%%%%%%%%%%%%
        %%%%%%%%%%%%%%%
        $\mathcal{A}(\overline{U}DW^+Z)$&$\sQbr{2}{4}\Sqb{43}\anb{\three 1} $&$v(\stw \Cfsixtysix-\ctw \Cfsixtyfive)$\\
        \cline{2-3}
        &$\sQbr{2}{3}\Sqb{24}\aNbr{1}{4} $&$i\sqrt{2}v\Cfsixtyfive$\\
        \cline{2-3}
        &$\Sqb{34}\Anb{34}(\asb{1\three 2}-\asb{1\four 2})$&$i\sqrt{2}\Cfninetysix$\\
        \hline
        %%%%%%%%%%%%%%%
         %%%%%%%%%%%%%%%
        \multirow{1}{*}{$\mathcal{A}(\overline{u}dW^+\gamma_+)$}&$\sQbl{3}{4}^2(\sab{1(\three-4) 2} )$&$\frac{1}{\sqrt{2}}(\ctw \Cftwentyseven +i\stw \Cffifteen)$\\
         \cline{2-3}
         &$\sqb{14}\aNbr{2}{3}(\sab{4(1-2)\three})$&$\frac{i}{\sqrt{2}}\stw \Cfseventeen$\\
         \cline{2-3}
         &\multirow{1}{*}{$\sQbr{1}{4}\sQbr{3}{4}\aNbr{2}{3} $}&$-\frac{v}{2}(\stw \Cffortyeight - 2 \ctw \Cffiftyfive)$\\
         \hline
         %%%%%%%%%%%%%%%
         $\mathcal{A}(\overline{u}dW^+\gamma_-)$&$\sQbr{1}{3}\anb{24}(\sab{\three (1-2) 4})$&$\frac{i}{\sqrt{2}}\stw \Cfseventeen$\\
         \hline
         %%%%%%%%%%%%%%%
         \multirow{1}{*}{$\mathcal{A}(\overline{d}uW^-\gamma_+)$}&$\sQbl{3}{4}^2(\sab{1(\three-4) 2} )$&$\frac{1}{\sqrt{2}}(\ctw \Cftwentyseven-i \stw \Cffifteen)$\\
         \cline{2-3}
         &$\sqb{14}\aNbr{2}{3}(\sab{4(1-2)\three})$&$-\frac{i}{\sqrt{2}}\stw \Cfseventeen$\\
         \cline{2-3}
         &\multirow{1}{*}{$\sqb{14}\sqb{4\three}\anb{\three 2} $}&$\frac{v}{2}(\stw \Cffortytwo + 2 \ctw \Cffiftythree)$\\
         \hline
         %%%%%%%%%%%%%%%
         $\mathcal{A}(\overline{d}uW^-\gamma_-)$&$\sQbr{1}{3}\anb{24}(\sab{\three (1-2) 4})$&$-\frac{i}{\sqrt{2}}\stw \Cfseventeen$\\
         \hline
         %%%%%%%%%%%%%%%
         $\mathcal{A}(\overline{U}DW^+\gamma_+)$&$\sqb{24}\sqb{4\three}\anb{\three 1} $&$2v(\stw \Cfsixtyfive+\ctw \Cfsixtysix)$\\
         \hline
         %%%%%%%%%%%%%%%
         $\mathcal{A}(\overline{U}DW^+\gamma_-)$&$0$&$-$\\
         \hline
         %%%%%%%%%%%%%%%
         $\mathcal{A}(\overline{D}UW^-\gamma_+)$&$\sqb{24}\sqb{4\three}\anb{\three 1} $&$2v(\stw \Cfhundredthirty + \ctw \Cfhundredthirtyone)$\\
         \hline
         %%%%%%%%%%%%%%%
         $\mathcal{A}(\overline{D}UW^-\gamma_-)$&0&-\\
         \hline
         %%%%%%%%%%%%%%%
         \multirow{1}{*}{$\mathcal{A}(\overline{u}^{\beta}d_{\alpha}W^+g_+)$}&$\sQbl{3}{4}^2(\sqb{1\three}\anb{\three 2} - \sab{142})(\lambda^A)_{\alpha}^{\;\beta}$&$\frac{1}{\sqrt{2}}\Cftwentyfive$ \\
         \cline{2-3}
         &$\sqb{14}\aNbr{2}{3}(\sab{4(1-2)\three})(\lambda^A)_{\alpha}^{\;\beta}$&$\frac{1}{\sqrt{2}}\Cftwentysix$\\
         \cline{2-3}
         &$\sqb{14}\sqb{4 \three}\anb{\three 2}(\lambda^A)_{\alpha}^{\;\beta} $&$v\Cffiftynine$\\
         \hline
         %%%%%%%%%%%%%%%
         $\mathcal{A}(\overline{u}^{\beta}d_{\alpha}W^+g_-)$&$\sQbr{1}{3}\anb{24}(\sab{\three (1-2) 4})(\lambda^A)_{\alpha}^{\;\beta}$&$\frac{1}{\sqrt{2}}\Cftwentysix$\\
         \hline
         %%%%%%%%%%%%%%%
         $\mathcal{A}(\overline{U}^{\beta}D_{\alpha}W^+g_+)$&$\sqb{24}\sqb{4\three}\anb{\three 1}\ellA  $&$v\Cfsixtyseven$\\
         \hline
         %%%%%%%%%%%%%%%
         \multirow{1}{*}{$\mathcal{A}(\overline{d}^{\beta}u_{\alpha}W^-g^A_+)$}&$\sQbl{3}{4}^2(\sab{1(\three-4) 2} )(\lambda^A)_{\alpha}^{\;\beta}$&$\frac{1}{\sqrt{2}}\Cftwentyfive$\\
        \cline{2-3}
        &$\sqb{14}\aNbr{2}{3}(\sab{4(1-2)\three})(\lambda^A)_{\alpha}^{\;\beta}$&$\frac{1}{\sqrt{2}}\Cftwentysix$\\
        \cline{2-3}
        &$\sqb{14}\sqb{4 \three}\anb{\three 2}(\lambda^A)_{\alpha}^{\;\beta} $&$v\Cffiftyseven$\\
        \hline
        %%%%%%%%%%%%%%%
        $\mathcal{A}(\overline{d}^{\beta}u_{\alpha}W^-g^A_-)$&$\sQbr{1}{3}\anb{24}(\sab{\three (1-2) 4})(\lambda^A)_{\alpha}^{\;\beta}$&$\frac{1}{\sqrt{2}}\Cftwentysix$\\
        \hline
        %%%%%%%%%%%%%%%
        $\mathcal{A}(\overline{D}^{\beta}U_{\alpha}W^-g^A_+)$&$\sqb{24}\sqb{4\three}\anb{\three 1}(\lambda^A)_{\alpha}^{\;\beta} $&$v\Cfhundredthirtytwo$\\
        \hline
        \caption{LE chirality-conserving CTs containing two fermions and two gauge bosons. Two-gluon states are omitted as they are in one-to-one correspondence with the HE amplitudes.}
        \label{ChiYes}
    \end{longtable}
\end{landscape}
%%%%%%%%%%%%%%%%%%%%%%%%%%%%%%%%%%%%%%%%%%%%%%%%%%%%%%%%%%%%%%%%%%%%%%%%%%%%%%%%%%%%%%%%%%%%%%%%%%%%%%%
%%%%%%%%%%%%%%%%%%%%%%%%%%%%%%%%%%%%%%%%%%%%%%%%%%%%%%%%%%%%%%%%%%%%%%%%%%%%%%%%%%%%%%%%%%%%%%%%%%%%%%%%%%%%%%%%%%%%%%%
\section{Coefficient counting, complex conjugation and CP}
\label{AppCount}
In this Appendix we discuss the 
relations between HE coefficients due to complex conjugation and discrete symmetries in
a few representative examples. 
The first example will also illustrate the mapping to the basis of operators. The third example  gives insight into the way that these HE relations between coefficients ensure that the LE amplitudes satisfy these same constraints. The bases  listed in Tables~\ref{table:vvvv4} and~\ref{Tferm} should be interpreted with these relations in mind.

Let's begin by examining the positive and negative helicity contributions to the $U^cUB_{\pm}2H$ amplitudes. Consider the basis,
\beqa
c_1 \sqb{23}\sab{3(4-5)1} + c_2 \sqb{23}\sab{3(4+5)1}
\label{eq16}
\eeqa
in which the action of charge conjugation on the Higgs fields is manifest. Complex conjugation takes the structure with, e.g,  $c_1$ to
\begin{equation}
    c_1^*\anb{13}\asb{3(4-5)1},
\end{equation}
which contributes to the amplitude $\mathcal{M}(1_{U}2_{U^c}3_{B_-}4_{H}5_{H^{\dagger}})$. 
We can relate this amplitude to the original one by relabeling $1\leftrightarrow 2$ and $4\leftrightarrow 5$ 
to obtain the structure
\begin{equation}
\label{eq18}
    -c_1^* \anb{23}\sab{3(4-5)1}.
\end{equation}
This structure, which na\"ively could be thought to  be independent, is related by complex conjugation to the first structure 
and as such comes with $-c_-^*$, instead of an independent complex coefficient. 
This means that instead of four 
independent real parameters we have two. 
Additionally, 
the two structures~\eqref{eq16} and~\eqref{eq18} are related by a CP transformation, under which the coefficients are unchanged. 
Demanding CP invariance in the theory would then imply $c_1=-c_1^*$.

It is instructive to repeat  this discussion in terms of operators.
By~\cite{Murphy:2020rsh}, we have four operators that contribute to the amplitude,
\beqa
c_-\overline{u}\gamma^{\nu}u H^{\dagger}\overset{\leftrightarrow}{D^{\mu}}H B_{\mu\nu}, \\
\Tilde{c}_-\overline{u}\gamma^{\nu}u H^{\dagger}\overset{\leftrightarrow}{D^{\mu}}H \Tilde{B}_{\mu\nu}, \\
c_+\overline{u}\gamma^{\nu}u D^{\mu}(H^{\dagger}H) B_{\mu\nu}, \\
\Tilde{c}_+\overline{u}\gamma^{\nu}u D^{\mu}(H^{\dagger}H) \Tilde{B}_{\mu\nu},
\eeqa
where we suppressed all SU(2) indices which are trivially contracted and where,
\begin{equation}
    \Tilde{B}_{\mu\nu} \equiv \frac{1}{2}\epsilon_{\mu\nu\alpha\beta}B^{\alpha\beta}.
\end{equation}
The two operators with tildes on their coefficients are CP-odd, and the other two are CP even. For a positive helicity $B$, these operators give rise to the amplitudes,
\beqa
\frac{c_{\pm}}{\sqrt{2}}\sqb{23}\asb{1(4\pm 5)3},\\
-i\frac{\tilde{c}_{\pm}}{\sqrt{2}}\sqb{23}\asb{1(4\pm 5)3},
\eeqa
and for a negative helicity $B$ one gets, 
\beqa
\frac{c_{\pm}}{\sqrt{2}}\anb{13}\asb{3(4\pm 5)2},\\
i\frac{\tilde{c}_{\pm}}{\sqrt{2}}\anb{13}\asb{3(4\pm 5)2}.
\eeqa
Two things are of note here. First, since the operators are all hermitian, the four coefficients $c_{\pm},\; \tilde{c}_{\pm}$ are all real. Second, this choice of basis does not generate amplitudes of definite helicity, and as such is not the most conducive basis for mapping to the amplitude picture. 
To make this connection manifest, we define the combinations,
\beqa
c^{(1)}_{\pm}\overline{u}\gamma^{\nu}u H^{\dagger}\overset{\leftrightarrow}{D^{\mu}}H (B_{\mu\nu}\pm i \tilde{B}_{\mu\nu}), \\
c^{(2)}_{\pm}\overline{u}\gamma^{\nu}u D^{\mu}(H^{\dagger}H) (B_{\mu\nu} \pm i \tilde{B}_{\mu\nu}),
\eeqa
with $B_{\mu\nu}\pm i \tilde{B}_{\mu\nu}$ generating $\pm$-helicity $B$-amplitudes. Under a CP transformation, 
\begin{equation}
    B_{\mu\nu} + i \tilde{B}_{\mu\nu} \underset{CP}{\leftrightarrow} B_{\mu\nu} - i \tilde{B}_{\mu\nu}.
\end{equation}
Imposing CP conservation would imply,
\begin{equation}
    c_+^{(i)} = c_-^{(i)},
\end{equation}
which is the same result as setting $\tilde{c}=0$ in the CP-eigenbasis. 

Some amplitudes are self conjugate, in which case the coefficients are real/imaginary regardless of the CP requirements. Take the $W_+W_+W_-W_-$ structure from Table~\ref{table:vvvv4},
\begin{equation}
    c^{2W_+2W_-}\sqb{12}^2\anb{34}^2.
\end{equation}
Complex conjugation will take this structure to
\begin{equation}
    \anb{12}^2\sqb{34}^2
    \label{cc}.
\end{equation}
What is its coefficient? From our discussion above, one might say $(c^{2W_+2W_-})^*$. However, note that the structure in Eq.~\eqref{cc} corresponds to $W_-W_-W_+W_+$, which is just a relabeling of the four $W$'s. This means that $c^{2W_+2W_-}$ is real regardless of CP considerations. 
Again, this is also clear when mapped to the operator basis. 
The operator that generates this amplitude is:
\begin{equation}
    (W_{\mu\nu}+i\tilde{W}_{\mu\nu})^2(W_{\mu\nu}-i\tilde{W}_{\mu\nu})^2,
\end{equation}
is Hermitian, meaning its coefficient is real.

We expect these types of relations to hold in the LE limit as well. 
In fact, as mentioned in the text, we do not show LE contributions that can be obtained by complex conjugation. 
For example, consider the amplitude $\overline{u}dW^+Z$, whose conjugate amplitude is $\overline{d}uW^-Z$. 
Let us show using a representative example that the HE relations between coefficients ensure that these two amplitudes are 
indeed complex conjugates of each other.
The most complicated HE structures that contribute to this amplitude are from the helicity amplitudes,
\beq
\mathcal{M}(Q^{c,j}Q_iW^a_{\pm}H^{\dagger,k}H_j),
\eeq
where $a=+,-$. The dimension-8 CT contributions to the positive and negative helicity amplitudes are, 
respectively,
\beqa
\sqb{13}\sab{342} X^{ajk}_{+, il}, \;\; \sqb{13}\sab{352} Y^{ajk}_{+, il},\\
\anb{23}\asb{341} X^{ajk}_{-il},\;\; \anb{23}\asb{351} Y^{ajk}_{-, il},\label{opphel}
\eeqa
where the tensors $X^{ajk}_{\pm,il},\;Y^{ajk}_{\pm,il}$ encode the Wilson coefficients and group-theory structures 
for each of the respective terms. 
A-priori, these structures seem unrelated. However, complex conjugation takes:
\begin{equation}
    \sqb{13}\sab{342}X^{ajk}_{+,il}\rightarrow \anb{31}\asb{342}(X^{ajk}_{+,il})^*,
\end{equation}
which contributes to the amplitude $\mathcal{M}(1_{Q}2_{Q^c}3_{W^a_-}4_{H}5_{H^{\dagger}})$. 
Again, by relabeling $(1\leftrightarrow 2),4\leftrightarrow 5)$, we find that
\beqa
\sqb{13}\sab{342}X^{ajk}_{+,il}\rightarrow -\anb{23}\asb{351}(X^{aik}_{+,jl})^*,
\eeqa
which in turn tells us that,
\beqa
Y_{-,il}^{ajk} = -\left(X^{ail}_{+,jk}\right)^*.
\label{conrats}
\eeqa
Thus, we only have 6 complex coefficients, in agreement with the counting in Ref.~\cite{Murphy:2020rsh}. 
To write the tensor coefficients explicitly, it is convenient to take a basis with well defined CP properties in which the 
relations~\eqref{conrats} are simply satisfied, 
\beqa
X^{ajk}_{+ il} &=& c_1 (\tau^a)_i^{\;j}\delta_l^k + c_2 (\tau^a)_l^{\;k}\delta_i^j + c_3 \epsilon^{abc}(\tau^a)_{i}^{\;j}(\tau^b)_l^{\;k},\\
Y^{ajk}_{+ il} &=& c_4 (\tau^a)_i^{\;j}\delta_l^k + c_5 (\tau^a)_l^{\;k}\delta_i^j + c_6 \epsilon^{abc}(\tau^a)_{i}^{\;j}(\tau^b)_l^{\;k}.
\eeqa
In this basis, we get,
\beqa
Y_{-,il}^{ajk} = c_1^* (\tau^a)_j^{\;i*}\delta_l^k + c_2^* (\tau^a)_l^{\;k*}\delta_j^i + c_3^* \epsilon^{abc}(\tau^a)_{j}^{\;i*}(\tau^b)_l^{\;k*}
\eeqa
Computing the contributions to both amplitudes using the coefficients defined yields,
\beqa
\mathcal{A}(\overline{u}dW^+_+Z)&:&\quad \frac{1}{4}(2ic_1+c_3 - 2i c_4-c_6),\\ 
\mathcal{A}(\overline{d}uW^-_-Z)&:&\quad\frac{1}{4}(-2ic_1 +c_3 +2i c_4 - c_6),
\eeqa 
meaning the contributions from the two helicity categories are related by complex conjugation, as expected. Note that, had we chosen a different basis, the conjugation relation would 
not be so trivial, but would still hold when considering together the contributions of different LE terms.

As above, we can verify that in a $CP$ conserving theory these CTs involve six real parameters.
Charge conjugation effectively exchanges the indices of the two fermions and the indices of the two scalars. 
Parity flips momenta and leaves the spin unchanged, which reverses helicities of particles. 
Under CP, 
\beqa
\sqb{13}\sab{142} &\leftrightarrow& -\anb{23}\asb{351}, \\
\sqb{13}\sab{152} &\leftrightarrow& -\anb{23}\asb{341},
\eeqa
which leads to the relations,
\begin{equation}
    X^{\pm jk}_{\pm,il} = - Y^{\mp il}_{\mp,jk}. 
    \label{relats}
\end{equation} 
Using the relations~\eqref{relats}, we find:
\beqa
X^{ajk}_{- il} &=& c_5 (\tau^a)_i^{\;j}\delta_l^k + c_4 (\tau^a)_l^{\;k}\delta_i^j + c_6 \epsilon^{abc}(\tau^a)_{j}^{\;i}(\tau^b)_k^{\;l}\\
-Y^{ajk}_{- il} &=& c_2 (\tau^a)_i^{\;j}\delta_l^k + c_1 (\tau^a)_l^{\;k}\delta_i^j + c_3 \epsilon^{abc}(\tau^a)_{j}^{\;i}(\tau^b)_k^{\;l},
\eeqa
and we see that we indeed have six additional constraints, meaning we are left with six real parameters. 
%%%%%%%%%%%%%%%%%%%%%%%%%%%%%%%%%%%%%%%%%%%%%%%%%%%%%%%%%%%%%%%%%%%%%%%%%%%
\section{Chirality-violating fermionic amplitudes}
\label{AppCVA}
\subsection{High-energy contact terms}
Although they do not interfere with the SM and can therefore be neglected at this order, we include here the  LE bases for the CTs of chirality-violating fermionic amplitudes with vector boson final states. The HE bases are collected in Table~\ref{TfermD}, and the Higgsed results are shown in Table~\ref{ChVio}. Differences between the HEFT and SMEFT for these amplitudes are shown in Table~\ref{T12}. 
\begin{longtable}[|c|]{|c|c|c|}
       \hline
       Amplitude & Massless Contact-Term & Wilson Coefficient \\
       \hline \hline
       \endfirsthead
       \hline
       \multicolumn{3}{|c|}{Continued from Table~\ref{TfermD}}\\
       \hline
       Amplitude & Massless Contact-Term & Wilson Coefficient \\
       \hline \hline
       \endhead
       \hline
       \endfoot
       \endlastfoot
\multicolumn{3}{|c|}{\textbf{$f^+f^+V^{\pm}s$}}\\
       \hline
       %%%%%%%%%%%%%%%
       \multirow{1}{*}{$Q^{c, i}UW^a_+H^{\dagger,j}$}&$\sqb{13}\sqb{23}s_{34}\epsilon^{ik}(\tau^a)_k^{\; j}$&$c_1^{Q^cUW_+H}$\\
       \cline{2-3}
       &$\sqb{13}\sqb{23}s_{14}\epsilon^{ik}(\tau^a)_k^{\; j}$&$c_2^{Q^cUW_+H}$\\
        \hline
        %%%%%%%%%%%%%%%
        \multirow{1}{*}{$Q^{c, i}UW^a_-H^{\dagger,j}$}&$\sqb{12}\anb{3143}\epsilon^{ik}(\tau^a)_k^{\; j}$&$c^{Q^cUW_-H}$\\
        \hline
        %%%%%%%%%%%%%%%
        $Q^{c, i}UB_+H^{\dagger,j}$&$\sqb{13}\sqb{23}s_{34}\epsilon^{ij}$&$c_1^{Q^cUB_+H}$\\
        \cline{2-3}
        &$\sqb{13}\sqb{23}s_{14}\epsilon^{ij}$&$c_2^{Q^cUB_+H}$\\
        \hline
        %%%%%%%%%%%%%%%
        $Q^{c,i}UB_-H^{\dagger,j}$&$\sqb{12}\anb{3143}\epsilon^{ij}$&$c^{Q^cUB_-H}$\\
        \hline
        %%%%%%%%%%%%%%%
        $Q^{c,\beta i}U_{\alpha }g^A_+H^{\dagger,j}$&$\sqb{13}\sqb{23}s_{34}(\lambda^A)_{\alpha}^{\;\beta}\epsilon_{ij}$&$c_1^{Q^cUg_+H}$\\
        \cline{2-3}
        &$\sqb{13}\sqb{23}s_{14}(\lambda^A)_{\alpha}^{\;\beta}\epsilon_{ij}$&$c_2^{Q^cUg_+H}$\\
        \hline
        %%%%%%%%%%%%%%%
        $Q^{c,\beta i}U_{\alpha}g^A_-H^{\dagger,j}$&$\sqb{12}\anb{3143}(\lambda^A)_{\alpha}^{\;\beta}\epsilon_{ij}$&$c^{Q^cUg_-H}$\\
        \hline
        %%%%%%%%%%%%%%%
        $Q^{c,j}DW^a_+H_i$&$\sqb{13}\sqb{23}s_{34} (\tau^a)^{\;j}_{i}$&$c_1^{Q^cDW_+H}$\\
        \cline{2-3}
        &$\sqb{13}\sqb{23}s_{14} (\tau^a)^{\;j}_{i}$&$c_2^{Q^cDW_+H}$\\
        \hline
        %%%%%%%%%%%%%%%
        $Q^{c,j}DW^a_-H_i$&$\sqb{12}\anb{3143} (\tau^a)^{\;j}_{i}$&$c^{Q^cDW_-H}$\\
        \hline
        %%%%%%%%%%%%%%%
        $Q^{c, j}DB_+H_i$&$\sqb{13}\sqb{23}s_{34} \delta_i^j$&$c_1^{Q^cDB_+H}$\\
        \cline{2-3}
        &$\sqb{13}\sqb{23}s_{14} \delta_i^j$&$c_2^{Q^cDB_+H}$\\
        \hline
        %%%%%%%%%%%%%%%
        $Q^{c,j}DB_-H_i$&$\sqb{12}\anb{3143} \delta_i^j$&$c^{Q^cDB_-H}$\\
        \hline
        %%%%%%%%%%%%%%%
        $Q^{c,\beta j}D_{\alpha }g^A_+H_i$&$\sqb{13}\sqb{23}s_{34}(\lambda^A)^{\;\beta}_{\alpha}\delta_i^j$&$c_1^{Q^cDg_+H}$\\
        \cline{2-3}
        &$\sqb{13}\sqb{23}s_{14}(\lambda^A)^{\;\beta}_{\alpha}\delta_i^j$&$c_2^{Q^cDg_+H}$\\
        \hline
        %%%%%%%%%%%%%%%
        $Q^{c,\beta j}D_{\alpha}g^A_-H_i$&$\sqb{12}\anb{3143}(\lambda^A)^{\;\beta}_{\alpha}\delta^j_i$&$c^{Q^cDg_-H}$\\
        \hline
        %%%%%%%%%%%%%%%
        \multicolumn{3}{|c|}{\textbf{$f^+f^+sss$}}\\
        \hline
        %%%%%%%%%%%%%%%
        \multirow{1}{*}{$Q^{c, j}UH^{\dagger,k}H^{\dagger,l}H_i$}&$(\sqb{1342}-\sqb{1432})\delta_i^j\epsilon^{kl}$&$c_1^{Q^cU3H}$\\
        \cline{2-3}
        &$(\sqb{1342}+\sqb{1432})(\delta_i^k\epsilon^{jl}+\delta_i^l\epsilon^{jk})$&$c_2^{Q^cU3H}$\\
        \cline{2-3}
        &$(\sqb{1352}-\sqb{1452})\delta^j_i\epsilon^{kl}$&$c_3^{Q^cU3H}$\\
        \cline{2-3}
        &$(\sqb{1352}+\sqb{1452})(\delta_i^k\epsilon^{jl}+\delta_i^l\epsilon^{jk})$&$c_4^{Q^cU3H}$\\
        \cline{2-3}
        &$(\sqb{1532}-\sqb{1542})\delta^j_i\epsilon^{kl}$&$c_5^{Q^cU3H}$\\
        \cline{2-3}
        &$(\sqb{1532}+\sqb{1542})(\delta_i^k\epsilon^{jl}+\delta^l_i\epsilon^{jk})$&$c_6^{Q^cU3H}$\\
        \hline
        %%%%%%%%%%%%%%%
        \multirow{1}{*}{$Q^{c, j}DH^{\dagger,k}H_iH_l$}&$(\sqb{1342}+\sqb{1352})(\delta_i^j\delta^k_l+\delta_i^k\delta_l^j)$&$c_1^{Q^cD3H}$\\
        \cline{2-3}
        &$(\sqb{1342}-\sqb{1352})\epsilon_{il}\epsilon^{jk}$&$c_2^{Q^cD3H}$\\
        \cline{2-3}
        &$(\sqb{1452}+\sqb{1542})(\delta_i^j\delta^k_l+\delta_i^k\delta_l^j)$&$c_3^{Q^cD3H}$\\
        \cline{2-3}
        &$(\sqb{1452}-\sqb{1542})\epsilon_{il}\epsilon^{jk}$&$c_4^{Q^cD3H}$\\
        \cline{2-3}
        &$(\sqb{1432}+\sqb{1532})(\delta_i^j\delta_l^k+\delta_i^k\delta_l^j)$&$c_5^{Q^cD3H}$\\
        \cline{2-3}
        &$(\sqb{1432}-\sqb{1532})\epsilon_{il}\epsilon^{jk}$&$c_6^{Q^cD3H}$\\
        \hline
        %%%%%%%%%%%%%%%

        \multicolumn{3}{|c|}{\textbf{$f^+f^+V^{\pm}V^{\pm}s$}}\\
        \hline
        %%%%%%%%%%%%%%%
        \multirow{1}{*}{$Q^{c,i}UW^a_+W^b_+H^{\dagger,j}$}&$\sqb{34}(\sqb{13}\sqb{24}+\sqb{14}\sqb{23})((\tau^a)_k^{\;i}(\tau^b)_l^{\;j}-a\leftrightarrow b)\epsilon^{kl}$&$c_1^{QU2W_+H}$\\
        \cline{2-3}
        &$\sqb{12}\sqb{34}^2((\tau^a)_k^{\;i}(\tau^b)_l^{\;j}+a\leftrightarrow b)\epsilon^{kl}$&$c_2^{QU2W_+H}$\\
        \hline
        %%%%%%%%%%%%%%%
        $Q^{c,i}UW^a_-W^b_-H^{\dagger,j}$&$\sqb{12}\anb{34}^2\delta^{ab}\epsilon^{ij}$&$c^{QU2W_-H}$\\
        \hline
        %%%%%%%%%%%%%%%
        \multirow{1}{*}{$Q^{c,j}UW^a_+B_+H^{\dagger,j}$}&$\sqb{13}\sqb{24}\sqb{34}\epsilon^{ik}(\tau^a)_k^{\;j}$ &$c_1^{QUW_+B_+H}$\\
        \cline{2-3}
        &$\sqb{14}\sqb{23}\sqb{34}\epsilon^{ik}(\tau^a)_k^{\;j}$&$c_2^{QUW_+B_+H}$\\
        \hline
        %%%%%%%%%%%%%%%
        \multirow{1}{*}{$Q^{c,i}UW^a_-B_-H^{\dagger,j}$}&$\sqb{12}\anb{34}^2\epsilon^{ik}(\tau^a)_k^{\;j}$&$c_1^{QUW_-B_-H}$\\
        \hline
        %%%%%%%%%%%%%%%
        $Q^{c,j}UB_+B_+H^{\dagger,j}$&$\sqb{12}\sqb{34}^2\epsilon^{ij}$&$c^{QU2B_+H}$\\
        \hline
        %%%%%%%%%%%%%%%
        $Q^{c,i}UB_-B_-H^{\dagger,j}$&$\sqb{12}\anb{34}^2\epsilon^{ij}$&$c^{QU2B_-H}$\\
        \hline
        %%%%%%%%%%%%%%%
        \multirow{1}{*}{$Q^{c,\beta i}U_{\alpha}g^A_+g^B_+H^{\dagger,j}$}&$\sqb{12}\sqb{34}^2\epsilon^{ij}d^{ABC}(\lambda^C)_{\alpha}^{\;\beta}$&$c_1^{QU2g_+H}$\\
        \cline{2-3}
        &$\sqb{12}\sqb{34}^2\epsilon^{ij}\delta^{ab}\delta^{\beta}_{\alpha}$&$c_2^{QU2g_+H}$\\
        \cline{2-3}
        &$\sqb{34}(\sqb{13}\sqb{24}+\sqb{14}\sqb{23})\epsilon^{ij}f^{ABC}(\lambda^C)_{\alpha}^{\;\beta}$&$c_3^{QU2g_+H}$\\
        \hline
        %%%%%%%%%%%%%%%
        \multirow{1}{*}{$Q^{c,\beta i}U_{\alpha}g^A_-g^B_-H^{\dagger,j}$}&$\sqb{12}\anb{34}^2\epsilon^{ij}d^{ABC}(\lambda^C)_{\alpha}^{\;\beta}$&$c_1^{QU2g_-H}$\\
        \cline{2-3}
        &$\sqb{12}\anb{34}^2\epsilon^{ij}\delta^{ab}\delta_{\beta}^{\alpha}$&$c_2^{QU2g_-H}$\\
        \hline
        %%%%%%%%%%%%%%%
        \multirow{1}{*}{$Q^{c,\beta i}U_{\alpha}W^a_+g^A_+H^{\dagger,j}$}&$\sqb{13}\sqb{24}\sqb{34}\epsilon^{ik}(\tau^a)_k^{\;j}(\lambda^A)_{\alpha}^{\;\beta}$&$c_1^{QUW_+g_+H}$\\
        \cline{2-3}
        &$\sqb{14}\sqb{23}\sqb{34}\epsilon^{ik}(\tau^a)_k^{\;j}(\lambda^A)_{\alpha}^{\;\beta}$&$c_2^{QUW_+g_+H}$\\
        \hline
        %%%%%%%%%%%%%%%
        \multirow{1}{*}{$Q^{c,\beta i}U_{\alpha}W^a_-g^A_-H^{\dagger,j}$}&$\sqb{12}\anb{34}^2\epsilon^{ik}(\tau^a)_k^{\;j}(\lambda^A)_{\alpha}^{\;\beta}$&$c_1^{QUW_-g_-H}$\\
        \cline{2-3}
        &$\sqb{12}\anb{34}^2\epsilon^{jk}(\tau^a)_k^{\;i}(\lambda^A)_{\alpha}^{\;\beta}$&$c_2^{QUW_-g_-H}$\\
        \hline
        %%%%%%%%%%%%%%%
        $Q^{c, \beta i}U_{\alpha}B_+g^A_+H^{\dagger,j}$&$\sqb{14}\sqb{23}\sqb{34} \epsilon^{ij}(\lambda^A)_{\alpha}^{\;\beta}$&$c_1^{QUB_+g_+H}$\\
        \cline{2-3}
        &$\sqb{13}\sqb{24}\sqb{34} \epsilon^{ij}(\lambda^A)_{\alpha}^{\;\beta}$&$c_2^{QUB_+g_+H}$\\
        \hline
        %%%%%%%%%%%%%%%
        $Q^{c,\beta i}U_{\alpha}B_-g^A_-H^{\dagger,j}$&$\sqb{12}\anb{34}^2 \epsilon^{ij}(\lambda^A)_{\alpha}^{\;\beta}$&$c^{QUB_-g_-H}$\\
        \hline
        %%%%%%%%%%%%%%%
        \multirow{1}{*}{$Q^{c,j}DW^a_+W^b_+H_i$}&$\sqb{34}(\sqb{13}\sqb{24}+\sqb{14} \sqb{23})\epsilon^{abc}(\tau^c)^{\;j}_{ i}$&$c_1^{QD2W_+H}$\\
        \cline{2-3}
        &$\sqb{12}\sqb{34}^2((\tau^a)^j_k(\tau^b)^k_i+a\leftrightarrow b)$&$c_2^{QD2W_+H}$\\
        \hline
        %%%%%%%%%%%%%%%
        $Q^{c,j}DW^a_-W^b_-H_i$&$\sqb{12}\anb{34}^2\delta^{ab}\delta_i^j$&$c^{QU2W_-H}$\\
        \hline
        %%%%%%%%%%%%%%%
        \multirow{1}{*}{$Q^{c,j}DW^a_+B_+H_i$}&$\sqb{13}\sqb{24}\sqb{34}(\tau^a)^{\;j}_{i}$ &$c_1^{QDW_+B_+H}$\\
        \cline{2-3}
        &$\sqb{14}\sqb{23}\sqb{34}(\tau^a)^{\;j}_{i}$&$c_2^{QDW_+B_+H}$\\
        \hline
        %%%%%%%%%%%%%%%
        \multirow{1}{*}{$Q^{c,j}DW^a_-B_-H_i$}&$\sqb{12}\anb{34}^2(\tau^a)^{\;j}_{i}$&$c^{QDW_-B_-H}$\\
        \hline
        %%%%%%%%%%%%%%%
        $Q^{c,j}DB_+B_+H_i$&$\sqb{12}\sqb{34}^2\delta_i^j$&$c^{QU2B_+H}$\\
        \hline
        %%%%%%%%%%%%%%%
        $Q^{c,j}DB_-B_-H_i$&$\sqb{12}\anb{34}^2\delta_i^j$&$c^{QD2B_-H}$\\
        \hline
        %%%%%%%%%%%%%%%
        \multirow{1}{*}{$Q^{c,\beta j}D_{\alpha}g^A_+g^B_+H_i$}&$\sqb{12}\sqb{34}^2\delta_i^jd^{ABC}(\lambda^C)_{\alpha}^{\;\beta}$&$c_1^{QD2g_+H}$\\
        \cline{2-3}
        &$\sqb{12}\sqb{34}^2\delta_i^j\delta^{ab}\delta_{\beta}^{\alpha}$&$c_2^{QD2g_+H}$\\
        \cline{2-3}
        &$\sqb{34}(\sqb{13}\sqb{24}+\sqb{14}\sqb{23})\delta_i^jf^{ABC}(\lambda^C)_{\alpha}^{\;\beta}$&$c_3^{QD2g_+H}$\\
        \hline
        %%%%%%%%%%%%%%%
        \multirow{1}{*}{$Q^{c,\beta j}D_{\alpha}g^A_-g^B_-H_i$}&$\sqb{12}\anb{34}^2\delta_i^jd^{ABC}(\lambda^C)_{\alpha}^{\;\beta}$&$c_1^{QD2g_-H}$\\
        \cline{2-3}
        &$\sqb{12}\anb{34}^2\delta_i^j\delta^{ab}\delta_{\beta}^{\alpha}$&$c_2^{QD2g_-H}$\\
        \hline
        %%%%%%%%%%%%%%%
        \multirow{1}{*}{$Q^{c,\beta j}D_{\alpha}W^a_+g^A_+H_i$}&$\sqb{13}\sqb{24}\sqb{34}(\tau^a)^{\;j}_{i}(\lambda^A)_{\alpha}^{\;\beta}$&$c_1^{QDW_+g_+H}$\\
        \cline{2-3}
        &$\sqb{14}\sqb{23}\sqb{34}(\tau^a)^{\;j}_{i}(\lambda^A)_{\alpha}^{\;\beta}$&$c_2^{QDW_+g_+H}$\\
        \cline{2-3}
        \hline
        %%%%%%%%%%%%%%%
        \multirow{1}{*}{$Q^{c,\beta j}D_{\alpha}W^a_-g^A_-H_i$}&$\sqb{12}\anb{34}^2(\tau^a)^{\;j}_{i}(\lambda^A)_{\alpha}^{\;\beta}$&$c^{QDW_-g_-H}$\\
        \hline
        %%%%%%%%%%%%%%%
        $Q^{c, \beta i}D_{\alpha}B_+g^A_+H_i$&$\sqb{14}\sqb{23}\sqb{34} \delta_i^j(\lambda^A)_{\alpha}^{\;\beta}$&$c^{QDB_+g_+H}$\\
        \cline{2-3}
        &$\sqb{13}\sqb{24}\sqb{34} \delta_i^j(\lambda^A)_{\alpha}^{\;\beta}$&$c^{QDB_+g_+H}$\\
        \hline
        %%%%%%%%%%%%%%%
        $Q^{c,\beta j}D_{\alpha}B_-g^A_-H_i$&$\sqb{12}\anb{34}^2 \delta_i^j(\lambda^A)_{\alpha}^{\;\beta}$&$c^{QDB_-g_-H}$\\
        \hline
        \caption{Dimension-8 HE structures for amplitudes with fermionic initial states. Amplitudes with $D^cD$ have the same structures as their $U^cU$ counterparts and are omitted for brevity. Their coefficients are obtained by replacing $U^cU$ in the respective coefficients with $D^cD$. When the color index is omitted, the color factor is just $\delta^{\beta}_{\alpha}$. The table is organized by helicity categories, which are listed as full lines and in bold, with helicity conserving amplitudes appearing first.}
        \label{TfermD}
        \end{longtable}
%%%%%%%%%%%%%%%%%%%%%%%%%%%%%%%%%%%%%%%%%%%%%%%%%%%%%%%%%%%
%%%%%%%%%%%%%%%%%%%%%%%%%%%%%%%%%%%%%%%%%%%%%%%%%%%%%%%%%%%%%
\begin{landscape}
\subsection{Low-energy contact terms}
\begin{longtable}{|c|c|>{\centering\arraybackslash}p{4.5in}|}  
\hline
    Amplitude & Massive Contact-Term & Wilson Coefficient \\
    \hline
    \endfirsthead
    \hline
    \multicolumn{3}{|c|}{Continued from Table~\ref{ChVio}}\\
    \hline
    Amplitude & Massive  Contact-Term & Wilson Coefficient \\
    \hline
    \endhead
    \hline
    \endfoot
    \endlastfoot
        \multirow{1}{*}{$\mathcal{A}(\overline{u}UW^+W^-)$} &$\Sqb{34}(\sqb{1\mathbf{4}}\sqb{2\mathbf{3}}+\sqb{1\mathbf{3}}\sqb{2\mathbf{4}}) $&$-\frac{v}{2\sqrt{2}}\Cfseventy$\\
        \cline{2-3}
        &$\sqb{12}\Sqb{34}^2 $&$-\frac{v}{2\sqrt{2}}\Cfseventyone$\\
        \cline{2-3}
        &$\sqb{12}\Anb{34}^2 $&$\frac{v}{\sqrt{2}}\Cfseventytwo$\\
        \cline{2-3}
        &$\sQbr{1}{3}\sQbr{2}{3}\sab{\four \three \four} $&$\Cfone$\\
        \cline{2-3}
        &$\sQbr{1}{3}\sQbr{2}{3}\sab{\four 1\four} $&$\Cftwo$\\
        \cline{2-3}
        &$\sqb{12}\asb{\three 1 \four}\Anb{34} $&$\Cfthree$\\
        \cline{2-3}
        &$\sQbr{1}{4}\Anb{34}\sQbr{2}{3} $&$2\sqrt{2}v\Cfthirtysixtag$\\
        \cline{2-3}
        &$\sQbr{1}{3}\Anb{34}\sQbr{2}{4} $&$-2\sqrt{2}v\Cfthirtysixtagtwo$\\
         \hline
        %%%%%%%%%%%%%%%
         \multirow{1}{*}{$\mathcal{A}(\overline{u}UZZ)$}&$(\sqb{1\mathbf{4}}\sqb{2\mathbf{3}}-\sQbr{2}{4}\sqb{1\mathbf{3}})\Sqb{34} $ &$\frac{v\stw \ctw}{2\sqrt{2}}(\Cfseventythree-\Cfseventyfive)$\\
         \cline{2-3}
         &$\sqb{12}\Sqb{34}^2 $&$\frac{v}{\sqrt{2}}\big(\stw^2 \Cfseventynine-\frac{\ctw^2}{2}\Cfseventyone\big)$\\
         \cline{2-3}
         &\multirow{1}{*}{$\sqb{12}\Anb{34}^2 $}&$\frac{v}{\sqrt{2}}\big(\ctw^2 \Cfseventytwo+\stw^2 \Cfeighty-\ctw \stw (\Cfseventyseven+\Cfseventyeight)\big)$\\
         \cline{2-3}
         &$\sqb{1\mathbf{3}}\sqb{2\mathbf{3}}\sab{\four \three \four}+(3\leftrightarrow 4) $&$\frac{i}{2}(2\stw \Cffour -\ctw \Cfone)$\\
         \cline{2-3}
         &$\sqb{1\mathbf{3}}\sqb{2\mathbf{3}}\sab{\four 1\four}+(3\leftrightarrow 4) $&$\frac{i}{2}(2\stw \Cffourtag -\ctw \Cftwo)$\\
         \cline{2-3}
         &$\sqb{12}\asb{\three 1 \four}\Anb{34}+(3\leftrightarrow 4)  $&$\frac{i}{2}(2\stw \Cffive - \ctw \Cfthree )$\\
         \cline{2-3}
         &$\Anb{34}(\sQbr{1}{3}\sQbl{4}{2}-\sQbr{1}{4}\sQbl{3}{2})  $&$-\sqrt{2}v(\Cfthirtyseven-2(\Cfthirtyseventag+\Cfthirtyseventagtwo))$\\
         \hline
         %%%%%%%%%%%%%%%
         
         \multirow{1}{*}{$\mathcal{A}(\overline{u}UZ\gamma_+)$}&\multirow{1}{*}{$\sqb{\mathbf{3}4}(\sqb{14}\sqb{2\mathbf{3}}+\sqb{1\three}\sqb{24}) $}&$-\frac{v}{2\sqrt{2}}( \Cfseventythree+ \Cfseventyfive)$\\
         \cline{2-3}
         &$\sqb{12}\sqb{\mathbf{3}4}^2 $&$-\frac{v}{\sqrt{2}}\big[ (\ctw^2 -\stw^2)(\Cfseventythree-\Cfseventyfive)-\ctw\stw\big( \Cfseventynine+\frac{1}{2}\Cfseventyone\big)\big]$\\
         \cline{2-3}
         &$\sqb{14}\sqb{24}\sab{\three 4\three} $&$\frac{i}{2}(\stw\Cfone +2\ctw \Cffour)$\\
         \cline{2-3}
         &$\sqb{14}\sqb{24}\sab{\three 1\three} $&$\frac{i}{2}(\stw \Cftwo +2\ctw \Cffourtag)$\\
         \hline
         %%%%%%%%%%%%%%%
\multirow{1}{*}{$\mathcal{A}(\overline{u}UZ\gamma_-)$}&$\sqb{12}\asb{41\three}\aNbl{3}{4} $&$-\frac{i}{2}(\stw\Cfthree+2\ctw \Cffive)$\\
         \cline{2-3}
         &\multirow{1}{*}{$\sqb{12}\aNbl{3}{4}^2 $}&$-\frac{v}{2\sqrt{2}}\big( \Cfseventyseven + \Cfseventyeight) +2 \ctw \stw (\Cfseventytwo- \Cfeighty)\big)$\\
        \hline
        %%%%%%%%%%%%%%%
        \multirow{1}{*}{$\mathcal{A}(\overline{u}^{\beta}U_{\alpha}Zg^A_+)$}&$\sQbr{1}{3}\sqb{24}\sQbl{3}{4}(\lambda^A)_{\alpha}^{\;\beta} $&$-\frac{v}{2\sqrt{2}}(\ctw \Cfeightysix-2\stw \Cfninetytwotag)$\\
        \cline{2-3}
        &$\sQbr{2}{3}\sqb{14}\sQbl{3}{4}(\lambda^A)_{\alpha}^{\;\beta}$&$-\frac{v}{2\sqrt{2}}(\ctw \Cfeightyeight-2\stw \Cfninetytwo)$\\
        \cline{2-3}
        &$\sqb{14}\sqb{24}\sab{\three 4\three}(\lambda^A)_{\alpha}^{\;\beta} $&$i\Cfsix$\\
        \cline{2-3}
        &$\sqb{14}\sqb{24}\sab{\three 1\three}(\lambda^A)_{\alpha}^{\;\beta} $&$i\Cfsixtwo$\\
        \hline
        %%%%%%%%%%%%%%%
        \multirow{1}{*}{$\mathcal{A}(\overline{u}^{\beta}U_{\alpha}Zg^A_-)$}&$\sqb{12}\aNbl{3}{4}^2(\lambda^A)^{\alpha}_{\beta} $&$-\frac{v}{2\sqrt{2}}(\ctw \Cfninety - 2 \stw \Cfninetythree)$\\
        \cline{2-3}
        &$\sqb{12}\asb{41\three}\aNbl{3}{4}(\lambda^A)_{\alpha}^{\;\beta} $&$-\sqrt{2}\Cfsixtag$\\
        \hline
        %%%%%%%%%%%%%%%
        \multirow{1}{*}{$\mathcal{A}(\overline{u}U\gamma_+\gamma_+)$}&$\sqb{12}\sqb{34}^2 $ &$\frac{v}{2\sqrt{2}}\big[(-\stw^2 \Cfseventyone + 2 \ctw^2 \Cfseventynine)+\ctw \stw (-\Cfseventythree+\Cfseventyfive)\big]$  \\
        \hline
        %%%%%%%%%%%%%%%
        \multirow{1}{*}{$\mathcal{A}(\overline{u}U\gamma_-\gamma_-)$}&\multirow{1}{*}{$\sqb{12}\anb{34}^2 $}&$\frac{v}{\sqrt{2}}(\stw^2 \Cfseventytwo + \stw^2 \Cfeighty)+\ctw \stw (\Cfseventyseven + \Cfseventyeight)$\\
        \hline
    %%%%%%%%%%%%%%%
        \multirow{1}{*}{$\mathcal{A}(\overline{u}^{\beta}U_{\alpha}\gamma_+g_+^A)$}&$\sqb{13}\sqb{24}\sqb{34}(\lambda^A)_{\alpha}^{\;\beta} $&$\frac{v}{2\sqrt{2}}(\stw\Cfeightysix+2\ctw \Cfninetytwotag)$\\
        \cline{2-3}
        &$\sqb{14}\sqb{23}\sqb{34}(\lambda^A)_{\alpha}^{\;\beta} $&$\frac{v}{2\sqrt{2}}(\stw \Cfeightyeight+2\ctw \Cfninetytwo)$\\
        \hline
        %%%%%%%%%%%%%%%
        $\mathcal{A}(\overline{u}^{\beta}U_{\alpha}\gamma_-g_-^A)$&$\sqb{12}\anb{34}^2(\lambda^A)_{\alpha}^{\;\beta} $&$\frac{v}{2\sqrt{2}}(\stw \Cfninety + 2 \ctw \Cfninetythree)$\\
        \hline 
        %%%%%%%%%%%%%%%
        \multirow{1}{*}{$\mathcal{A}(\overline{d}DW^+W^-)$} &$\sqb{12}\Sqb{34}^2 $ & $\frac{v}{2\sqrt{2}}\Cfhundredtwelve$\\
        \cline{2-3}
        &$(\sQbr{1}{4}\sQbr{2}{3}+\sQbr{2}{4}\sQbr{1}{3})\Sqb{34} $&$-\frac{iv}{2\sqrt{2}}\Cfhundredeleven$\\
        \cline{2-3}
        &$\sqb{12}^2\Anb{34}^2 $&$\Cfhundredthirteen$\\
        \cline{2-3}
        &$\sQbr{1}{4}\sQbr{2}{4}\sab{\three \four \three} $&$\Cfseven$\\
        \cline{2-3}
        &$\sQbr{1}{4}\sQbr{2}{4}\sab{\three 1 \three} $&$\Cfseventwo$\\
        \cline{2-3}
        &$\sqb{12}\asb{\four 1\three}\Anb{34} $&$-\Cfseventag$\\
        \cline{2-3}
        &$\sQbr{1}{4}\Anb{34}\sQbr{2}{3} $&$2\sqrt{2}v(\Cfthirtyeight+\Cfthirtynine)$\\
        \cline{2-3}
        &$\sQbr{1}{3}\Anb{34}\sQbr{2}{4} $&$-2\sqrt{2}v(\Cfthirtyeighttwo+\Cfthirtyninetwo)$\\
         \hline
         %%%%%%%%%%%%%%%
         \multirow{1}{*}{$\mathcal{A}(\overline{d}DZZ)$}&$\sqb{12}\Sqb{34}^2 $ &$\frac{v}{2\sqrt{2}}(\ctw^2 \Cfhundredtwelve + 2 \stw^2 \Cfhundredseventeen)$\\
         \cline{2-3}
         &$(\sQbr{1}{4}\sQbr{2}{3}-\sQbr{1}{3}\sQbr{2}{4})\Sqb{34} $&$\frac{v}{2\sqrt{2}}\ctw\stw (\Cfhundredfourteen-\Cfhundredfifteen)$\\
         \cline{2-3}
         &$\sqb{12}\Anb{34}^2 $&$\frac{v}{\sqrt{2}}(\ctw^2 \Cfhundredthirteen - \ctw\stw \Cfhundredsixteen + \stw^2 \Cfhundredeighteen)$\\
         \cline{2-3}
         &$\sQbr{1}{4}\sQbr{2}{4}\sab{\four \three \four}+(3\leftrightarrow 4) $&$\frac{i}{2}(\ctw \Cfseven-2\stw \Cfeight)$\\
         \cline{2-3}
         &$\sQbr{1}{4}\sQbr{2}{4}\sab{\three 1 \three}+(3\leftrightarrow 4) $&$\frac{i}{2}(\ctw \Cfseventwo-2\stw \Cfeighttag)$\\
         \cline{2-3}
         &$\sqb{12}\asb{\four 1 \three}\Anb{34}+(3\leftrightarrow 4) $&$-\frac{i}{2}(\ctw \Cfseventag-2\stw \Cfnine)$\\
         \cline{2-3}
         &$\sqb{12}\asb{\three 1 \four}\Anb{34}+(3\leftrightarrow 4) $&$-\frac{i}{2}(\ctw \Cfseventag-2\stw \Cfnine)$\\
         \cline{2-3}
         &$\sQbr{1}{3}\Anb{34}\sQbr{2}{4}+(3\leftrightarrow 4)  $&$-\sqrt{2}v(2\Cfthirtyeight-\Cfthirtyeighttag+2\Cfthirtyeighttwo)$\\
         \hline
         %%%%%%%%%%%%%%%
\multirow{1}{*}{$\mathcal{A}(\overline{d}DZ\gamma_+)$}&$\sqb{12}\sQbl{3}{4}^2 $&$-\frac{v}{2\sqrt{2}}\big[(\ctw^2-\stw^2)(\Cfhundredfourteen-\Cfhundredfifteen)\ctw \stw (\Cfhundredtwelve-2\Cfhundredseventeen)\big]$\\
    \cline{2-3}
    &$(\sqb{24}\sQbr{1}{3}+\sqb{24}\sqb{1\three})\sQbl{3}{4}$&$-\frac{v}{2\sqrt{2}}(\Cfhundredfourteen+ \Cfhundredfifteen)$\\
    \cline{2-3}
    &$\sqb{14}\sqb{24}\sab{\three 4 \three} $&$-\frac{i}{2}(\stw \Cfseven+2\ctw \Cfeight)$\\
    \cline{2-3}
    &$\sqb{14}\sqb{24}\sab{\three 1 \three} $&$-\frac{i}{2}(\stw \Cfseventwo+2\ctw \Cfeighttag)$\\
        \hline
        %%%%%%%%%%%%%%%
        \multirow{1}{*}{$\mathcal{A}(\overline{d}DZ\gamma_-)$}&\multirow{1}{*}{$\sqb{12}\aNbl{3}{4}^2 $}&$\frac{v}{2\sqrt{2}}(-\ctw^2 \Cfhundredsixteen + \stw^2 \Cfhundredsixteen + 2 \ctw\stw (\Cfhundredeighteen-\Cfhundredthirteen))$\\
        \cline{2-3}
        &$\sqb{12}\asb{41\three}\aNbl{3}{4} $&$\frac{i}{2}(\stw \Cfseventag+2\ctw \Cfnine)$\\
        \hline
        %%%%%%%%%%%%%%%
        \multirow{1}{*}{$\mathcal{A}(\overline{d}^{\beta}D_{\alpha}Zg_+)$}&$\sQbr{1}{3}\sqb{24}\sQbl{3}{4}\ellA$&$-\frac{v}{2\sqrt{2}}(\ctw \Cfhundredtwentyfour -2\stw \Cfhundredtwentyseventag)$\\
        \cline{2-3}
        &$\sQbr{2}{3}\sqb{14}\sQbl{3}{4}\ellA$&$-\frac{v}{2\sqrt{2}}(\ctw \Cfhundredtwentyfive-2\stw \Cfhundredtwentyseventag)$\\
        \cline{2-3}
        &$\sqb{14}\sqb{24}\sab{\three 4\three}\ellA$&$-i\Cften$\\
        \cline{2-3}
        &$\sqb{14}\sqb{24}\sab{\three 1 \three}\ellA$&$-i\Cftentag$\\
         \hline
         %%%%%%%%%%%%%%%
         \multirow{1}{*}{$\mathcal{A}(\overline{d}^{\beta}D_{\alpha}Zg_-)$}&$\sqb{12}\aNbl{3}{4}^2\ellA$&$-\frac{v}{2\sqrt{2}}(\ctw \Cfhundredtwentysix-2\stw \Cfhundredtwentyeight)$\\
         \cline{2-3}
         &$\sqb{12}\asb{41\three}\aNbl{3}{4}\ellA$&$i\Cfeleven$\\
         \hline
         %%%%%%%%%%%%%%%
         \multirow{1}{*}{$\mathcal{A}(\overline{d}D\gamma_+ \gamma_+)$}&$\sqb{12}\sqb{34}^2 $&$\frac{v}{2\sqrt{2}}(\stw^2 \Cfhundredtwelve+2\ctw^2 \Cfhundredseventeen)$\\
        \cline{2-3}
        &$\sqb{34}(\sqb{13}\sqb{24}-\sqb{14}\sqb{23}) $&$\frac{v}{2\sqrt{2}}\ctw \stw (\Cfhundredfourteen-\Cfhundredfifteen)$\\
         \hline
         %%%%%%%%%%%%%%%
         $\mathcal{A}(\overline{d}D\gamma_- \gamma_-)$&$\sqb{12}\anb{34}^2 $&$\frac{v}{\sqrt{2}}(\ctw^2 \Cfhundredthirteen + \ctw\stw \Cfhundredsixteen+\ctw^2 \Cfhundredeighteen)$\\
         \hline
         %%%%%%%%%%%%%%%
         \multirow{1}{*}{$\mathcal{A}(\overline{d}D\gamma_+ g_+)$}&$\sqb{13}\sqb{24}\sqb{34}\ellA  $&$\frac{v}{2\sqrt{2}}(\stw \Cfhundredtwentyfour+2\ctw \Cfhundredtwentyseventag)$\\
         \cline{2-3}
         &$\sqb{14}\sqb{23}\sqb{34}\ellA  $&$\frac{v}{2\sqrt{2}}(\stw \Cfhundredtwentyfive+2\ctw \Cfhundredtwentyseven)$\\
         \hline
         %%%%%%%%%%%%%%%
         $\mathcal{A}(\overline{d}D\gamma_- g_-)$&$\sqb{12}\anb{34}^2\ellA  $&$\frac{v}{2\sqrt{2}}(\stw \Cfhundredtwentysix+2\ctw \Cfhundredtwentyeight)$\\
         \hline
         %%%%%%%%%%%%%%%
         \multirow{1}{*}{$\mathcal{A}(\overline{u}DW^+Z)$}&$\sqb{12}\Sqb{34}^2 $&$\frac{v}{4}\ctw \Cfhundredtwelve$\\
        \cline{2-3}
        &$(\sQbr{1}{3}\sQbr{2}{4}+\sQbr{2}{3}\sQbr{1}{4})\Sqb{34} $ & $\frac{v}{2}(\stw \Cfhundredfourteen - i\ctw \Cfhundredeleven)$ \\
        \cline{2-3}
        &$\sqb{12}\Anb{34}^2  $&$\frac{v}{2}\stw \Cfhundredsixteen$\\
        \cline{2-3}
        &$(\sQbr{1}{4}\sQbr{2}{4}+(3\leftrightarrow 4))\Anb{34}\Sqb{34} $&$\frac{1}{\sqrt{2}}(\ctw \Cfseven + 2\stw \Cfeight)$\\
        \cline{2-3}
        &$\sQbr{1}{4}\sQbr{2}{4}\aNbr{1}{3}\sQbr{1}{3}+(3\leftrightarrow 4) $&$\frac{1}{\sqrt{2}}(\ctw \Cfseventwo + 2\stw \Cfeighttag)$\\
        \cline{2-3}
        &$\sqb{12}\asb{\three 1 \four}\Anb{34} $&$\frac{i}{\sqrt{2}}\Cfseventag$\\
        \cline{2-3}
        &$\sqb{12}\asb{\four 1 \three}\Anb{34} $&$-\frac{1}{\sqrt{2}}(\ctw \Cfseventag +2\stw \Cfnine)$\\
        \cline{2-3}
        &$\sQbr{1}{3}\Anb{34}\sQbr{2}{4} $&$iv(2\Cfthirtyeight-\Cfthirtyeighttag+2\Cfthirtynine+\Cfthirtyninetag)$\\
        \cline{2-3}
        &$\sQbr{1}{4}\Anb{34}\sQbr{2}{3} $&$iv(\Cfthirtyeighttag-2\Cfthirtyeighttwo+\Cfthirtyninetag-2\Cfthirtyninetwo)$\\
        \cline{2-3}
        \hline
        %%%%%%%%%%%%%%%
\multirow{1}{*}{$\mathcal{A}(\overline{u}DW^+\gamma_+)$}&$\sqb{12}\sQbl{3}{4}^2 $ &$-\frac{-v}{4}\Cfhundredtwelve$ \\
         \cline{2-3}
         &$(\sQbr{1}{3}\sqb{24}+\sQbr{2}{3}\sqb{14})\sQbl{3}{4} $&$\frac{v}{2}(\Cfhundredfourteen+i\Cfhundredeleven)$\\
         \cline{2-3}
         &$\sqb{14}\sqb{24}\sab{\three 4 \three} $&$\frac{1}{\sqrt{2}}(2\ctw \Cfeight-\stw \Cfseven)$\\
         \cline{2-3}
         &$\sqb{14}\sqb{24}\sab{\three 1 \three} $&$\frac{1}{\sqrt{2}}(2\ctw \Cfeight-\stw \Cfseven)$\\
         \hline
         %%%%%%%%%%%%%%%
         $\mathcal{A}(\overline{u}DW^+\gamma_-)$&$\sqb{12}\aNbl{3}{4}^2 $&$\frac{v}{2}\ctw \Cfhundredsixteen$\\
         \cline{2-3}
         &$\sqb{12}\asb{41\three}\aNbl{3}{4} $&$\frac{1}{\sqrt{2}}\stw \Cfseventag-\sqrt{2}\ctw \Cfnine$\\
         \hline
         %%%%%%%%%%%%%%%
$\mathcal{A}(\overline{u}^{\beta}D_{\alpha}W^+g_+^A)$&$\sqb{\three 4}\sqb{1\three }\sqb{24}\ellA$&$\frac{v}{2}\Cfhundredtwentyfour$\\
         \cline{2-3}
         &$\sqb{\three 4}\sqb{14}\sqb{2\three}\ellA$&$\frac{v}{2}\Cfhundredtwentyfive$\\
         \cline{2-3}
         &$\sqb{14}\sqb{24}\sab{\three 4 \three}(\lambda^A)_{\alpha}^{\;\beta}$&$\sqrt{2}\Cften$\\
         \cline{2-3}
         &$\sqb{14}\sqb{24}\sab{\three 1 \three}(\lambda^A)_{\alpha}^{\;\beta}$&$\sqrt{2}\Cftentag$\\
         \hline
         %%%%%%%%%%%%%%%
\multirow{1}{*}{$\mathcal{A}(\overline{u}^{\beta}D_{\alpha}W^+g_-^A)$}&$\sqb{12}\asb{41\three}\aNbl{3}{4}(\lambda^A)_{\alpha}^{\;\beta}$&$-\Cfeleven$\\
         \cline{2-3}
         &$\sqb{12}\anb{\three 4}^2\ellA$& $\frac{v}{2}\Cfhundredtwentysix$\\
         \hline
         %%%%%%%%%%%%%%%
         \multirow{1}{*}{$\mathcal{A}(\overline{d}UW^-Z)$}&\multirow{1}{*}{$\sQbr{1}{3}\sQbr{2}{4}\Sqb{34} $}&$-\frac{iv}{2}(\ctw\Cfseventy+\stw \Cfseventythree)$\\
         \cline{2-3}
         &\multirow{1}{*}{$\sQbr{2}{3}\sQbr{1}{4}\Sqb{34} $}&$\frac{iv}{2}(\ctw\Cfseventy+\stw \Cfseventyfive)$\\
         \cline{2-3}
         &$\sqb{12}\Anb{34}^2 $&$-\frac{v}{2}\stw (\Cfseventyseven+\Cfseventyeight)$\\
         \cline{2-3}
         &$\sQbr{1}{4}\sQbr{2}{4}\sab{\four \three \four} $&$-\frac{1}{\sqrt{2}}(\ctw \Cfone+2\stw \Cffour)$\\
         \cline{2-3}
         &$\sQbr{1}{4}\sQbr{2}{4}\sab{\three 1 \three} $&$-\frac{1}{\sqrt{2}}(\ctw \Cftwo+2\stw \Cffourtag)$\\
         \cline{2-3}
         &$\sQbr{1}{3}\sQbr{2}{3}\sab{\four \three \four} $&$-\frac{i}{\sqrt{2}}\Cfone$\\
         \cline{2-3}
         &$\sQbr{1}{3}\sQbr{2}{3}\sab{\four 1\four} $&$-\frac{i}{\sqrt{2}}\Cftwo$\\
         \cline{2-3}
         &$\sqb{12}\asb{\four1\three}\Anb{34} $&$\frac{1}{\sqrt{2}}(\ctw \Cfthree +2\stw \Cffive)$\\
         \cline{2-3}
         &$\sqb{12}\asb{\three 1 \four}\Anb{34} $&$-\frac{i}{\sqrt{2}}\Cfthree$\\
         \cline{2-3}
         &$\sQbr{1}{4}\Anb{34}\sQbr{2}{3} $&$i\sqrt{2}v (\Cfthirtysix+2\Cfthirtysixtag+\Cfthirtyseven-2\Cfthirtyseventagtwo)$\\
         \cline{2-3}
         &$\sQbr{1}{3}\Anb{34}\sQbr{2}{4} $&$-i\sqrt{2}v (\Cfthirtysix-2\Cfthirtysixtag-\Cfthirtyseven+2\Cfthirtyseventag)$\\
         \hline
         %%%%%%%%%%%%%%%
         
         \multirow{1}{*}{$\mathcal{A}(\overline{d}UW^-\gamma_+)$}&\multirow{1}{*}{$\sQbr{1}{3}\sqb{24}\sQbl{3}{4} $}&$\frac{v}{2\sqrt{2}}(\stw \Cfseventy - \ctw \Cfseventythree)$\\
         \cline{2-3}
         &\multirow{1}{*}{$\sqb{14}\sQbr{2}{3}\sQbl{3}{4} $}&$\frac{v}{2\sqrt{2}}(\stw \Cfseventy - \ctw \Cfseventyfive)$\\
         \cline{2-3}
         &$\sqb{14}\sqb{24}\sab{\three 4 \three} $&$\frac{1}{\sqrt{2}}(\stw \Cfone - 2 \ctw \Cffour)$\\
         \cline{2-3}
         &$\sqb{14}\sqb{24}\sab{\three 1 \three} $&$\frac{1}{\sqrt{2}}(\stw \Cftwo - 2 \ctw \Cffourtag)$\\
         \hline
         %%%%%%%%%%%%%%%
         \multirow{1}{*}{$\mathcal{A}(\overline{d}UW^-\gamma_-)$}&$\sqb{12}\aNbl{3}{4}^2 $&$-\frac{1}{\sqrt{2}}\ctw (\Cfseventyseven+\Cfseventyeight)$\\
         \cline{2-3}
         &$\sqb{12}\anb{14}\sQbr{1}{3}\aNbl{3}{4} $&$-\frac{1}{\sqrt{2}}(\stw\Cfthree- 2 \ctw \Cffive)$\\
         \hline
         %%%%%%%%%%%%%%%
         \multirow{1}{*}{$\mathcal{A}(\overline{d}^{\beta}U_{\alpha}W^-g^A_+)$}&$\sQbr{1}{3}\sqb{24}\sQbl{3}{4}(\lambda^A)_{\alpha}^{\;\beta} $&$-\frac{v}{2}\Cfeightysix$\\
        \cline{2-3}
        &$\sqb{14}\sQbr{2}{3}\sQbl{3}{4}(\lambda^A)_{\alpha}^{\;\beta} $&$-\frac{v}{2}\Cfeightyeight$\\
        \cline{2-3}
        &$\sqb{14}\sqb{24}\sab{\three 4 \three}(\lambda^A)_{\alpha}^{\;\beta} $&$-\sqrt{2}\Cfsix$\\
        \cline{2-3}
        &$\sqb{14}\sqb{24}\sab{\three 1 \three}(\lambda^A)_{\alpha}^{\;\beta} $&$-\sqrt{2}\Cfsixtwo$\\
        \hline
        %%%%%%%%%%%%%%%
        \multirow{1}{*}{$\mathcal{A}(\overline{d}^{\beta}U_{\alpha}W^-g^A_-)$}&$\sqb{12}\aNbl{3}{4}^2(\lambda^A)_{\alpha}^{\;\beta} $&$-\frac{v}{2}\Cfninety$\\
        \cline{2-3}
        &$\sqb{12}\asb{14\three}\aNbl{3}{4}(\lambda^A)_{\alpha}^{\;\beta} $&$\sqrt{2}\Cfsixtag$\\
        \hline
        %%%%%%%%%%%%%%%
        \caption{LE chirality-violating CTs containing two fermions and two gauge bosons. Two-gluon states are omitted as they are in one-to-one correspondence with the HE amplitudes.}
\label{ChVio}
\end{longtable}
\end{landscape}
%%%%%%%%%%%%%%%%%%%%%%%%%%%%%%%%%%%%%%%%%%%%%%%%%%%%%%%%%
%%%%%%%%%%%%%%%%%%%%%%%%%%%%%%%%%%%%%%%%%%%%%%%%%%%%%%%%%%
\begin{longtable}{|c|c|c|c|c|c|}
\hline
Structure & HEFT & $\overline{u}uW^+W^-$ & $\overline{d}dW^+W^-$&$\overline{u}dW^-Z$&$\overline{d}uW^+Z$\\
         \hline \hline
         \endfirsthead
         \hline
         \multicolumn{6}{|c|}{Continued from Table~\ref{T12}}\\
         \hline
         Structure & HEFT & $\overline{u}uW^+W^-$ & $\overline{d}dW^+W^-$&$\overline{u}dW^-Z$&$\overline{d}uW^+Z$\\
         \hline \hline
         \endhead
         \hline
         \endfoot
         \endlastfoot
$\sqb{12}\asb{\four 1 \three}\Anb{34}$&$\checkmark(7)$&$-$&$\checkmark$&$\checkmark$&$\checkmark$\\
         \hline
         %%%%%%%%%%%%%%%
         $\sqb{12}\asb{\three 1 \four}\Anb{34}$&$\checkmark(7)$&$\checkmark$&$-$&$\checkmark$&$\checkmark$\\
         \hline
         %%%%%%%%%%%%%%%
         $\sQbr{1}{4}\sQbr{2}{4}\{\sab{\three 1 \three}\;; \sab{\three \four \three}\}$&$\checkmark(7)$&$-$&$\checkmark$&$\checkmark$&$-$\\
         \hline
         %%%%%%%%%%%%%%%
         $\sQbr{1}{3}\sQbr{2}{3}\{\sab{\four 1 \four}\;; \sab{\four \three \four}\}$&$\checkmark(7)$&$\checkmark$&$-$&$-$&$\checkmark$\\
         \hline
         \caption{%Discrepancies between 
         Differences between the HEFT and SMEFT chirality-violating CTs in $\overline{f}fW^+W^-,\;f^cf'W^{\pm}Z$. 
         Parentheses indicate the minimal dimension at which structures are generated in the HEFT. 
         A check mark means a structure is generated for the listed amplitude, while a dash means it is not.}
         \label{T12}
         \end{longtable}
%%%%%%%%%%%%%%%%%%%%%%%%%%%%%%%%%%%%%%%%%%%%%%%%%%%%%%%%%%
%%%%%%%%%%%%%%%%%%%%%%%%%%%%%%%%%%%%%%%%%%%%%%%%%%%%%%%%%%%%%%%%%%%%
%
\bibliographystyle{JHEP}
\bibliography{references}
\end{document}